\documentclass[12pt, oneside]{article} 
\usepackage[margin=1.25in]{geometry}
\usepackage{graphicx}
\usepackage{float}
\newsavebox{\figurepanel}
\usepackage{amssymb}

\usepackage{amsmath}
\usepackage{bbm}
\usepackage{setspace}
\usepackage{amsthm}
\usepackage{color}
\usepackage[svgnames]{xcolor}
\usepackage{booktabs}
\usepackage{array} 
\usepackage{caption}
\usepackage[authoryear,round,longnamesfirst,numbers]{natbib}
\usepackage{diagbox}
\usepackage{enumitem}
\usepackage[]{todonotes}
\usepackage{comment}

\usepackage{eurosym}

\usepackage{makecell}
\newcolumntype{C}{>{\centering\arraybackslash $}p{1.3cm}<{$}}
\newcolumntype{D}{>{\centering\arraybackslash $}p{1.4cm}<{$}}
\newcolumntype{S}{>{\centering\arraybackslash $}p{0.8cm}<{$}}

\usepackage{subcaption}
\usepackage{hyperref}
\usepackage[capitalize,nameinlink]{cleveref}
\makeatletter
\renewcommand*\caption@@xlabel[2]{%
  \caption@@@xlabel
  \def\caption@labelname{#2}%
  \cref@label#1{#2}}
\makeatother
\hypersetup{
colorlinks,breaklinks,naturalnames,
citecolor=DarkBlue,urlcolor=red,linkcolor=DarkBlue
}

\usepackage{pgf}
\usepackage{tikz}
\usetikzlibrary{decorations.markings}
\usetikzlibrary{patterns,hobby}
\usetikzlibrary{shapes.misc, positioning}
\usetikzlibrary{positioning,shapes.geometric,arrows.meta, arrows,calc, shadows,decorations.pathreplacing}
\usepackage{pgfplots}

\usepackage[customcolors]{hf-tikz}
\usepgfplotslibrary{fillbetween} 
\pgfplotsset{compat=1.18}

\newtheorem{lemma}{Lemma}
\newtheorem{preliminarylemma}{Lemma}[section]
\newtheorem{proposition}{Proposition}
\newtheorem{corollary}{Corollary}
\newtheorem{theorem}{Theorem}

\newtheorem{assumption}{Assumption}

\makeatletter
\DeclareRobustCommand{\iscircle}{\mathord{\mathpalette\is@circle\relax}}
\newcommand\is@circle[2]{%
  \begingroup
  \sbox\z@{\raisebox{\depth}{$\m@th#1\bigcirc$}}%
  \sbox\tw@{$#1\square$}%
  \resizebox{!}{\ht\tw@}{\usebox{\z@}}%
  \endgroup
}
\makeatother

\renewcommand{\epsilon}{\varepsilon}

\definecolor{Burgundy}{RGB}{206,5,40}
\definecolor{Rust}{RGB}{200,10,10}
\definecolor{UCSDblue}{RGB}{55,80,200}

\title{How to Mismatch Consumers\footnote{I am indebted to Nageeb Ali and Nima Haghpanah for their guidance and support. I thank Alessandro Bonatti, Kalyan Chatterjee, Yuhta Ishii, Vijay Krishna, Ran Shorrer, Ron Siegel, and Alex Smolin for helpful discussions, as well as seminar and conference participants for their comments.}}

\usepackage{needspace}

\begin{document}

\author{Wenhao Li\thanks{HKU Business School (e-mail: \href{mailto: wenhao.econ@gmail.com}{wenhao.econ@gmail.com}).} }

\date{this version: November 7, 2026}


\maketitle

\begin{abstract}
Should a platform show all consumers all products, or at least those they like most? Or could it raise consumer surplus by mismatching some consumers: showing them only less-preferred products or even excluding them from trade? I show that in both monopoly and duopoly, mismatching can raise aggregate consumer surplus by inducing lower prices. In duopoly, consumer-optimal matching uses a cutoff policy: along each diagonal of equal valuation differences, consumers with lower valuations are matched only to less-preferred products, and some may subsequently be excluded. With independent uniform valuations, this raises aggregate consumer surplus by about 41\% over full matching.
\end{abstract}

\thispagestyle{empty}
\clearpage
\setcounter{page}{1}

\section{Introduction} \label{section: introduction}
\begingroup\emergencystretch=1em
Consider a platform that knows consumers' values for a range of products and chooses which products to show them. Should the platform reveal all the products or the consumer's favorite products, or should it mismatch some consumers revealing only those products that the consumer likes less? One's first intuition may be that if the platform prioritizes consumer surplus, it ought to reveal all the products or at least the consumer's favorite. Revealing all products, after all, would seem to be the best way to intensify market competition as no firm would then presume to have a captive consumer. Perhaps surprisingly, my analysis shows that the best way to raise consumer surplus may actually involve mismatchings: revealing to a consumer only those products she likes less. Such distortions can trigger price discounts that raise consumer surplus.

The basic economic force can be seen even in a monopolistic market. Consider a single firm interacting with a continuum of consumers whose valuations are uniformly distributed on $[0,1]$. Suppose for simplicity that the cost of production is zero. The platform observes each consumer's valuation and decides whether to grant her access to the firm's product. Consumers who are excluded obtain a payoff of zero.

First, suppose that the platform grants all consumers access to the monopoly, as in Panel (a) in Figure~\ref{fig:IntroImprove}. As a result, the monopoly sets a price at $1/2$. Only those consumers who value the product above this price will make a purchase and thus obtain a positive surplus.

Now, look at Panel (b) in the same figure. Here, the platform excludes consumers whose values lie between $13/20$ and $15/20$. By doing so---excluding $2/20$ of consumers---the platform prompts the monopoly to lower its price to $9/20$. The intuition is that by removing these customers, the firm has less to lose (relative to if it accessed all consumers) from lowering prices below $1/2$.\footnote{Why this price reduction? When the aforementioned set of consumers is excluded, the monopoly has no incentives to set prices within that range. Breaking it down: For price $p$ under $13/20$, the monopoly's profit would be $p(1-p-2/20)$, taking into account the excluded $2/20$ consumers. This function peaks at $9/20$ with a maximum $81/400$. For prices above $15/20$, the optimal profit is $(15/20)(1-15/20)=75/400$, which is realized at precisely $15/20$.} As a result of this price reduction, consumers in the two blue regions benefit from the price drop. Those in the mint region can now afford to trade with the monopoly, thereby gaining a positive surplus. On the other hand, the excluded consumers between $13/20$ and $15/20$ are worse off. But taking everything into account, the net gain in consumer surplus is strictly positive.\footnote{The increase in consumer surplus is $(1/2-2/20)(1/20)+(1/20)(9.5/20-9/20)$. The decrease, due to excluding certain consumers, is $(2/20)(14/20-1/2)$. The net gain in consumer surplus ends up being $(1/20)(0.5/20)$. In fact, in this uniform example, for each cutoff $c\in(1/2,3/4)$, there exists sufficiently small $\epsilon$ so that by excluding consumers in interval $[c,c+\epsilon]$, the platform can induce a strictly lower monopoly price and a strictly larger consumer surplus.}

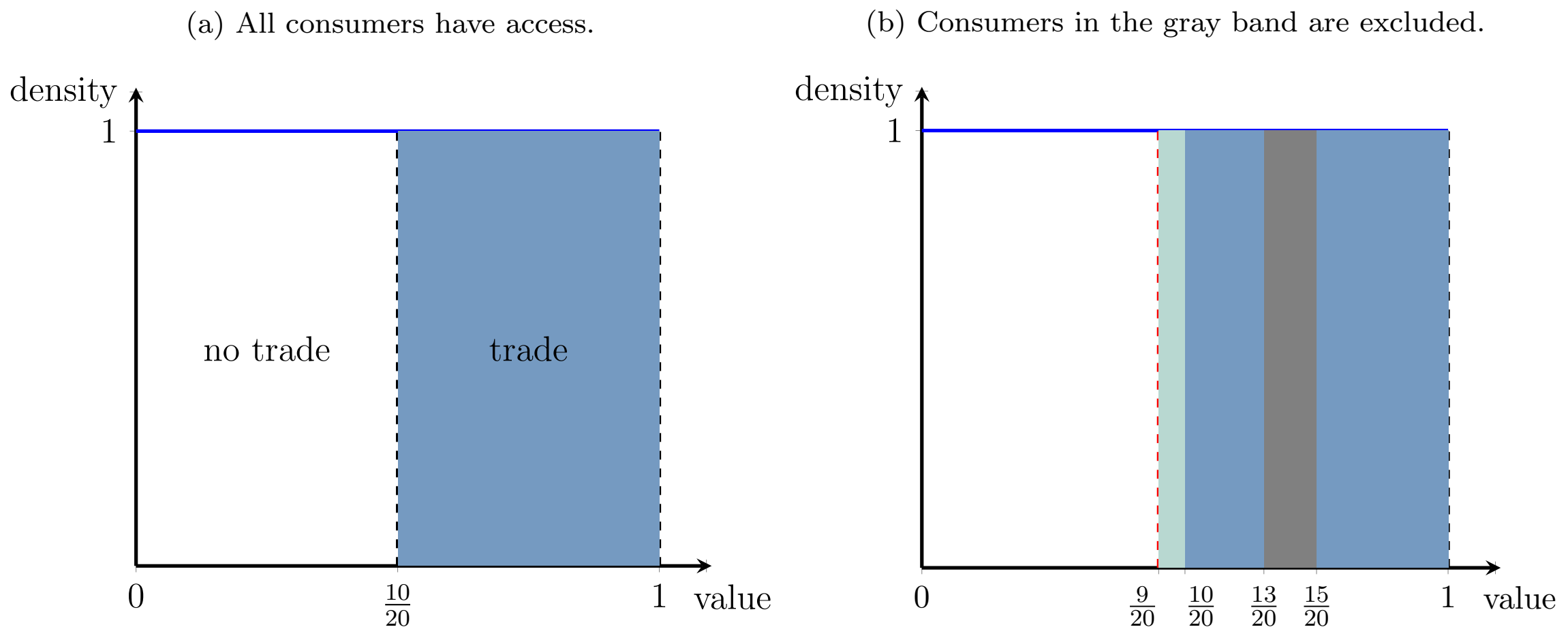
\begin{figure}[!htbp]
\centering
\definecolor{IntroBoth}{HTML}{B8D8D1}
\definecolor{IntroFirmOne}{HTML}{759AC1}
\captionsetup[subfigure]{position=top,font=scriptsize}

\begin{subfigure}{0.5\textwidth}
\centering
\caption{All consumers have access.}
\begin{lrbox}{\figurepanel}
\begin{tikzpicture}
\begin{axis}[
    axis lines = left,
    axis line style={very thick},
    xmin=0, xmax=1.1,
    ymin=0, ymax=1.1,
    xtick=\empty,
    ytick=\empty,
    xtick={0,0.5,1,1.09},       
    ytick={1, 1.09},       
    xticklabels={$0$, $\frac{10}{20}$,$1$, \hspace{1.5em}value},  
    yticklabels={$1$, density},  
]


\addplot[ dashed, very thick] coordinates {(0.5,0) (0.5,1)};
\addplot[dashed, very thick] coordinates {(1,0) (1,1)};
\addplot[blue,very thick] coordinates {(0,1) (1,1)};

\node[] at (axis cs:0.25,0.5) {no trade};


\fill[IntroFirmOne] (axis cs:0.5,0) rectangle (axis cs:1,1);

\node[] at (axis cs:0.75,0.5) {trade};

\end{axis}
\end{tikzpicture}
\end{lrbox}%
\resizebox{\linewidth}{!}{\usebox{\figurepanel}}
\end{subfigure}%
\begin{subfigure}{0.5\textwidth}
\centering
\caption{Consumers in the gray band are excluded.}
\begin{lrbox}{\figurepanel}
\begin{tikzpicture}
\begin{axis}[
    axis lines = left,
    axis line style={very thick},
    xmin=0, xmax=1.1,
    ymin=0, ymax=1.1,
    xtick=\empty,
    ytick=\empty,
    xticklabel style={font=\small},
    xtick={0,0.45, 0.5,0.65, 0.75, 1,1.09},       
    ytick={1, 1.09},       
    xticklabels={$0$,\makebox[0pt][r]{$\frac{9}{20}$}, \makebox[0pt][l]{$\frac{10}{20}$},$\frac{13}{20}$, $\frac{15}{20}$,  $1$, \hspace{1.5em}value},  
    yticklabels={$1$, density},  
]


\addplot[blue,very thick] coordinates {(0,1) (1,1)};
\addplot[dashed,very thick] coordinates {(0.5,0) (0.5,1)};
\addplot[dashed, red ,very thick] coordinates {(0.45,0) (0.45,1)};
\addplot[dashed,very thick] coordinates {(1,0) (1,1)};

\fill[gray] (axis cs:0.65,0) rectangle (axis cs:0.75,1);

\fill[IntroBoth] (axis cs:0.45,0) rectangle (axis cs:0.5,1);

\fill[IntroFirmOne] (axis cs:0.5,0) rectangle (axis cs:0.65,1);
\fill[IntroFirmOne] (axis cs:0.75,0) rectangle (axis cs:1,1);



\end{axis}
\end{tikzpicture}
\end{lrbox}%
\resizebox{\linewidth}{!}{\usebox{\figurepanel}}
\end{subfigure}

\caption{Excluding Consumers May Raise Consumer Surplus}
\label{fig:IntroImprove}
\end{figure}

The example shows that excluding some consumers can increase aggregate consumer surplus. But what is the consumer-optimal exclusion policy? The consumer-optimal policy excludes intermediate-value consumers in the interval $[0.443,0.731]$. However, instead of excluding these consumers with probability one, the platform chooses their exclusion probabilities so that the monopolist is willing to reduce its price to approximately $0.443$ and is indifferent between this price and higher prices in the interval. Figure~\ref{fig:IntroMonopolyOptimal} illustrates the resulting demand curve, with price on the vertical axis and demand on the horizontal axis. Under full access, the monopolist faces the linear demand curve shown in blue. Under the consumer-optimal exclusion policy, the induced demand curve, shown in red, coincides with the unit-elastic demand curve between $0.443$ and $0.731$, generating constant monopoly profit of approximately $0.197$. The policy increases consumer surplus from $1/8$ to approximately $0.134$.

\begin{figure}[H]
\centering
\includegraphics{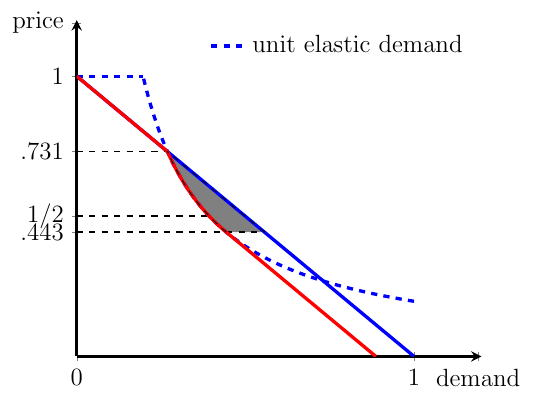}
\caption{Consumer-Optimal Access Policy}
\label{fig:IntroMonopolyOptimal}
\end{figure}

The consumer-optimal exclusion policy has a demand-preserving property: among all exclusion policies that leave the monopolist with the same maximum profit, it generates the greatest demand at every price. I show that such a demand-preserving exclusion policy exists for arbitrary valuation distributions and any targeted reduction in the monopolist's profit. Moreover, it maximizes consumer surplus among all policies yielding the same monopoly profit. The platform then only needs to choose how much profit to leave the monopolist. Consequently, a consumer-optimal exclusion policy can always be chosen to be demand-preserving.

The example in Figure~\ref{fig:IntroImprove} seems to suggest that improving aggregate consumer surplus requires sacrificing some consumers: those excluded lose surplus, even though others benefit from lower prices. But this is not necessary. Instead of excluding selected consumers with certainty, the platform can spread exclusion probabilistically across a larger group of consumers. Such a policy can still induce a lower monopoly price, while ensuring that the benefit from the price reduction outweighs each consumer's expected loss from possible exclusion. As a result, the platform can achieve an interim Pareto improvement: every consumer is weakly better off in expected surplus conditional on her valuation, and some consumers are strictly better off.

More generally, which exclusion policies are interim Pareto efficient for consumers? Given any exclusion policy and its induced monopoly price, can the platform find another exclusion policy inducing a different price that makes every consumer weakly better off and some strictly better off? At that induced price, guaranteeing that no consumer is worse off is equivalent to imposing an upper bound on each consumer's exclusion probability. The problem can therefore be formulated as a capacity-constrained monopoly problem: choose exclusion probabilities subject to these individual bounds while inducing the target price. I solve this problem by generalizing the idea of demand-preserving exclusion: among all exclusion policies satisfying the capacity constraints and leaving the monopolist with the same maximum profit, a demand-preserving exclusion policy exists, and it maximizes consumer surplus. Consequently, to determine whether an exclusion policy is interim Pareto efficient, we only need to compare it with demand-preserving exclusion policies.

Although improving consumer surplus need not come at the expense of any consumer, it generally comes at the cost of social surplus and trade quantity. Under log-concave value distributions, full access maximizes both social surplus and trade quantity. Any exclusion policy that strictly improves consumer surplus necessarily reduces both.

In duopoly, however, improving consumer surplus need not come at the expense of social surplus or trade quantity.

I consider the following duopoly setting. Two firms each sell a differentiated product to a continuum of consumers with unit-demand and heterogeneous valuations. For simplicity, marginal costs are zero. Before firms choose prices, the platform commits to a matching rule that determines which products each consumer can access: both products, either product alone, or neither. Firms observe the rule, but not individual valuations, and then choose uniform prices. Consumers learn their exact values for the products they can access and choose among them. They can always choose not to purchase any product and obtain zero payoff, but cannot independently access omitted products. I assume a symmetric, log-concave distribution of valuations and restrict attention to symmetric matchings and symmetric pure-strategy equilibria.

Unlike in monopoly, the platform can induce lower prices without excluding consumers. Suppose the platform wants to induce a symmetric equilibrium price $p$ slightly below the smallest full-matching equilibrium price $p^*$. Under full matching, a firm would prefer to deviate to a price higher than $p$ when its rival charges $p$.\footnote{Under log-concavity, each firm's full-matching best response is unique and continuous. Since $p^*$ is the smallest symmetric full-matching equilibrium price, the best response to any $p\in(0,p^*)$ exceeds $p$.} To deter such upward deviations, the platform can mismatch selected consumers by showing them only their less-preferred products.

Figure~\ref{fig:IntroMismatchDemand} illustrates how the platform can use mismatching to deter upward price deviations and induce the lower symmetric equilibrium price $p=1/3$ under independent uniform valuations. In the left panel, the platform matches consumers in the blue region only to firm 1 and those in the yellow region only to firm 2, even though these consumers prefer the other firm's product. The blue region is the reflection of the yellow region across the diagonal $v_1=v_2$.\footnote{The blue region is defined by $0\le v_1-p\le 1/16$ and $1/4\le v_2-v_1\le 7/12$.} All other consumers retain access to both products. The right panel shows how this mismatching policy shapes firm $i$'s demand when it sells at price $q$ and its rival charges $p$. Relative to full matching, mismatching leaves demand unchanged at $q=p$ and weakly reduces it at every higher price.

To understand why, consider firm 1. At price $p$, relative to full matching, firm 1 loses consumers in the yellow region but gains those in the blue region. Since the two regions have equal mass, its sales at $p$ remain unchanged. Under full matching, a consumer $v=(v_1,v_2)$ in the yellow region would continue buying from firm 1 after a price increase of up to $v_1-v_2$. By contrast, her mirror type $v'=(v_2,v_1)$ in the blue region, who is matched only to firm 1, stops purchasing when the price increase exceeds $v'_1-p=v_2-p$. By construction, $v_1-v_2>v_2-p$. Thus, at the target price $p$, compared with full matching, firm 1 loses less price-sensitive consumers and gains more price-sensitive consumers, making upward price deviations less attractive. Indeed, one can verify that no unilateral price deviation is profitable under this mismatching policy, so $p$ is a symmetric equilibrium price. Paradoxically, mismatching makes these consumers captive to individual firms yet intensifies price competition, because it makes firms' demand more sensitive to price increases.

\begin{figure}[H]
\centering
\includegraphics[width=0.98\textwidth]{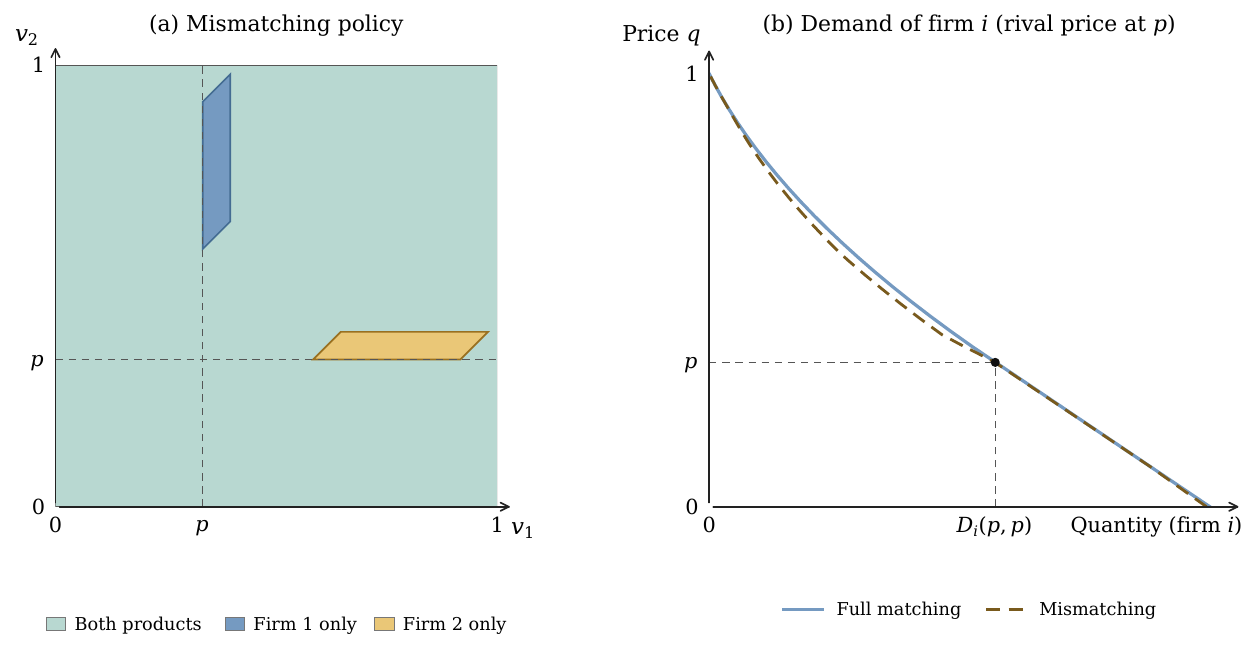}
\caption{Mismatching can shape demand and induce the target equilibrium price $p=1/3$.}
\label{fig:IntroMismatchDemand}
\end{figure}

This explains why, in duopoly, improving consumer surplus need not come at the expense of social surplus or trade quantity: the platform can induce lower equilibrium prices through mismatching rather than exclusion. I show that when the target price $p$ is sufficiently close to $p^*$, mismatching only a small mass of suitably selected consumers is enough to sustain the lower price as an equilibrium. Moreover, I show that a suitably randomized mismatching policy can simultaneously increase social surplus and trade quantity while generating an interim Pareto improvement for consumers.

How would a consumer-oriented platform design its matching policy? I show that there exists a consumer-optimal matching that combines such symmetric mismatching with exclusion, inducing an equilibrium price $p^o$ below the full-matching equilibrium price $p^*$. This matching can be constructed in two steps. First, omit products valued below $p^o$. Then, among consumers who value both products above $p^o$, apply a cutoff policy along each diagonal of equal valuation differences: consumers below the cutoff are matched only with their less-preferred product, while those above it retain access to both products. Second, exclude selected consumers who now consider only one firm. These may be consumers who initially valued only one product above $p^o$ or consumers mismatched in the first step.

In the second step, excluding consumers who consider only one firm changes that firm's demand without affecting its rival's demand. Thus, this exclusion problem reduces to a monopoly problem with capacity constraints: exclusion is limited to consumers who consider only that firm after Step 1. Therefore, I can apply the demand-preserving exclusion rule to solve this problem, completing the construction of the consumer-optimal matching.

Figure~\ref{fig:IntroOptimal} illustrates the two-step characterization under independent uniform valuations on $[0,1]$. In Step 1, consumers in the gray region are excluded because they value both products below $p^o$. The blue region represents consumers shown only firm 1: some value only firm 1's product above $p^o$, while others value both products above $p^o$ but are mismatched to their less-preferred product. The yellow region is symmetric, representing consumers shown only firm 2. The mint-green region represents consumers who retain access to both products. The curved boundaries between the blue and mint-green regions and between the yellow and mint-green regions illustrate the cutoff structure: along each diagonal of equal valuation differences, consumers with lower valuations are mismatched, while those with higher valuations retain access to both products. In Step 2, additional consumers shown only one product are excluded according to the demand-preserving exclusion rule, as indicated by the additional gray regions in the right panel.

This matching yields consumer surplus about 41 percent above the full-matching equilibrium. Trade quantity also increases,\footnote{Trade quantity rises from about 82.8\% under full matching to 88.0\% under the constructed matching.} despite the exclusion of some consumers: the lower price enables enough additional purchases to offset those lost through exclusion.

\begin{figure}[H]
\centering
\includegraphics[width=0.96\textwidth]{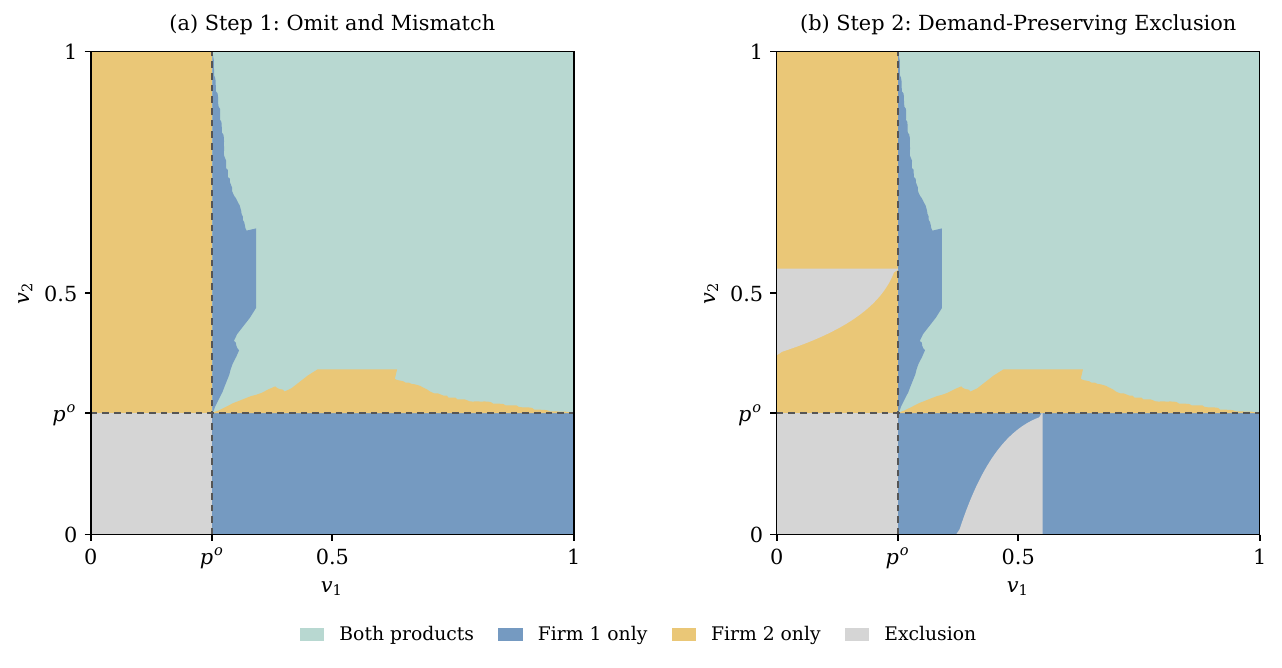}
\caption{Two-Step Structure of Consumer-Optimal Matching}
\label{fig:IntroOptimal}
\end{figure}

The cutoff has a simple economic interpretation. Among consumers with the same preference difference, mismatching entails the same loss in match quality. However, mismatching a consumer with lower valuations gives the assigned firm a customer with a lower willingness to pay, making price increases less attractive. Thus, at each preference difference, consumers with lower valuations are mismatched in Step 1 and may subsequently be excluded in Step 2, while those with higher valuations retain access to both products. Showing only the preferred product is never needed when both values exceed the equilibrium price $p^o$, even though such an assignment sacrifices no consumer surplus at that price and can help deter downward deviations.

Besides consumer surplus, a benevolent platform may seek to maximize social surplus or trade quantity. I show that, for either objective, an optimal matching takes an even simpler form if values are independent: it uses only the first step of the consumer-optimal construction, without subsequent exclusion. The trade-quantity-optimal matching takes a particularly simple form. Denote the resulting symmetric equilibrium price by $p^T$. The cutoff for the lower valuation is simply $|v_i-v_j|+p^T$. A consumer of type $v$ with $v_i>v_j>p^T$ is matched only to product $j$ if $v_j<v_i-v_j+p^T$; otherwise, she retains access to both products.

Figure~\ref{fig:IntroSocialTrade} illustrates the social-optimal matching in panel (a) and the trade-optimal matching in panel (b) under independent uniform valuations. In both panels, consumers in the gray region are excluded because they value both products below the equilibrium price. Consumers in the blue and yellow regions are shown only firm 1 and firm 2, respectively, while those in the mint-green region retain access to both products. In panel (a), the curved boundaries between the mint-green region and the blue and yellow regions illustrate the cutoff structure along each diagonal of equal valuation differences. In panel (b), the analogous boundaries are simply straight lines. Neither policy excludes consumers who would otherwise purchase at the resulting equilibrium price. The social-optimal matching induces a symmetric equilibrium price of $p^S\approx0.311$ and increases social surplus by approximately $3.35\%$ relative to full matching. The trade-optimal matching induces $p^T=2-\sqrt{3}\approx0.268$, increasing trade quantity by approximately $12.0\%$ relative to full matching.

\begin{figure}[H]
\centering
\includegraphics[width=0.98\textwidth]{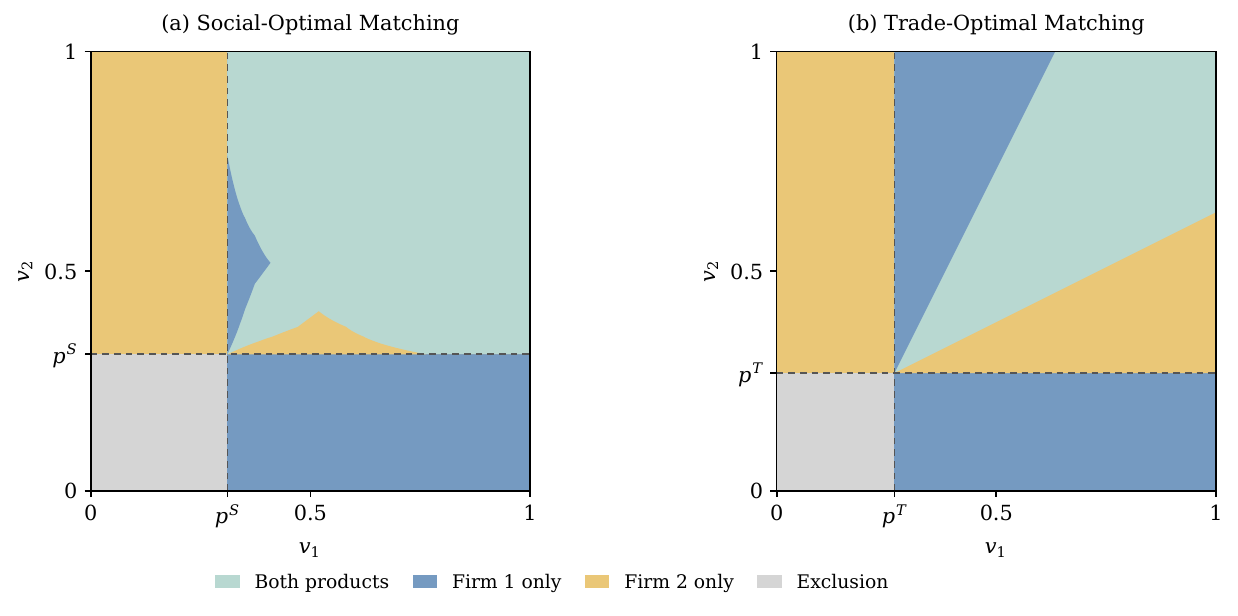}
\caption{Social-Optimal Matching and Trade-Optimal Matching}
\label{fig:IntroSocialTrade}
\end{figure}

So far, I have considered a platform that chooses a matching policy to pursue its own objectives. An alternative interpretation is that a single consumer (instead of a continuum of consumers) commits in advance, before her valuation is realized, to a valuation-dependent attention or awareness policy that restricts her future choices. Such a commitment can benefit her by inducing firms to offer lower prices. This paper characterizes the optimal commitment policy.

Throughout the paper, matching is independent of firms' prices. If the platform could freely condition access on posted prices, it could implement marginal-cost pricing when firms' costs are known: simply grant full access when both firms price at marginal cost and completely exclude any firm that deviates. With privately known marginal costs, however, allowing matching to depend on prices fundamentally changes the problem. This is especially clear in monopoly, where the platform effectively offers a menu of price--access policy pairs from which the firm chooses, turning the problem into one of screening. The present model instead isolates a different mechanism: how a platform can use exclusion and mismatching to influence firms' prices, thereby improving consumer surplus, social surplus, and trade quantity, and even generating interim Pareto improvements for consumers.

The analysis can be viewed as a normative exercise in which a benevolent platform seeks to maximize consumer surplus, social surplus, or trade quantity. Whether a revenue-maximizing platform pursues similar objectives depends on how it earns revenue. For example, consumer access fees can allow the platform to capture expected consumer surplus, commissions on sellers' revenues can capture part of producer surplus, and ancillary revenue from transactions can reward greater trade. Competition between platforms can also create incentives to increase consumer surplus to attract and retain consumers. Thus, a revenue-maximizing platform may find it profitable to restrict product access for the same reasons that a benevolent platform would.

\paragraph{Related literature.} This paper contributes to a literature showing that giving consumers more information or better product choices need not benefit them, because firms respond by changing their prices. \citet{armstrong2022consumer} show that consumer-optimal information may induce purchases of less-preferred products to intensify price competition. \citet{armstrong2022patterns} study how patterns of consumer consideration affect competition and show that greater awareness need not benefit consumers. \citet{condorelli2022buyer} characterize buyer-optimal matching, where distorting matches can benefit buyers by influencing post-match prices. My analysis builds on these insights but studies a different problem: how a platform should optimally design consumers' consideration sets when firms charge uniform prices. I show that matching consumers to less-preferred products can intensify competition, even though it makes these consumers captive to individual firms, and characterize the resulting optimal matching policies.

A related literature examines how platforms can influence sellers' pricing incentives by conditioning product exposure on their prices. \citet*{dinerstein2018consumer} study the trade-off between helping consumers discover preferred products and encouraging sellers to lower prices. \citet{ichihashi2025recommendations} study optimal price-contingent recommendations with privately informed sellers and establish a connection to mechanism design. In contrast, the platform here commits to matching independently of prices. Rather than rewarding price reductions with greater exposure, it influences firms' pricing incentives by shaping the demand they face.

Another approach to improving consumer welfare is to design the information available to firms, facilitating price discrimination. \citet*{bergemann2015limits} characterize the surplus outcomes achievable through monopoly market segmentation, while \citet*{ali2020voluntary} show how voluntary disclosure can benefit consumers under personalized pricing. \citet*{elliott2022market} characterize information design under price competition and establish that, when information can be freely designed, restricting product access cannot further improve the consumer-optimal outcome. The present paper instead studies uniform pricing, where firms cannot tailor prices to individual consumers. In this setting, restricting product access provides a distinct instrument for improving consumer welfare.

The monopoly analysis relates to \citet{roesler2017buyer} and \citet{condorelli2020information}, whose optimal designs generate unit-elastic demand through consumer learning and the choice of valuation distributions, respectively. Here the valuation distribution is fixed, and consumers learn their values exactly when granted access. The platform can only exclude consumers, yet a consumer-optimal exclusion policy also generates unit-elastic demand over a range of prices.

\endgroup

\section{Monopoly}\label{sec:monopoly}

\subsection{Environment}\label{sec:monopoly-environment}

A monopolist sells a product at zero marginal cost to a unit mass of consumers, each demanding at most one unit. Consumers have valuations $v$ distributed according to a cumulative distribution function $F$ on $[\underline v,\bar v]\subseteq\mathbb R_+$. Let $f$ denote its density. A consumer with valuation $v$ who purchases at price $p$ obtains utility $v-p$; otherwise, she obtains zero.

We introduce a platform that controls consumers' access to the product. Before the monopolist sets its price, the platform publicly commits to a measurable matching policy $\phi:[\underline v,\bar v]\to[0,1]$, where $\phi(v)$ is the probability that a consumer with valuation $v$ is granted access. The monopolist observes the matching policy but not individual valuations and chooses a uniform price. The platform then observes consumers' valuations and implements the matching policy. Consumers who are granted access learn their valuations and decide whether to purchase; excluded consumers cannot purchase the product and receive a payoff of zero.

Given a matching policy $\phi$, demand at price $p$ is $D_\phi(p)=\int_p^{\bar v}\phi(v)\,dF(v)$, and monopoly profit is $\Pi_\phi(p)=pD_\phi(p)$. An optimal price maximizes $\Pi_\phi$. When several prices are optimal, we evaluate the matching at the lowest optimal price.\footnote{Since $F$ is atomless and has compact support,
$\Pi_\phi$ is continuous on $[0,\bar v]$ for every
matching $\phi$. Its set of maximizers is therefore
nonempty and compact, so a lowest optimal price exists.}

The interim consumer surplus of a consumer with valuation $v$ is $u(v;\phi,p)=\phi(v)(v-p)_+$. Aggregate consumer surplus is
\[
CS(\phi,p)=\int_{\underline v}^{\bar v}u(v;\phi,p)\,dF(v).
\]
Social surplus is $SS(\phi,p)=\int_p^{\bar v}v\phi(v)\,dF(v)$, while trade quantity is simply $D_\phi(p)$.

Under full matching, denoted by $\phi^{\mathrm{full}}$, every consumer has access to the product, so $\phi^{\mathrm{full}}(v)=1$ for every $v$. Let $D(p)=1-F(p)$ denote full-matching demand and $\Pi^F(p)=pD(p)$ the corresponding monopoly profit. Write $p^M$ for the lowest profit-maximizing price under full matching.

The platform chooses a matching policy to influence the monopolist's price and the resulting market outcome. We first ask whether restricting access can improve consumer surplus relative to full matching.

\subsection{Exclusion Can Benefit Consumers}
\label{sec:monopoly-improvement}

At a fixed price, excluding consumers cannot increase consumer
surplus. But exclusion changes the monopolist's demand and may
induce it to lower its price, benefiting consumers who continue
to purchase. Can this price effect outweigh the consumer-surplus
loss from exclusion?

\begin{assumption}[Regular Monopoly Benchmark]\label{ass:monopoly-improvement}
The density $f$ is continuously differentiable and positive
on the interior of its support. The full-matching monopoly
profit $\Pi^F$ has a unique maximizer $p^M>\underline v$ and
$(\Pi^F)''(p^M)<0$.
\end{assumption}

Full matching is \emph{improvable} if some matching yields strictly higher consumer surplus than full matching under our lowest-optimal-price convention.

\begin{proposition}
\label{prop:fullisImprovable}
Under Assumption~\ref{ass:monopoly-improvement}, full matching is improvable.
\end{proposition}

\begin{figure}[H]
\centering

\begin{subfigure}{0.5\textwidth}
\centering
\begin{lrbox}{\figurepanel}
\begin{tikzpicture}
\begin{axis}[
    axis lines = middle,
    axis line style={very thick},
    xmin=0, xmax=1.05,
    ymin=-0.18, ymax=0.36,
    xtick=\empty,
    ytick=\empty,
    xtick={0.4,0.5,0.8,1},       
    ytick={-0.001, 0.16, 0.355},       
    xticklabels={\makebox[0pt][r]{$p_\varepsilon$},\makebox[0pt][l]{$p^M$},$c_\varepsilon$, price},  
    yticklabels={$o$,$\widehat\pi_\varepsilon$, profit},  
]

\addplot [
    domain=0:1, 
    samples=100, 
    color=blue,
    very thick
]
{x*(1-x)};
\addplot[dashed] coordinates {(0.5,0) (0.5,0.25)};
\addplot[dashed] coordinates {(0.4,0) (0.4,0.24)};

\node[font=\footnotesize, scale=1] at (axis cs:0.45,0.02) {$\varepsilon$};
\draw[->,shorten >=1mm] (axis cs:0.435,0.02) -- (axis cs:0.385,0.02);
\draw[->,shorten >=1mm] (axis cs:0.465,0.02) -- (axis cs:0.515,0.02);

\node at (axis cs:0.4,0.26) {A};

\addplot [
    domain=0:0.8, 
    samples=100, 
]
{0.24+0.2*(x-0.4)};
\addplot[dashed] coordinates {(0,0.16) (0.8,0.16)};

\addplot[dashed] coordinates {(0.8,0) (0.8,0.16)};

\addplot[red, very thick] coordinates {(0,0.16) (0.8,0.16)};

\end{axis}
\end{tikzpicture}
\end{lrbox}%
\resizebox{\linewidth}{!}{\usebox{\figurepanel}}
\caption{}
\end{subfigure}%
\begin{subfigure}{0.5\textwidth}
\centering
\begin{lrbox}{\figurepanel}
\begin{tikzpicture}
\begin{axis}[
    axis lines = left,
    axis line style={very thick},
    xmin=0, xmax=1,
    ymin=0, ymax=1.1,
    xtick=\empty,
    ytick=\empty,
    xtick={0,0.8,0.9,1},       
    ytick={1, 1.09},       
    xticklabels={$0$, \makebox[0pt][r]{$c_\varepsilon$}, \makebox[0pt][l]{$c'_\varepsilon$}, \hspace{1.5em}value},  
    yticklabels={$1$, match prob.},  
]

\addplot [
    domain=0:0.8, 
    samples=100, 
    color=blue,
    very thick
]
{1};

\addplot[dashed] coordinates {(0.8,0) (0.8,1)};
\addplot[dashed] coordinates {(0.9,0) (0.9,1)};
\addplot[blue,very thick] coordinates {(0.9,1) (1,1)};

\draw[->,shorten >=1mm] (axis cs:0.85,0.7) -- (axis cs:0.7,0.65);
\node[font=\footnotesize, scale=0.9] at (axis cs:0.5,0.65) {exclude mass $m_\varepsilon$ 
};
\node[font=\footnotesize, scale=0.9] at (axis cs:0.5,0.6) {of types right above $c_\varepsilon$};
\end{axis}
\end{tikzpicture}
\end{lrbox}%
\resizebox{\linewidth}{!}{\usebox{\figurepanel}}
\caption{}
\end{subfigure}

\caption{Constructing a matching that lowers the monopoly
price. Panel (a) uses the tangent to the full-matching profit
curve at $p^M-\varepsilon$ to determine an exclusion cutoff.
Panel (b) excludes a mass $m_\varepsilon$ of consumers
immediately above that cutoff.}
\label{fig:monopoly-improvement}
\end{figure}
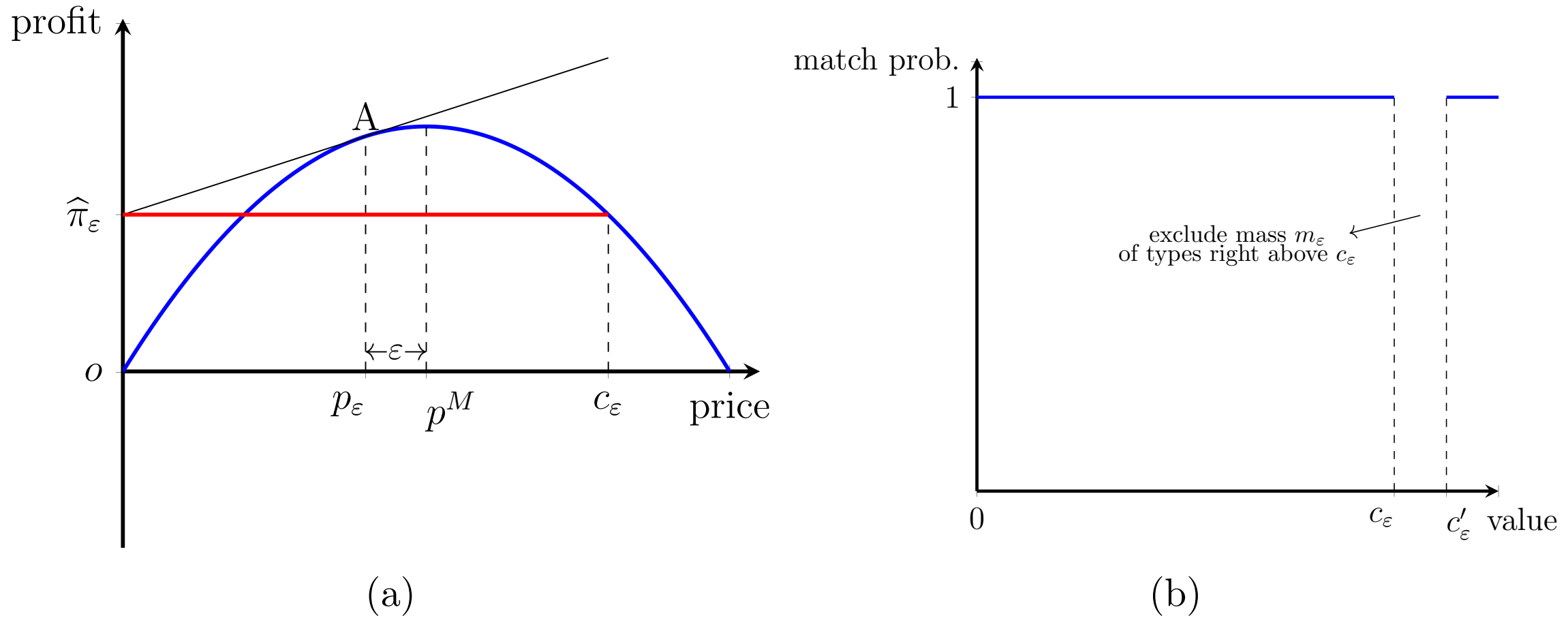

I prove Proposition~\ref{prop:fullisImprovable} by constructing a matching that improves upon full matching.
Figure~\ref{fig:monopoly-improvement} illustrates the
construction for the case in which the full-matching profit
function is globally strictly concave. Fix a small $\varepsilon>0$ and consider
inducing the price $p_\varepsilon=p^M-\varepsilon$.
In Panel (a), the line tangent to the full-matching profit
curve at $p_\varepsilon$ has positive slope
$m_\varepsilon=(\Pi^F)'(p_\varepsilon)$ and vertical intercept
$\widehat\pi_\varepsilon$. Let $c_\varepsilon>p^M$ be the price
at which the full-matching profit curve falls to
$\widehat\pi_\varepsilon$. As shown in Panel (b), the platform
excludes consumers with valuations in an interval
$[c_\varepsilon,c'_\varepsilon]$, chosen so that the excluded
mass equals $m_\varepsilon$.

Why does this exclusion induce the lower price? At any price
$q<c_\varepsilon$, all excluded consumers would otherwise
purchase, so the monopolist's profit becomes
$\Pi^F(q)-qm_\varepsilon$. By concavity, the tangent line in
Panel (a) lies above the full-matching profit curve.
Subtracting $qm_\varepsilon$ from this bound shows that the
monopolist's profit cannot exceed $\widehat\pi_\varepsilon$,
which is attained at $p_\varepsilon$. Prices above
$c_\varepsilon$ are also unprofitable, since exclusion cannot
raise profit and the full-matching profit is already below
$\widehat\pi_\varepsilon$. Thus, $p_\varepsilon$ is optimal
under the constructed matching. Appendix~\ref{app:monopoly-improvement} shows that the
same construction works under the weaker assumptions of
Proposition~\ref{prop:fullisImprovable}.

The price reduction benefits consumers who retain access,
while the excluded consumers lose little surplus because
their valuations approach $p^M$ as $\varepsilon$ tends to zero.
The gain from the price reduction is of first order in
$\varepsilon$, whereas the loss from exclusion is
$o(\varepsilon)$. Hence aggregate consumer surplus strictly
increases for sufficiently small $\varepsilon$.

The proposition establishes that full matching need not
maximize consumer surplus. We next ask how to choose the
exclusion policy optimally.

\subsection[Consumer-Optimal Matching and the Demand-Preserving Principle]{Consumer-Optimal Matching and the\\Demand-Preserving Principle}
\label{sec:monopoly-consumer-optimal}

Proposition~\ref{prop:fullisImprovable} establishes that
exclusion can increase consumer surplus, but it does not
tell us how to choose the exclusion policy optimally.
We approach this problem by separating two decisions:
how much profit to leave the monopolist and which
consumers to exclude.

A matching is \textbf{consumer-optimal} if it maximizes aggregate
consumer surplus, with each matching evaluated at its lowest
optimal monopoly price.

Let $\bar\pi=\max_p\Pi^F(p)$ denote monopoly profit
under full matching. Fix a profit level
$\pi\in(0,\bar\pi]$. A matching is
\textbf{demand-preserving} at $\pi$ if it induces
weakly greater demand at every price than any other
matching that leaves the monopolist the same maximum
profit $\pi$.

The following result shows that a demand-preserving
matching always exists and maximizes consumer surplus
among matchings that leave the monopolist the same profit.

\begin{theorem}[Demand-Preserving Matching]
\label{theorem:demand-preserving-monopoly}
For every $\pi\in(0,\bar\pi]$, there exists a
demand-preserving matching $\phi_\pi$ that maximizes
consumer surplus among all matchings leaving the
monopolist profit $\pi$. Consequently, a consumer-optimal
matching exists and can be chosen to be demand-preserving.
\end{theorem}

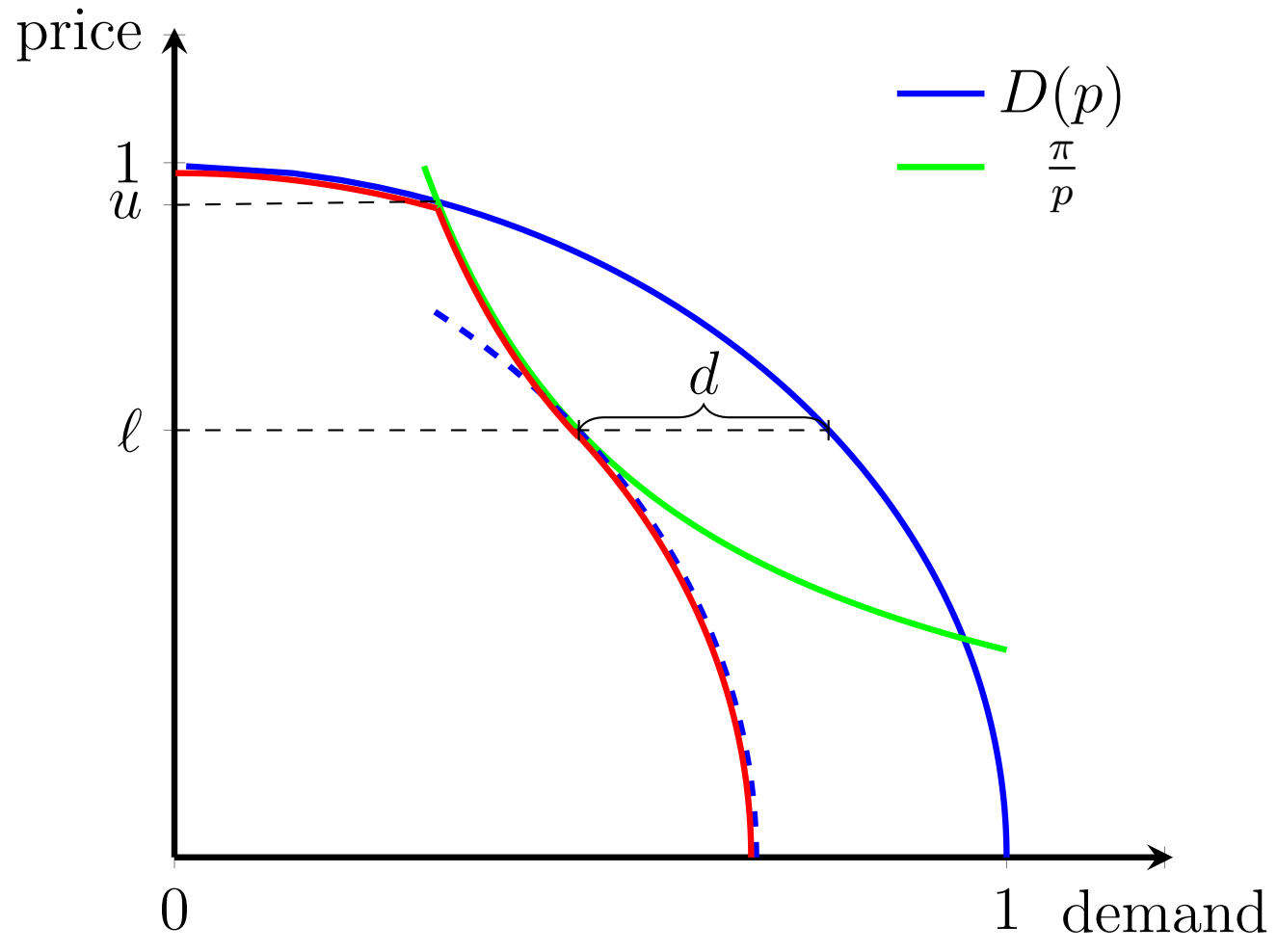
\begin{figure}[H]

\centering

\begin{tikzpicture}
\begin{axis}[
    axis lines=left,
    axis line style={very thick},
    legend style={draw=none},
    xmin=0, xmax=1.2,
    ymin=0, ymax=1.2,
    xtick=\empty,
    ytick=\empty,
     xtick={0,1,1.19},       
    ytick={0.618,0.944, 1.005, 1.19},       
    xticklabels={$0$, $1$, demand},  
    yticklabels={$\ell$,$u$,$1$, price},  
]
    \addplot[blue, very thick, domain=0:1, samples=100] ({sqrt(1-x^2)},{x});
    \addlegendentry{$D(p)$}
    \addplot[green, very thick, domain=0.3:1, samples=100] ({0.3/x},{x});
    \addlegendentry{$\frac{\pi}{p}$}
    \addplot[red, very thick, domain=0.949:1, samples=100] ({sqrt(1-x^2)},{x-0.01});
    \addplot[blue,dashed, very thick, domain=0:0.8, samples=100] ({sqrt(1-x^2)-0.301},{x});
    
    \addplot[dashed] coordinates {(0,0.618) (0.786,0.618)};
    \addplot[dashed] coordinates {(0,0.944) (0.316,0.949)};

    \addplot[red, very thick, domain=0.618:0.949, samples=100] ({0.3/x},{x-0.01});

     \addplot[red, very thick, domain=0:0.618, samples=100] ({sqrt(1-x^2)-0.301-0.006},{x});

     \addplot[mark=|] coordinates {(0.486, 0.618)}; 
    \addplot[mark=|] coordinates {(0.786, 0.618)}; 
    
    \draw[decorate,decoration={brace,amplitude=5pt}] (axis cs:0.486, 0.618) -- (axis cs:0.786, 0.618)
    node[midway,above=3pt] {$d$};

\end{axis}
\end{tikzpicture}

\caption{Constructing a demand-preserving matching.
The red curve is the pointwise upper bound on demand
among matchings that leave the monopolist profit $\pi$.
It combines full-matching demand, the equal-revenue
curve $\pi/p$, and a horizontal shift of full-matching
demand.}
\label{fig:monopoly-demand-preserving}
\end{figure}

We explain the construction using
Figure~\ref{fig:monopoly-demand-preserving}, assuming
for illustration that full-matching demand $D$ is
concave. Appendix~\ref{app:capacity} establishes the theorem without
this assumption.

Fix $\pi\in(0,\bar\pi)$. In
Figure~\ref{fig:monopoly-demand-preserving}, prices
are on the vertical axis and quantities on the horizontal
axis. The blue curve represents full-matching demand
$D(p)$, while the green curve represents $\pi/p$.

Consider any matching that leaves the monopolist profit
$\pi$. Its induced demand cannot exceed full-matching
demand, since the platform can only exclude consumers.
Nor can it exceed $\pi/p$ at any price $p>0$, since
otherwise the monopolist could earn more than $\pi$.
Thus, the induced demand curve must lie to the left
of both the blue and green curves.

These two bounds are not sufficient, however, because
exclusion affects demand at multiple prices. The
horizontal distance between full-matching demand and
induced demand at price $p$ measures the mass of excluded
consumers whose valuations exceed $p$:
\[
D(p)-D_\phi(p)
=\int_p^{\bar v}[1-\phi(v)]\,dF(v).
\]
This distance cannot decrease as the price falls,
since lowering the price only adds consumers to the
population whose valuations exceed it.

Figure~\ref{fig:monopoly-demand-preserving} shows
how this restriction determines the maximum feasible
demand curve. Starting from the upper intersection
$u$ of the blue and green curves, the horizontal
gap between them increases as the price falls,
reaching its maximum $d$ at price $\ell$.
Any matching that leaves the monopolist profit
$\pi$ must therefore exclude at least mass $d$
of consumers whose valuations exceed $\ell$.
The same exclusion must also reduce demand by
at least $d$ at every lower price.

Consequently, below $\ell$, induced demand cannot
exceed the dashed blue curve, which is full-matching
demand shifted horizontally to the left by $d$.
Combining this restriction with the original two
bounds gives the red curve in
Figure~\ref{fig:monopoly-demand-preserving}:
it coincides with full-matching demand above $u$,
follows the green curve between $\ell$ and $u$,
and follows the dashed blue curve below $\ell$.

The key observation is that the red curve is attainable
by a single matching. To see this, note that the slope
of an induced demand curve determines the probability
of granting access to each valuation type. The red
curve's slope coincides with that of full-matching demand outside
$(\ell,u)$, so the platform grants full access to
those consumers. Between $\ell$ and $u$, it induces
demand $\pi/p$, which requires access probability
\[
\phi_\pi(v)=\frac{\pi}{v^2f(v)},
\qquad \ell<v<u.
\]
Concavity of $D$ ensures that this probability lies
between zero and one. Thus, the platform partially
excludes consumers with intermediate valuations
while preserving full access for the remaining types.

By construction, the induced demand curve coincides
with $\pi/p$ between $\ell$ and $u$, so every price
in this interval yields monopoly profit $\pi$.
No other price yields more. Under our lowest-optimal-price
convention, the monopolist charges $\ell$.

Finally, why does this matching maximize consumer
surplus at profit $\pi$? Consumer surplus at price
$p$ equals $\int_p^{\bar v}D_\phi(v)\,dv$.
The demand-preserving matching induces the pointwise
greatest demand. Moreover, no matching leaving the
monopolist profit $\pi$ can induce an optimal price
below $\ell$, as
Figure~\ref{fig:monopoly-demand-preserving} shows.
At any price $p<\ell$, even the greatest feasible
demand is strictly below $\pi/p$, so the monopolist
cannot earn profit $\pi$ at that price. Thus, the
constructed matching achieves both the lowest possible
optimal price and the greatest demand at every price,
maximizing consumer surplus.

Appendix~\ref{app:capacity} proves that a demand-preserving matching
exists for general value distributions without assuming
concavity. The platform's problem therefore reduces
to choosing the profit level $\pi$, rather than searching
over all possible exclusion policies.

\subsection{Interim Pareto Improvements and Capacity-Constrained Exclusion}
\label{sec:monopoly-pareto}

So far, we have shown that exclusion can increase aggregate consumer surplus, but it may make excluded consumers worse off. Must
some consumers lose whenever the platform uses exclusion to lower
the monopoly price? The answer is no. By randomizing exclusion,
the platform can increase every consumer type's interim surplus weakly, and strictly increase it for some types.

We say that a matching $\phi$ \emph{interim Pareto improves} another
matching $\widetilde\phi$ if almost every consumer type obtains weakly
greater interim surplus under $\phi$, and a positive measure of
types obtains strictly greater interim surplus.
A matching is \emph{interim Pareto improvable} if another matching
interim Pareto improves it.

\begin{proposition}
\label{prop:monopoly-pareto}
Under Assumption~\ref{ass:monopoly-improvement}, full matching is interim Pareto improvable.
\end{proposition}

To establish the result, we modify the construction in Figure~\ref{fig:monopoly-improvement}.
There, the platform excludes a small group of consumers entirely
to induce a lower price. But the monopolist's pricing incentives
depend on the total mass excluded above the cutoff
$c_\varepsilon$, not on whether particular consumers are excluded
with certainty. We can therefore spread the same exclusion mass
across a larger group, assigning each consumer a small probability
of exclusion.

A consumer with valuation $v>p^M$ obtains surplus $v-p^M$ under
full matching. If the price falls to $p_\varepsilon=p^M-\varepsilon$,
her expected surplus becomes
$\phi_\varepsilon(v)(v-p_\varepsilon)$.
Thus, protecting her interim surplus requires
\[
v-p^M\leq\phi_\varepsilon(v)(v-p_\varepsilon),
\]
or, equivalently,
\[
1-\phi_\varepsilon(v)
\leq\frac{\varepsilon}{v-p_\varepsilon}.
\]
Consumers with valuations close to the original monopoly price
can tolerate relatively large exclusion probabilities, since
losing access sacrifices little surplus. Appendix~\ref{app:monopoly-improvement} shows that
these individual bounds leave enough room to exclude the mass
required to induce the lower price. Indeed, the probabilities
can be chosen strictly below the bounds, so every consumer who
purchases under full matching strictly benefits in expectation.

Thus, a platform can increase aggregate consumer surplus without
making any consumer type worse off, even though some consumers
occasionally lose access to the product.\footnote{\citet{haghpanah2023pareto} establish Pareto-improving segmentation in multiproduct monopoly markets, allowing different consumer segments to face different prices and product menus. Here, the monopolist charges a uniform price, and Pareto improvements arise through restrictions on consumer access.}

More generally, suppose the platform wishes to maximize aggregate consumer
surplus while guaranteeing every consumer at least a specified
level of interim surplus. How should it design the matching?
Let $h(v)\geq 0$ denote the interim surplus guaranteed to type
$v$. This could be her interim surplus under any benchmark matching, such as full matching, or the payoff she receives from an outside option. We consider
the problem
\begin{equation}
\label{eq:interim-guarantee-problem}
\begin{aligned}
\max_{\phi,p}\quad &CS(\phi,p)\\
\text{subject to}\quad
&p\in\arg\max_q\Pi_\phi(q),\\
&u(v;\phi,p)\geq h(v)
\quad\text{for almost every }v.
\end{aligned}
\end{equation}

The key observation is that interim welfare guarantees
translate into constraints on exclusion. To see this, fix
a target monopoly price $p$. A consumer with valuation
$v>p$ receives interim surplus $\phi(v)(v-p)$. Guaranteeing
her at least $h(v)$ requires
\[
\phi(v)(v-p)\geq h(v),
\]
or equivalently,
\[
1-\phi(v)\leq 1-\frac{h(v)}{v-p}.
\]
Thus, protecting consumers' interim surplus places an upper
bound on their exclusion probabilities. For this to be
possible, we must have $h(v)\leq(v-p)_+$ for almost every
$v$; otherwise, some consumers cannot obtain their
guaranteed surplus even with full access at price $p$.
Whenever this condition holds, define
\[
k_{h,p}(v)=
\begin{cases}
\displaystyle 1-\frac{h(v)}{v-p},&v>p,\\[6pt]
1,&v\leq p.
\end{cases}
\]
The interim guarantees are then equivalent to requiring
that the exclusion probability of each type $v$ be no
greater than $k_{h,p}(v)$. Consumers below $p$ can be
excluded freely, since they obtain zero surplus at the
target price, although such exclusion may affect the
monopolist's incentives to lower its price. Write
\[
K_{h,p}(q)
=\int_q^{\bar v}k_{h,p}(v)\,dF(v).
\]
This function measures the mass of consumers eligible
for exclusion above each price, taking into account their
individual welfare guarantees. Hence, at a fixed target
price, problem~\eqref{eq:interim-guarantee-problem}
becomes a monopoly problem with limited exclusion capacity.
We next solve this more general problem, which will also
arise in the duopoly analysis.

\subsubsection{Capacity-Constrained Monopoly Problem}
\label{sec:monopoly-capacity}

Consider the same monopoly environment as in Section~\ref{sec:monopoly-environment}.
Some consumers are eligible for exclusion; $K(q)$ denotes demand at $q$ from these eligible consumers.
Assume $K$ is continuous, $K(\bar v)=0$, and
\[
0\leq K(a)-K(b)\leq D(a)-D(b)
\qquad(0\leq a<b\leq\bar v).
\]
\begingroup\emergencystretch=3em
Given a target price $p$, the
\emph{capacity-constrained monopoly problem},
denoted by $\mathrm{CM}(D,K,p)$, is to choose the exclusion probability for each eligible consumer to maximize consumer surplus,
subject to $p$ being an optimal price for the monopolist.
The problem need not be feasible: there may be no way to exclude eligible consumers so that $p$ is an optimal monopoly price. Whenever it is feasible, a solution exists (Appendix~\ref{app:capacity-fixed-price}).
\par\endgroup

We solve $\mathrm{CM}(D,K,p)$ by first relaxing the
requirement that the induced monopoly price be $p$.
The relaxed problem involves two choices: how much profit
to leave the monopolist and which consumers to exclude.
We first fix monopoly profit and solve the latter problem,
extending the demand-preserving principle from
Section~\ref{sec:monopoly-consumer-optimal}. We then return to the
target-price problem and solve it by choosing the highest
feasible profit consistent with $p$.

Write $\widehat D$ for demand after exclusion and
$M=D-\widehat D$ for the resulting demand loss.

\begin{lemma}[Feasible Demand with Capacity]
\label{lem:capacity-feasible-demand}
A demand curve $\widehat D$ can be induced by a
capacity-feasible exclusion rule if and only if
$M(\bar v)=0$ and
\begin{equation}
\label{eq:capacity-increments}
0\leq M(a)-M(b)\leq K(a)-K(b)
\qquad(0\leq a<b\leq\bar v).
\end{equation}
\end{lemma}

The first inequality says that demand loss can only
increase as price falls; the second says that it cannot
increase faster than exclusion capacity. In particular,
feasible demand lies between $D-K$ and $D$.

A capacity-feasible exclusion rule is
\emph{demand-preserving} at profit $\pi$ if it induces
pointwise weakly greater demand than every capacity-feasible
exclusion rule that leaves the monopolist the same maximum
profit $\pi$. Define
\[
\underline\pi_K
:=\max_{q\in[0,\bar v]}q[D(q)-K(q)].
\]
The monopolist's maximum profit is $\underline\pi_K$ under
maximal exclusion and $\bar\pi$ without exclusion. Every
profit between these bounds is attainable by continuously
scaling maximal exclusion to no exclusion.

\begin{proposition}[Capacity-Constrained Demand Preservation]
\label{prop:capacity-demand-preserving}
For every $\pi\in[\underline\pi_K,\bar\pi]$ with
$\pi>0$, there exists a demand-preserving exclusion
rule $\rho_{\pi,K}$ that maximizes consumer surplus among
all capacity-feasible exclusion rules leaving the
monopolist profit $\pi$.
\end{proposition}

The proof is in Appendix~\ref{app:capacity}. We explain the construction
here. To keep maximum profit below $\pi$, demand at price
$q$ cannot exceed $\pi/q$. Thus the required demand loss
at $q>0$ is at least
\[
g_\pi(q)=[D(q)-\pi/q]_+,
\]
with $g_\pi(0)=0$.

Now fix a price $q$ at which a loss $g_\pi(q)$ is required,
and consider what this implies for the loss at another
price $c$. If $c<q$, demand loss cannot decrease as price
falls, so $M(c)\geq g_\pi(q)$. If $c>q$, at most
$K(q)-K(c)$ mass of consumers with values between $q$
and $c$ is eligible for exclusion. Hence
\[
M(c)+K(q)-K(c)\geq g_\pi(q).
\]
Thus, each price $q$ imposes a lower bound on demand loss
at every price $c$. Combining the two cases, this lower
bound is $g_\pi(q)-[K(q)-K(c)]_+$. Since the requirements
generated by all prices $q$ must hold simultaneously,
demand loss must lie above their pointwise envelope:
\begin{equation}
\label{eq:capacity-envelope}
M(D,K,\pi)(c)
=
\max_{q\in[0,\bar v]}
\left\{
g_\pi(q)-[K(q)-K(c)]_+
\right\}.
\end{equation}
We write the arguments $D,K,\pi$ to emphasize the dependence of the demand-loss envelope on these primitives.

The key result is that this envelope is itself feasible:
a single exclusion rule attains it at every price.
Its induced demand is $D_{\pi,K}=D-M(D,K,\pi)$.
Why does this construction also maximize consumer surplus
at profit $\pi$? Consider any capacity-feasible exclusion
rule with maximum profit $\pi$, and let $p$ be its lowest
optimal price. Because the constructed rule is
demand-preserving, its demand at $p$ is weakly higher.
Hence its profit at $p$ is at least $\pi$. But its maximum
profit is exactly $\pi$, so its profit at $p$ must equal
$\pi$, and $p$ is also optimal under the constructed rule.
At the same price, the constructed rule has weakly greater
demand at every price above $p$, and therefore weakly
greater consumer surplus. Evaluating it at its lowest
optimal price can only increase consumer surplus further.

We now return to the capacity-constrained monopoly problem
$\mathrm{CM}(D,K,p)$, in which the target price $p$ is
fixed. Proposition~\ref{prop:capacity-demand-preserving}
allows us to solve it by choosing the appropriate profit
level that induces $p$.

\begin{corollary}[Optimal Exclusion at a Fixed Price]
\label{cor:capacity-fixed-price}
Whenever $\mathrm{CM}(D,K,p)$ is feasible, the
demand-preserving exclusion rule at the highest profit
compatible with $p$ solves $\mathrm{CM}(D,K,p)$.
\end{corollary}

Appendix~\ref{app:capacity-fixed-price} proves the corollary, characterizes when the problem is feasible, and gives an explicit formula for the highest feasible profit.
Thus, for any feasible target price, the optimal exclusion
rule can be constructed using the demand-preserving
envelope. However, solving the fixed-price problem is not
enough to solve our motivating problem with interim
welfare guarantees: we must also optimize over feasible
target prices.

\subsubsection{Welfare Guarantees and Interim Pareto Efficiency}
\label{sec:monopoly-welfare-guarantees}

We now return to problem~\eqref{eq:interim-guarantee-problem}.
For a target price $p$ to be feasible, two conditions must
hold. First, the guarantees must satisfy
$h(v)\leq(v-p)_+$ for almost every $v$; otherwise,
some consumers cannot obtain their guaranteed surplus even
with full access. Second, given the exclusion capacity
$K_{h,p}$ determined by these guarantees, the problem
$\mathrm{CM}(D,K_{h,p},p)$ must be feasible.
Let $\mathcal P(h)$ denote the set of target prices
satisfying both conditions.

\begin{corollary}[Optimal Matching with Interim Welfare Guarantees]
\label{cor:interim-guarantees}
The welfare guarantees $h$ are feasible if and only if $\mathcal P(h)$ is nonempty. Whenever feasible, there exists $p^o\in\mathcal P(h)$ such that a solution to $\mathrm{CM}(D,K_{h,p^o},p^o)$ maximizes aggregate consumer surplus subject to $h$.
\end{corollary}

Corollary~\ref{cor:interim-guarantees} reduces the original
problem to an optimization over feasible target prices.
For each target price $p$, the interim welfare guarantees
determine the exclusion capacity $K_{h,p}$, while
Corollary~\ref{cor:capacity-fixed-price} identifies the highest
compatible monopoly profit $\pi_p$. The corresponding
demand-preserving matching maximizes aggregate consumer
surplus at that price. The platform therefore only needs
to choose the target price yielding the greatest consumer
surplus, with the optimal matching determined by the
demand-preserving construction.

The preceding characterization also allows us to determine
whether a given matching is interim Pareto efficient.
A matching is \emph{interim Pareto efficient for consumers}
if no other matching interim Pareto improves it. Consider
any matching $\widetilde\phi$, let $\widetilde p$ denote
its lowest induced monopoly price, and define
$h^{\widetilde\phi}(v)=u(v;\widetilde\phi,\widetilde p)$. To determine
whether $\widetilde\phi$ is interim Pareto efficient,
we could search over all alternative matching policies
and ask whether one benefits some consumers without
harming anyone. The following result makes this unnecessary.

\begin{corollary}[Interim Pareto Efficiency]
\label{cor:interim-pareto-efficiency}
The matching $\widetilde\phi$ is interim Pareto efficient for consumers if and only if, for every $p\in\mathcal P(h^{\widetilde\phi})$, the solution to $\mathrm{CM}(D,K_{h^{\widetilde\phi},p},p)$ yields consumer surplus weakly below $CS(\widetilde\phi,\widetilde p)$.
\end{corollary}

If the solution to $\mathrm{CM}(D,K_{h^{\widetilde\phi},p},p)$ yields strictly higher consumer surplus than the benchmark for some $p\in\mathcal P(h^{\widetilde\phi})$, it provides an interim Pareto improvement. Otherwise, the benchmark is interim Pareto efficient. Thus, instead of searching over all possible matching policies, we only need to compare the benchmark with the demand-preserving matching associated with each feasible target price.

\subsection{Welfare Consequences of Exclusion}
\label{sec:monopoly-welfare}

Exclusion can increase aggregate consumer surplus, and
Proposition~\ref{prop:monopoly-pareto} shows that it can
even generate an interim Pareto improvement for consumers.
These benefits, however, come at the cost of efficiency.
The platform lowers the monopoly price by restricting
access to selected consumers, potentially preventing
mutually beneficial transactions. Under log-concavity,
this reduction in trade necessarily dominates the
additional purchases generated by the lower price.

\begin{proposition}[Welfare Consequences of Exclusion]
\label{prop:monopoly-welfare}
Suppose the valuation density $f$ is log-concave.
Full matching maximizes social surplus and trade quantity.
Moreover, any matching that strictly increases aggregate
consumer surplus relative to full matching strictly
reduces both social surplus and trade quantity.
\end{proposition}

The result follows from a bound on demand at the induced
monopoly price. If a matching $\phi$ induces price $p_\phi$,
the monopolist's incentive not to raise its price slightly implies
$D_\phi(p_\phi)\leq p_\phi f(p_\phi)$. Under log-concavity, $pf(p)$ is
strictly increasing on the valuation support below the full-matching monopoly
price $p^M$. Consequently, whenever exclusion induces
a lower price, trade quantity remains strictly below
$D(p^M)$. A matching inducing a higher price cannot
increase trade quantity either. Since full matching
serves the consumers with the highest valuations, it
also maximizes social surplus.

Thus, although exclusion can benefit consumers through
lower prices, under log-concavity it cannot improve efficiency in monopoly.
As we will see in the duopoly analysis, this conclusion
changes when the platform can match consumers only to their less-preferred products rather than simply excluding them.

\section{Duopoly}
\label{sec:duopoly}

\subsection{Environment}
\label{sec:duopoly-environment}

We now consider two firms, $N=\{1,2\}$, supplying differentiated
products at zero marginal cost to a unit mass of unit-demand
consumers. Each consumer has valuations $v=(v_1,v_2)$, drawn
from a symmetric, log-concave distribution $\mu$ with continuously
differentiable density $f$, positive on the interior of a
compact, convex, symmetric support
$V\subseteq\mathbb R_+^2$ of positive area. Write
$\bar v=\max_{v\in V}\max_i v_i$. Firms know the distribution
but not individual consumers' valuations.

Before prices are chosen, a platform publicly commits to a
measurable matching rule
\[
\phi:V\longrightarrow\Delta(2^N),
\]
where $\phi(S\mid v)$ is the probability of displaying
exactly the products in $S$ to a consumer with valuations $v$.
Firms observe the matching rule and simultaneously choose
nonnegative uniform prices. Consumer valuations are then
realized and observed by the platform, which implements the
matching rule. Consumers discover only the displayed products
and learn their values exactly. Each consumer chooses a displayed
product to purchase. The consumer can always choose not to
purchase anything and obtain zero payoff. Thus, $S=\varnothing$
means the consumer is excluded from trade. The \emph{full matching},
denoted by $\phi^{\mathrm{full}}$, displays both products
to every consumer. At prices
$\mathbf p=(p_1,p_2)$, aggregate consumer surplus is
\[
CS(\phi,\mathbf p)
=
\int_V\sum_{S\subseteq N}
\phi(S\mid v)
\max\left\{0,\max_{i\in S}(v_i-p_i)\right\}
\,d\mu(v),
\]
where the inner maximum over $S=\varnothing$ is omitted.

We restrict attention to symmetric matchings and symmetric
pure-strategy price equilibria. Let $\sigma$ exchange firm
identities and valuation coordinates. A matching is
\emph{symmetric} if
$\phi(S\mid v)=\phi(\sigma(S)\mid\sigma(v))$
almost everywhere. Given a matching $\phi$, firm $i$'s
demand when it charges $q$ and its rival $j$ charges $r$ is
\[
\begin{aligned}
D_i(q,r\mid\phi)
={}&\int_V\phi(\{i\}\mid v)
\mathbf 1\{v_i\ge q\}\,d\mu(v)\\
&+\int_V\phi(\{1,2\}\mid v)
\mathbf 1\{v_i-q\ge\max(0,v_j-r)\}\,d\mu(v).
\end{aligned}
\]
Its profit is $\Pi_i(q,r\mid\phi)=qD_i(q,r\mid\phi)$.
A matching $\phi$ supports $(p,p)$ as a symmetric price equilibrium if
\[
\Pi_i(p,p\mid\phi)\ge\Pi_i(q,p\mid\phi)
\qquad\text{for every }q\ge0,\quad i=1,2.
\]

Let $p^*$ denote the smallest symmetric equilibrium price
under full matching, whose existence is guaranteed by
log-concavity. For each $p\ge0$, let $\mathcal M(p)$ denote
the symmetric matchings supporting $(p,p)$ as an equilibrium.

\subsection{Mismatching can improve upon full matching}
\label{sec:mismatching-benefits}

In monopoly, a platform can lower prices by excluding consumers, but under log-concavity this necessarily reduces social surplus and trade quantity. Can competition avoid this tradeoff? It can. Rather than excluding consumers, the platform can show some of them only the product they like less. This changes the composition of firms' customers and can make price increases less attractive, even though it reduces some consumers' match quality.

\begin{proposition}
\label{prop:local-improvement}
Suppose $(p^*,p^*)\in\operatorname{int}V$.\footnote{The condition rules out nearly homogeneous Bertrand environments in which consumers' valuations are bounded away from zero. In such environments, price competition already drives the full-matching price close to zero, leaving little room for further consumer gains from lower prices.} For every $p<p^*$ sufficiently close to $p^*$, there exists a symmetric matching without exclusion that supports $(p,p)$ as an equilibrium and, relative to full matching, yields an interim Pareto improvement for consumers and strictly increases both social surplus and trade quantity.
\end{proposition}

The intuition is illustrated in Figure~\ref{fig:IntroMismatchDemand} in the Introduction. By symmetrically mismatching selected consumers, the platform replaces consumers less sensitive to price increases with more price-sensitive ones, without changing either firm's sales at the target price. Although these consumers become captive to individual firms, they are more likely to stop purchasing after a price increase. This makes upward price deviations less attractive and allows the platform to sustain a lower equilibrium price.

But can the price reduction compensate consumers for being mismatched? Suppose the equilibrium price falls from $p^*$ to $p=p^*-\varepsilon$, and a selected consumer is mismatched with probability $\rho$. If her preference difference is $d$, her expected loss in match quality is $\rho d$, while the price reduction saves her $\varepsilon$. Thus, she is better off whenever $\rho d<\varepsilon$. The proof in Appendix~\ref{app:local-improvement} constructs a matching that satisfies this condition while deterring all unilateral price deviations. Every consumer type is then weakly better off, with some strictly better off.

Moreover, every mismatched consumer continues to purchase at the lower price. Hence mismatching preserves trade at $p$, and the price reduction brings additional consumers into the market. Thus, trade quantity strictly increases relative to full matching. Finally, the platform mismatches only a small mass of consumers who have close valuations for the two products. The resulting loss in match quality can be made arbitrarily small relative to the first-order social-surplus gain from additional transactions induced by lower prices. Thus, social surplus increases as well.

\subsection{Consumer-Optimal Matching}

Proposition~\ref{prop:local-improvement} shows that mismatching can improve upon full matching. But how should a consumer-oriented platform choose which consumers to mismatch or exclude? We approach this problem by first fixing a symmetric equilibrium price $p$ and characterizing the consumer-optimal matching that supports it. We then optimize over the equilibrium price.

A matching $\phi^*\in\mathcal M(p)$ is \emph{symmetric consumer-optimal supporting $(p,p)$} if it maximizes $CS(\phi,(p,p))$ over $\phi\in\mathcal M(p)$.

Given a symmetric deterministic matching $\phi$, let $\widetilde V_i(\phi)=\{v:\phi(\{i\}\mid v)=1\}$ denote the consumers shown only firm $i$'s product, and let $K_i(q\mid\phi)=\int_{\widetilde V_i(\phi)}\mathbf 1\{v_i\ge q\}\,d\mu(v)$ denote demand at $q$ from these consumers. Firm $i$'s total demand when its rival charges $p$ is $D_i(q,p\mid\phi)$, as defined in Section~\ref{sec:duopoly-environment}.

\Needspace{10\baselineskip}
\begin{theorem}[Symmetric Consumer-Optimal Matching]\label{theorem:sequential-mismatching}
For every $0<p<p^*$, there exists a symmetric consumer-optimal matching supporting $(p,p)$ that can be constructed in the following two steps.
\begin{enumerate}
\item \textbf{Omit low-value products and mismatch by a cutoff.} Start from full matching. Product $i$ is omitted from a type-$v$ consumer's consideration set whenever $v_i<p$. There exists a cutoff $\tau^*:[0,\bar v-p]\to[p,\bar v]$ such that a consumer with $v_i>v_j>p$ is matched only with firm $j$ if $v_j<\tau^*(v_i-v_j)$; otherwise, she continues to consider both firms. Call this deterministic matching $\widetilde\phi$.
\item \textbf{Exclusion.} The problem $\mathrm{CM}(D_i(\cdot,p\mid\widetilde\phi),K_i(\cdot\mid\widetilde\phi),p)$ is feasible. Let $\rho^*$ be a solution. For $i=1,2$, exclude each $v\in\widetilde V_i(\widetilde\phi)$ with probability $\rho^*(v_i)$.
\end{enumerate}
\end{theorem}

The consumer-optimal matching can also be chosen deterministic. Conditional on $v_i=x$, any required fraction of consumers in $\widetilde V_i(\widetilde\phi)$ can be excluded using a cutoff in the other valuation.

Theorem~\ref{theorem:sequential-mismatching} substantially simplifies the platform's problem. An arbitrary symmetric matching specifies probabilities over four consideration sets at every valuation pair. Instead, for any fixed $p<p^*$, a symmetric consumer-optimal matching can be constructed using a single cutoff function, followed by the demand-preserving exclusion rule developed in Section~\ref{sec:monopoly}.

The cutoff has a simple economic interpretation. Among consumers with the same preference difference, mismatching entails the same loss in match quality. But consumers with lower valuations are more sensitive to price increases. Thus, along each diagonal of equal valuation differences, the platform mismatches consumers with lower valuations, while those with higher valuations retain access to both products. An additional implication is that showing consumers only their preferred product is unnecessary when both products are valued above $p$. Although such assignments can help deter downward price deviations without sacrificing consumer surplus at the target price, an optimum can always be attained without them.

The second step exploits the monopoly analysis. Once the cutoff matching is determined, any additional exclusion applies only to consumers shown a single product. Excluding these consumers changes the demand of their assigned firm without affecting the rival's demand. The platform can therefore solve the capacity-constrained monopoly problem separately for each firm, using the demand-preserving exclusion rule characterized in Section~\ref{sec:monopoly}.

\begin{proof}[Proof of \Cref{theorem:sequential-mismatching}]
Fix $0<p<p^*$. Consider a relaxed version of our original problem in which the downward-deviation constraints are ignored and everything else remains unchanged. Take an arbitrary optimum of this relaxed problem. An optimum exists by Appendix~\ref{app:downward-deviations}. We first normalize it. Remove products valued below $p$, and replace preferred-only assignments for consumers with both values above $p$ by full consideration. These changes preserve demand at $p$ when the rival also charges $p$, preserve consumer surplus at $(p,p)$, and weakly reduce demand from upward deviations. Thus, the resulting matching, denoted by $\phi^R$, continues to be an optimum of the relaxed problem.

\Needspace{8\baselineskip}
\begin{lemma}\label{claim:upward-exactness}
Neither firm has a profitable downward deviation under $\phi^R$.
\end{lemma}

The idea behind the proof of Lemma~\ref{claim:upward-exactness} is as follows. We show that log-concavity bounds the demand gained from price cuts. Thus, a price cut cannot be profitable if demand at $p$, when the rival also charges $p$, is above a certain level. We then show that if this demand were below that level, a significant mass of consumers must have been excluded. Suitably restoring some of these consumers would increase consumer surplus without violating the upward incentive constraints, contradicting optimality of $\phi^R$.\footnote{There is an appealing alternative argument if one also optimizes the relaxed problem over the target price. Let $\hat p$ be an optimizer and $\hat\phi$ a corresponding optimal matching. If a profitable price cut existed at $\hat p$, a best response to $\hat p$ would lie below $\hat p$, whereas a best response to zero is strictly positive. If best responses varied continuously with the rival's price, the same matching $\hat\phi$ would therefore admit a symmetric equilibrium at some lower price $r<\hat p$. The lower common price would increase consumer surplus under $\hat\phi$, contradicting optimality of $\hat p$. Under an arbitrary matching, however, best responses need not vary continuously and may jump across the diagonal. The direct argument in Appendix~\ref{app:downward-deviations} avoids this difficulty and proves the stronger result for every fixed $p$.}

The lemma is proved in Appendix~\ref{app:downward-deviations}. Thus $\phi^R$ supports $(p,p)$ as an equilibrium and hence is symmetric consumer-optimal supporting $(p,p)$. We now obtain its two-step representation.

We first establish Step~1. Its first part is already delivered by the normalization: products valued below $p$ are omitted. The normalization also shows that, for consumers with both values above $p$, preferred-only matching is never used. Thus, for such consumers, the remaining possibilities are full consideration, the less-preferred product only, or exclusion. It remains to establish the cutoff structure among these assignments.

Fix a preference difference $d>0$ and consider consumers with values $(p+s+d,p+s)$, together with their mirror types. At $(p,p)$, a consumer with full consideration buys her preferred product and obtains consumer surplus $s+d$, whereas a consumer shown only her less-preferred product buys it and obtains consumer surplus $s$. An excluded consumer does not purchase and obtains zero consumer surplus.

Suppose a lower-$s$ group receives full consideration while a higher-$s$ group of equal conditional probability mass receives only the less-preferred product. Exchange their assignments, together with those of their mirror types. Demand at $(p,p)$ is unchanged: under either arrangement, each firm sells to one group that prefers its product and an equal conditional probability mass that prefers the other product. Consumer surplus at $(p,p)$ is also unchanged: every purchase contributes the base surplus $s$, while full consideration contributes an additional $d$; the same conditional probability mass receives full consideration before and after the exchange.

The exchange nevertheless relaxes the upward incentive constraints. Given the rival's price at $p$, a consumer's reservation price under full consideration is $p+d$, independent of $s$, whereas under less-preferred-only matching it is $p+s$. The exchange therefore reduces reservation prices for the same conditional probability mass of consumers, but by a larger amount, so demand from upward deviations weakly falls. The exchanged matching therefore continues to be an optimum of the relaxed problem. Since it continues to satisfy the normalization, Lemma~\ref{claim:upward-exactness} implies that it also supports $(p,p)$ as an equilibrium. Hence it is again symmetric consumer-optimal supporting $(p,p)$.

We can make the same exchange between full consideration and exclusion, together with the corresponding mirror types. The only difference is that the exchange now strictly increases consumer surplus: full consideration is moved from the lower-$s$ group to an equal conditional probability mass of higher-$s$ consumers. Hence, at an optimum, full consideration cannot be assigned below exclusion along the same preference difference.

These exchanges yield a symmetric consumer-optimal matching, denoted by $\widetilde\phi^R$, in which, for each preference difference $d$, full consideration is assigned to an upper tail of $s$. Hence there is a cutoff $\tau^*(d)$ in the lower valuation $p+s$ such that consumers above the cutoff receive full consideration, while those below it receive their less-preferred product or the outside option.

We now construct the matching $\widetilde\phi$ in Step~1 of the theorem from $\widetilde\phi^R$. When both values exceed $p$, replace every outside-option assignment below the cutoff by the less-preferred product only. This views any exclusion in this region as first assigning the less-preferred product in Step~1 and then excluding the consumer in Step~2. When exactly one value exceeds $p$, the normalization leaves either that product with value above $p$ alone or the outside option; replace the latter by that product alone, again postponing exclusion to Step~2. When both values are below $p$, the normalization leaves the consumer with the outside option. Thus $\widetilde\phi$ is exactly the deterministic matching described in Step~1 of the theorem.

We now establish Step~2. By construction, $\widetilde\phi^R$ is obtained from $\widetilde\phi$ by excluding some consumers in $\widetilde V_i(\widetilde\phi)$, for $i=1,2$.

Consider firm $i$. Under the exclusion policy inherited from $\widetilde\phi^R$, $p$ is optimal for firm $i$ when firm $j$ charges $p$, because $\widetilde\phi^R$ supports $(p,p)$ as an equilibrium. Now undo all exclusions from $\widetilde V_j(\widetilde\phi)$, replacing the outside option by $\{j\}$. This does not change firm $i$'s demand at any price, since these consumers never consider firm $i$. Hence $p$ remains optimal for firm $i$. The remaining exclusion policy applies only to consumers in $\widetilde V_i(\widetilde\phi)$, exactly as allowed in $\mathrm{CM}(D_i(\cdot,p\mid\widetilde\phi),K_i(\cdot\mid\widetilde\phi),p)$. Thus this monopoly problem is feasible.

Let $\rho^*$ be a solution and apply it to both firms' singleton populations. Exclusions from $\widetilde V_i(\widetilde\phi)$ affect only firm $i$'s demand, so the two exclusion problems are independent. Therefore, after consumers in $\widetilde V_j(\widetilde\phi)$ are excluded according to $\rho^*$, $p$ remains optimal for firm $i$; symmetrically, $p$ remains optimal for firm $j$. Hence $(p,p)$ remains an equilibrium.

Finally, $\rho^*$ maximizes consumer surplus in the monopoly problem among all feasible exclusion rules. Excluding a consumer $v\in\widetilde V_i(\widetilde\phi)$ sacrifices the same consumer surplus in the monopoly problem and in the duopoly problem, namely $v_i-p$. Since the exclusion rule inherited from $\widetilde\phi^R$ is feasible, replacing it by $\rho^*$ for both firms weakly increases consumer surplus. Because $\widetilde\phi^R$ is already symmetric consumer-optimal supporting $(p,p)$, the resulting matching is also symmetric consumer-optimal supporting $(p,p)$. This establishes Step~2 and completes the proof.
\end{proof}

\Needspace{7\baselineskip}
Finally, we optimize over the target equilibrium prices. A matching $\phi^o$ is \emph{symmetric consumer-optimal} if there exists $p^o\ge0$ such that $\phi^o\in\mathcal M(p^o)$ and
\[
CS(\phi^o,(p^o,p^o))\ge CS(\phi,(p,p))
\]
for every $p\ge0$ and every $\phi\in\mathcal M(p)$.

\begin{corollary}
\label{cor:symmetric-consumer-optimum}
If $(p^*,p^*)\in\operatorname{int}V$, there exists a symmetric consumer-optimal matching that can be constructed as in Theorem~\ref{theorem:sequential-mismatching}.
\end{corollary}

A standard compactness argument gives existence of a symmetric consumer-optimal matching $\phi^o$ and an associated equilibrium price $p^o$. Any $p^o\geq p^*$ is dominated in consumer surplus by full matching at $p^*$, which is itself ruled out by Proposition~\ref{prop:local-improvement}. Nor can $p^o=0$, since any equilibrium at zero yields zero consumer surplus. Hence $p^o\in(0,p^*)$, and the result follows from Theorem~\ref{theorem:sequential-mismatching}.

\subsection{Social-Surplus and Trade Maximization}
\label{sec:social-trade-optimal}

Proposition~\ref{prop:local-improvement} shows that, unlike in monopoly, mismatching in duopoly can increase social surplus and trade quantity. These are also important objectives for a benevolent platform or regulator: social surplus measures the total value generated by trade, while trade quantity measures the number of transactions. We now ask how a platform should design matching to maximize either objective.

The two objectives respond differently to mismatching. Lower prices do not directly increase social surplus, since they are transfers between consumers and firms. They can nevertheless increase social surplus by enabling additional valuable transactions, while mismatching reduces the value generated by some transactions. Trade quantity, by contrast, depends only on whether consumers purchase, not on which products they buy.

A matching $\phi^S$ is \emph{symmetric social-optimal} if there exists $p^S\geq0$ such that $\phi^S\in\mathcal M(p^S)$ and
\[
SS(\phi^S,(p^S,p^S))\geq SS(\phi,(p,p))
\]
for every $p\geq0$ and $\phi\in\mathcal M(p)$. A \emph{symmetric trade-optimal} matching $\phi^T$ is defined analogously, replacing social surplus with trade quantity.

\begin{proposition}
\label{prop:social-trade-two-step}
If $(p^*,p^*)\in\operatorname{int}V$, there exist a symmetric social-optimal matching and a symmetric trade-optimal matching that can each be constructed in the two steps of Theorem~\ref{theorem:sequential-mismatching}.
\end{proposition}

Thus, the two-step construction applies to all three objectives, although the optimal equilibrium prices and cutoffs may differ. But is the second step, which excludes additional consumers, ever necessary? Exclusion can weaken firms' incentives to raise prices, but it also prevents potentially valuable transactions. The following result shows that, under i.i.d.\ log-concave valuations, both social surplus and trade quantity can be maximized without additional exclusion.

For trade quantity, we obtain an even sharper characterization. A matching is called \emph{maximal mismatching} at price $p$ if it omits every product valued below $p$, shows only firm $j$'s product to each consumer with $v_i>v_j>p$ and $v_i-v_j>v_j-p$, and shows every other consumer all products valued above $p$. We denote this matching by $\phi_p^{\max}$.

Omitting products valued below $p$ does not affect purchases at the target price. As illustrated by the mismatching mechanism in Figure~\ref{fig:IntroMismatchDemand} in the Introduction, symmetrically mismatching consumers preserves each firm's sales at $p$ while replacing consumers less sensitive to price increases with more price-sensitive ones, making upward price deviations less attractive.

\Needspace{9\baselineskip}
\begin{theorem}[Social- and Trade-Optimal Matching]\label{theorem:social-trade-optimal}
Suppose, in addition to the assumptions of Section~\ref{sec:duopoly-environment}, that $v_1$ and $v_2$ are i.i.d.\ with common support $[0,\bar v]$. Then:
\begin{enumerate}[label=(\alph*)]
\item There exists a symmetric social-optimal matching that can be constructed using only Step~1 of Theorem~\ref{theorem:sequential-mismatching}.
\item There exists an equilibrium price $p^T$ such that maximal mismatching at price $p^T$ is a symmetric trade-optimal matching.
\end{enumerate}
\end{theorem}

The trade-optimal result follows from the fact that mismatching preserves trade quantity at a given price while weakening firms' incentives to raise prices. Maximal mismatching exploits this effect fully, allowing the platform to sustain a lower equilibrium price and expand trade without sacrificing transactions through exclusion.

Under i.i.d.\ log-concavity, we show that using exclusion to induce an even lower price cannot further increase trade quantity. To establish the result, we first derive two upper bounds on trade quantity under any symmetric matching and equilibrium price. The first is the quantity traded under full matching at the same price: restricting access cannot create additional transactions at a fixed price. The second follows from firms' incentives not to raise their prices slightly. These bounds intersect at a unique price $p^T$. Below $p^T$, the incentive bound is increasing in price; above $p^T$, the full-matching bound is decreasing. Their common value at $p^T$ therefore bounds trade quantity across all symmetric equilibria. Finally, maximal mismatching at $p^T$ supports that price as an equilibrium and attains this bound.

Social-surplus maximization involves an additional consideration: match quality. Although mismatching can induce lower prices and expand trade, it also matches consumers only to their less-preferred products, reducing the value generated by their purchases. Thus, unlike trade-optimal matching, social-optimal matching need not use maximal mismatching. Moreover, exclusion may appear desirable because it can discourage price increases without incurring these mismatch losses. Theorem~\ref{theorem:social-trade-optimal} shows, however, that exclusion is unnecessary: its benefit in weakening firms' pricing incentives cannot justify the loss of valuable transactions.

The main difficulty in proving part (a) is ruling out exclusion. Exclusion may be necessary to sustain very low equilibrium prices, and even at higher prices, it may be less costly than matching consumers only to their less-preferred products. Nevertheless, we show that a social-optimal matching can be chosen without exclusion. The challenge is that restoring excluded consumers increases social surplus and firms' equilibrium sales, but also makes upward price deviations more attractive. To compare these effects, we consider a relaxed problem with finitely many upward-deviation constraints. Its Lagrangian implies that exclusion is suboptimal whenever a weighted sum of the incentive-constraint multipliers is below one. We establish this bound by jointly perturbing the equilibrium price and matching: we raise the target price slightly while transferring the original matching assignments to consumers with slightly higher valuations. This adjustment limits the loss of trade caused by the price increase. Although the resulting matching need not satisfy the incentive constraints, the Lagrangian accounts for their violations. Comparing the Lagrangian at the two prices then yields the required multiplier bound, ruling out exclusion.

\begingroup\emergencystretch=1em
Together with Proposition~\ref{prop:monopoly-welfare}, Theorem~\ref{theorem:social-trade-optimal} reveals a sharp contrast between monopoly and duopoly. Under monopoly, full matching maximizes both social surplus and trade quantity. Under duopoly, restricting consumers to their less-preferred products can improve both objectives by intensifying price competition. Yet while consumer-optimal matching may additionally rely on exclusion, social- and trade-optimal matching do not require it.
\par\endgroup

Whether the no-exclusion conclusions extend to dependent valuations remains open. The two-step characterization continues to hold under the symmetric, jointly log-concave distribution in Section~\ref{sec:duopoly-environment}. However, for neither social surplus nor trade quantity have we found a proof that Step~2 can always be omitted, or a counterexample showing that exclusion is necessary at a global optimum.

\section{Discussion and Conclusion}\label{sec:conclusion}

\paragraph{Competitive fringe.}
Our restriction to one or two firms can be interpreted as a restriction to one or two firms with market power, rather than to the total number of sellers in the market. Suppose many fringe firms supply an identical product at a common marginal cost, and the platform makes at least two of these firms available to every consumer. Bertrand competition then supports pricing at marginal cost among the fringe firms. Their products effectively provide each consumer with an outside option whose payoff equals the value of the fringe product net of marginal cost (or zero, if the consumer prefers not to purchase). The platform's problem of designing access to one or two additional firms with market power can therefore be studied using our monopoly or duopoly framework, with valuations measured relative to this outside option.

\paragraph{Conclusion.}
Consider a platform that knows consumers' values for different products and chooses which products to show them. If the platform seeks to benefit consumers, it might seem natural to show everyone all products, or at least their favorites. Showing all products, after all, would seem to be the best way to intensify competition, since no firm could presume to have captive consumers. Yet the analysis shows that the platform can increase aggregate consumer surplus by excluding selected consumers or showing them only products they like less. Such restrictions can induce firms to lower their prices, and the resulting price reductions can more than compensate for the losses from exclusion and mismatching.

How should the platform choose which consumers to exclude or mismatch? In monopoly, the platform can only influence prices by excluding consumers. I show that, among all exclusion policies leaving the monopolist the same profit, there exists one whose induced demand is at least as large as the demand induced by any other rule at every price. This demand-preserving policy maximizes consumer surplus at that profit level, so the platform only needs to choose how much profit to leave the monopolist. The same analysis applies when exclusion is limited to a subset of consumers: a generalized demand-preserving exclusion policy is again consumer-optimal.

With two competing firms, I consider symmetric, log-concave valuation distributions and restrict my analysis to symmetric matchings and symmetric pure-strategy price equilibria. The consumer-optimal matching combines mismatching with exclusion and induces an equilibrium price below the full-matching benchmark. It can be constructed in two steps. First, omit products valued below the equilibrium price. Among consumers who value both products above this price, apply a cutoff rule along each diagonal of equal valuation differences: consumers with lower valuations are matched only to their less-preferred product, while those with higher valuations retain access to both products. Second, further exclude selected consumers who, after the first step, have access to only one firm's product. This exclusion follows the demand-preserving rule developed for monopoly.

In both monopoly and duopoly, improving aggregate consumer surplus need not come at the expense of any consumer. By suitably randomizing exclusion or mismatching, the platform can achieve an interim Pareto improvement over full matching: every consumer type is weakly better off in expected surplus conditional on her valuations, and a positive measure of types is strictly better off. The implications for social surplus and trade quantity, however, differ sharply. Under log-concavity, any improvement in consumer surplus through exclusion in monopoly necessarily reduces both social surplus and trade quantity. With competing firms, by contrast, mismatching can increase all three simultaneously. Moreover, under i.i.d.\ log-concave valuations, both social surplus and trade quantity can be maximized using the same two-step structure as consumer-optimal matching, but with an important simplification: the second step of exclusion is unnecessary.

\clearpage
\appendix
\begin{center}
{\Large\bfseries Appendix}
\end{center}

Appendix~\ref{app:monopoly} contains the monopoly proofs. Section~\ref{app:monopoly-improvement} establishes consumer-surplus and interim Pareto improvements through exclusion. Section~\ref{app:capacity} proves the capacity-constrained demand-preserving construction. It also proves Theorem~\ref{theorem:demand-preserving-monopoly} as a special case without capacity constraints. Section~\ref{app:interim-guarantees} applies this characterization to interim welfare guarantees and Pareto efficiency. Section~\ref{app:monopoly-welfare} proves that full matching maximizes social surplus and trade quantity under log-concavity.

Appendix~\ref{app:duopoly} contains the duopoly proofs. Section~\ref{app:local-improvement} proves that mismatching without exclusion can simultaneously generate an interim Pareto improvement for consumers and strictly increase social surplus and trade quantity. Section~\ref{app:downward-deviations} establishes a key step in proving the consumer-optimal matching characterization in Theorem~\ref{theorem:sequential-mismatching}: an optimal matching subject only to upward-deviation constraints also rules out profitable downward deviations. Section~\ref{app:social-trade-two-step} extends the two-step characterization to social-surplus and trade-quantity maximization. Section~\ref{app:iid-case} proves that, under i.i.d.\ log-concave valuations, both social surplus and trade quantity can be maximized without the second step of exclusion, and characterizes the trade-optimal matching as maximal mismatching. The supporting technical lemmas are collected at the end of the appendix.

\section{Monopoly}\label{app:monopoly}
\subsection[Consumer-Surplus Improvements and Pareto Improvements]{Consumer-Surplus Improvements\texorpdfstring{\\}{ }and Pareto Improvements}

\label{app:monopoly-improvement}

\begin{proof}[Proof of Proposition~\ref{prop:fullisImprovable}]

Fix a sufficiently small $\varepsilon>0$.
As in Figure~\ref{fig:monopoly-improvement}, let
$p_\varepsilon=p^M-\varepsilon$,
$m_\varepsilon=(\Pi^F)'(p_\varepsilon)>0$, and
$\widehat\pi_\varepsilon
=\Pi^F(p_\varepsilon)-p_\varepsilon m_\varepsilon$.
Choose $c_\varepsilon>p^M$ satisfying
$\Pi^F(c_\varepsilon)=\widehat\pi_\varepsilon$, and exclude
consumers in $[c_\varepsilon,c'_\varepsilon]$, where
$F(c'_\varepsilon)-F(c_\varepsilon)=m_\varepsilon$.
These cutoffs exist for sufficiently small $\varepsilon$
and converge to $p^M$, by the local negative curvature of
$\Pi^F$ and the positivity of $f$.

We verify that the construction remains valid without global
concavity. Choose a neighborhood of $p^M$ on which $\Pi^F$
is strictly concave. For sufficiently small $\varepsilon$,
the cutoffs and $p_\varepsilon$ lie in this neighborhood.
The tangent-line argument in
Section~\ref{sec:monopoly-improvement} then rules out all
deviations within the neighborhood: for prices below
$c_\varepsilon$, exclusion removes exactly $m_\varepsilon$
customers, while for higher prices, profit cannot exceed
$\Pi^F(c_\varepsilon)=\widehat\pi_\varepsilon$.
Outside the neighborhood, continuity and the uniqueness of
$p^M$ imply a uniform strict profit gap relative to
$\Pi^F(p^M)$. Since
$\widehat\pi_\varepsilon\to\Pi^F(p^M)$, these deviations are
also unprofitable for sufficiently small $\varepsilon$.
Hence $p_\varepsilon$ is optimal.

Finally, retained consumers with valuations above $p^M$
gain $\varepsilon$, while each excluded consumer loses at
most $c'_\varepsilon-p^M$. Thus,
\[
\begin{aligned}
&CS(\phi_\varepsilon,p_\varepsilon)-CS(\phi^{\mathrm{full}},p^M)\\
&\quad\ge
[D(p^M)-m_\varepsilon]\varepsilon
-m_\varepsilon(c'_\varepsilon-p^M).
\end{aligned}
\]
Since
$m_\varepsilon=-(\Pi^F)''(p^M)\varepsilon+o(\varepsilon)$
and $c'_\varepsilon\to p^M$, the first term is
$D(p^M)\varepsilon+o(\varepsilon)$, whereas the second is
$o(\varepsilon)$. The difference is therefore strictly
positive for sufficiently small $\varepsilon$.
Under our lowest-optimal-price convention, the monopolist selects a price
weakly below $p_\varepsilon$, which can only increase consumer surplus.

\end{proof}

\begin{proof}[Proof of Proposition~\ref{prop:monopoly-pareto}]
Fix sufficiently small $\varepsilon>0$. As established in the
proof of Proposition~\ref{prop:fullisImprovable}, the target price
$p_\varepsilon=p^M-\varepsilon$ can be supported by excluding
a mass
\[
m_\varepsilon=(\Pi^F)'(p_\varepsilon)>0
\]
of consumers with valuations above $c_\varepsilon>p^M$,
where $c_\varepsilon\to p^M$. The pricing argument depends
only on the total mass excluded above $c_\varepsilon$.

Choose a fixed $b\in(p^M,\bar v)$ sufficiently close to $p^M$
that $f(v)\geq\kappa>0$ on $[p^M,b]$. For sufficiently small
$\varepsilon$, we have $c_\varepsilon<b$. Consider the
exclusion probabilities
\[
\rho_\varepsilon(v)=
\begin{cases}
\displaystyle
\lambda_\varepsilon
\frac{\varepsilon}{v-p_\varepsilon},
&v\in[c_\varepsilon,b],\\[6pt]
0,&\text{otherwise},
\end{cases}
\]
where $\lambda_\varepsilon$ is chosen so that the total
excluded mass equals $m_\varepsilon$. Specifically, let
\[
A_\varepsilon
=\int_{c_\varepsilon}^{b}
\frac{\varepsilon}{v-p_\varepsilon}\,dF(v),
\qquad
\lambda_\varepsilon=\frac{m_\varepsilon}{A_\varepsilon}.
\]

We show that $0<\lambda_\varepsilon<1$ for sufficiently small
$\varepsilon$. Since $f(v)\geq\kappa$ on
$[c_\varepsilon,b]$,
\[
\begin{aligned}
A_\varepsilon
&=\int_{c_\varepsilon}^{b}
\frac{\varepsilon f(v)}{v-p_\varepsilon}\,dv\\
&\geq\kappa\varepsilon
\int_{c_\varepsilon}^{b}
\frac{dv}{v-p_\varepsilon}\\
&=\kappa\varepsilon
\log\frac{b-p_\varepsilon}
{c_\varepsilon-p_\varepsilon}.
\end{aligned}
\]
As $\varepsilon\downarrow0$, the logarithm diverges because
$c_\varepsilon\to p^M$ and
$p_\varepsilon\to p^M$, while $b>p^M$ is fixed. Meanwhile,
Assumption~\ref{ass:monopoly-improvement} implies
\[
m_\varepsilon
=-(\Pi^F)''(p^M)\varepsilon+o(\varepsilon).
\]
Consequently, $A_\varepsilon/m_\varepsilon\to\infty$,
so $0<\lambda_\varepsilon<1$ for sufficiently small
$\varepsilon$. Since $v-p_\varepsilon>\varepsilon$ for
$v\in[c_\varepsilon,b]$, we also have
$0<\rho_\varepsilon(v)<1$ on this interval.

By construction,
$\int\rho_\varepsilon(v)\,dF(v)=m_\varepsilon$.
Moreover, all exclusion occurs above $c_\varepsilon$.
The same pricing argument as in the proof of Proposition~\ref{prop:fullisImprovable}
therefore establishes that $p_\varepsilon$ is an optimal
monopoly price under
$\phi_\varepsilon=1-\rho_\varepsilon$.

It remains to verify the interim Pareto improvement.
For every $v>p^M$,
\[
\begin{aligned}
u(v;\phi_\varepsilon,p_\varepsilon)
&=(1-\rho_\varepsilon(v))(v-p_\varepsilon)\\
&=v-p^M+\varepsilon
  -\rho_\varepsilon(v)(v-p_\varepsilon)\\
&\geq v-p^M+(1-\lambda_\varepsilon)\varepsilon\\
&>u(v;\phi^{\mathrm{full}},p^M).
\end{aligned}
\]
Types with $v\in(p_\varepsilon,p^M]$ are never excluded
and strictly gain from the price reduction. Types with
$v\leq p_\varepsilon$ continue to obtain zero surplus.
Thus every consumer type is weakly better off, and a
positive measure of types is strictly better off.
Under our lowest-optimal-price convention, the monopolist selects
a price weakly below $p_\varepsilon$, which can only increase
every consumer's interim surplus.
\end{proof}

\subsection{Capacity-Constrained Demand Preservation}
\label{app:capacity}\label{app:demand-preserving-monopoly}

\begin{proof}[Proof of Lemma~\ref{lem:capacity-feasible-demand}]
Suppose first that an exclusion rule $\rho$ is feasible, and let $M$ be its induced demand loss. Since exclusion is drawn from the eligible population, $M(\bar v)=0$, and for $0\le a<b\le\bar v$,
\[
0\le M(a)-M(b)\le K(a)-K(b).
\]

Conversely, suppose $M$ satisfies these conditions. Since $K$ is continuous, so is $M$. The tails $K$ and $M$ define finite measures $\nu^e$ and $\lambda$. The upper increment bound implies $\lambda\le\nu^e$, so the Radon--Nikodym derivative $\rho=d\lambda/d\nu^e$ exists and lies in $[0,1]$ almost everywhere. Thus $\rho$ is a feasible exclusion rule inducing $M$.
\end{proof}

\begin{proof}[Proof of Proposition~\ref{prop:capacity-demand-preserving}]
Every capacity-feasible demand lies between $D-K$ and $D$, so its maximum profit lies in $[\underline\pi_K,\bar\pi]$.

Fix $\pi\in[\underline\pi_K,\bar\pi]$ with $\pi>0$, and let $M=M(D,K,\pi)$ be the demand-loss envelope defined in \eqref{eq:capacity-envelope}. It remains to verify that $M$ is feasible. For each $q$, let
\[
\psi_q(c)=g_\pi(q)-[K(q)-K(c)]_+.
\]
For $a<b$,
\[
0\le\psi_q(a)-\psi_q(b)\le K(a)-K(b).
\]
Taking the pointwise maximum over $q$ preserves these common increment bounds, so $M$ satisfies the increment restrictions in \eqref{eq:capacity-increments}. Moreover, $\pi\ge\underline\pi_K$ implies $g_\pi(q)=[D(q)-\pi/q]_+\le K(q)$ for every $q>0$, while $g_\pi(0)=0$. Since $K(\bar v)=0$,
\[
M(\bar v)=\max_q\{g_\pi(q)-K(q)\}=0.
\]
Hence Lemma~\ref{lem:capacity-feasible-demand} implies that $M$ is induced by a capacity-feasible exclusion rule.

The argument in the main text then establishes that this rule is demand-preserving and maximizes consumer surplus among capacity-feasible exclusion rules with maximum profit $\pi$.
\end{proof}

\begin{proof}[Proof of Theorem~\ref{theorem:demand-preserving-monopoly}]
We now prove Theorem~\ref{theorem:demand-preserving-monopoly} by taking $K=D$.
For $K=D$, $\underline\pi_K=0$, so the construction applies to every $\pi\in(0,\bar\pi]$. Setting $\phi_\pi=1-\rho_{\pi,D}$ gives a matching that attains the pointwise demand bound and maximizes consumer surplus at profit $\pi$. In this case the envelope reduces to
\[
M(D,D,\pi)(c)=\max_{q\in[c,\bar v]}g_\pi(q),
\]
because $D(q)\ge D(c)$ and $\pi/q\ge\pi/c$ imply $g_\pi(q)-D(q)+D(c)\le g_\pi(c)$ for $0<q<c$, while the term at $q=0$ is nonpositive.

To establish existence, consider pairs $(M,p)$ such that $M$ is a feasible demand loss and $p\in[0,\bar v]$ is optimal for $D-M$. These pairs form a nonempty compact set\footnote{
The technical subtlety is that the space of measurable
matchings is not directly compact. Instead, feasible
induced demand curves form a compact set under uniform
convergence: their increments are bounded by those
of the continuous full-matching demand curve.
Every such curve is implementable by a matching.
The optimal-pricing constraint is closed, and
consumer surplus is continuous.
}, and consumer surplus is continuous on it. Hence a consumer-optimal matching exists.

The resulting monopoly profit at a consumer optimum must be strictly positive. Otherwise, demand is zero at every positive price and consumer surplus is zero, whereas full matching yields strictly positive consumer surplus. Hence the preceding construction also gives a demand-preserving consumer-optimal matching, proving Theorem~\ref{theorem:demand-preserving-monopoly}.

\end{proof}

\subsubsection{The Target-Price Problem}
\label{app:capacity-fixed-price}

For $p=0$, the problem is feasible only if maximal exclusion eliminates all demand, in which case maximal exclusion is optimal and consumer surplus is zero. Henceforth, let $p>0$.

Define
\begin{equation}\label{eq:cutoff-upper}
\overline\pi_p
:=\min\left\{pD(p),\ \inf_{p<q\le\bar v}\frac{pq}{q-p}[D(p)-D(q)]\right\},
\end{equation}
and
\begin{equation}\label{eq:cutoff-lower}
\underline\pi_p
:=\max\left\{\underline\pi_K,\ \sup_{0<q<p}\frac{pq}{p-q}[D(q)-D(p)-K(q)+K(p)]\right\}.
\end{equation}
The problem $\mathrm{CM}(D,K,p)$ is feasible if and only if $\underline\pi_p\le\overline\pi_p$.

\begin{proof}[Proof of the feasibility bounds and Corollary~\ref{cor:capacity-fixed-price}]
Suppose a capacity-feasible exclusion rule induces demand loss $M$ and makes $p$ optimal with profit $\pi$. Then $\pi\le pD(p)$. For $q>p$, $M(q)\le M(p)$, so the price incentive constraint implies
\[
q[D(q)-M(p)]\le\pi.
\]
Substituting $M(p)=D(p)-\pi/p$ and rearranging gives
\[
\pi\le\frac{pq}{q-p}[D(p)-D(q)].
\]
Hence $\pi\le\overline\pi_p$.

Maximal exclusion gives $\pi\ge\underline\pi_K$. For $0<q<p$, capacity implies $M(q)\le M(p)+K(q)-K(p)$, so the price incentive constraint implies
\[
q[D(q)-M(p)-K(q)+K(p)]\le\pi.
\]
Substituting for $M(p)$ and rearranging gives
\[
\pi\ge\frac{pq}{p-q}[D(q)-D(p)-K(q)+K(p)].
\]
Hence $\pi\ge\underline\pi_p$, proving necessity.

Conversely, fix $0<\pi\in[\underline\pi_p,\overline\pi_p]$, and take $M=M(D,K,\pi)$. By Proposition~\ref{prop:capacity-demand-preserving}, $M$ is feasible and caps profit at $\pi$. Write $m=D(p)-\pi/p\ge0$. Using the functions $\psi_q$ defined above,
\[
M(p)=\max_{q\in[0,\bar v]}\psi_q(p),
\qquad
\psi_p(p)=g_\pi(p)=m.
\]
For $q>p$, the upper bound on $\pi$ implies $D(q)-\pi/q\le m$. Since $K(q)\le K(p)$ and $m\ge0$, it follows that $\psi_q(p)=g_\pi(q)\le m$. For $0<q<p$, the lower bound on $\pi$ implies
\[
D(q)-\pi/q-[K(q)-K(p)]\le m.
\]
Since $K(q)-K(p)\ge0$ and $m\ge0$, the same bound holds after replacing $D(q)-\pi/q$ by its positive part. Thus $\psi_q(p)\le m$. Finally, $\psi_0(p)=-[K(0)-K(p)]_+\le0\le m$. Therefore $M(p)=m$, so
\[
p[D(p)-M(p)]=\pi.
\]
Profit at $p$ attains the ceiling $\pi$, and $p$ is optimal. If the interval includes zero, then $\underline\pi_K=0$, so $D-K=0$ and maximal exclusion attains zero profit at $p$. This proves sufficiency.

The highest feasible profit is therefore $\overline\pi_p$. If it is positive, the constructed demand-preserving rule at that profit supports $p$. The envelope is nonincreasing in $\pi$, and the pointwise lower-bound argument in the main text implies that its demand loss is no greater than that of any other capacity-feasible exclusion rule supporting $p$, whose maximum profit cannot exceed $\overline\pi_p$. It therefore maximizes consumer surplus at $p$. If $\overline\pi_p=0$, every feasible rule has zero sales and consumer surplus, and maximal exclusion is optimal. This proves Corollary~\ref{cor:capacity-fixed-price}.
\end{proof}

\subsection{Welfare Guarantees and Interim Pareto Efficiency}
\label{app:interim-guarantees}

\begin{proof}[Proof of Corollary~\ref{cor:interim-guarantees}]
Fix a target price $p$. The interim welfare guarantees
can be satisfied at $p$ only if
$h(v)\leq(v-p)_+$ almost everywhere. When this condition
holds, the guarantees are equivalent to the exclusion
constraints determined by $k_{h,p}$. Thus the fixed-price
problem is exactly $\mathrm{CM}(D,K_{h,p},p)$, which is
feasible if and only if $p\in\mathcal P(h)$.

By Corollary~\ref{cor:capacity-fixed-price}, for each such
price the demand-preserving exclusion rule at profit
$\pi_p$ maximizes consumer surplus. Optimizing the
resulting consumer surplus over $\mathcal P(h)$ therefore
solves problem~\eqref{eq:interim-guarantee-problem}.

Existence follows by the same compactness argument as in the proof of Theorem~\ref{theorem:demand-preserving-monopoly}. The only difference is that the exclusion capacity $K_{h,p}$ now depends on the target price $p$, but it varies continuously with $p$, so the same argument applies.
\end{proof}

\begin{proof}[Proof of Corollary~\ref{cor:interim-pareto-efficiency}]
Let $h^{\widetilde\phi}(v)=u(v;\widetilde\phi,\widetilde p)$.
The benchmark matching is feasible for the problem
in Corollary~\ref{cor:interim-guarantees}, so the
optimal value is at least
$CS(\widetilde\phi,\widetilde p)$.

If the inequality is strict, the demand-preserving
matching characterized by
Corollary~\ref{cor:interim-guarantees} weakly improves
almost every consumer type's interim surplus and strictly
improves it for a positive measure of types. It
therefore interim Pareto improves $\widetilde\phi$.

Conversely, any interim Pareto improvement of
$\widetilde\phi$ satisfies the guarantees $h^{\widetilde\phi}$ and
strictly increases aggregate consumer surplus.
Consequently, the optimum in
Corollary~\ref{cor:interim-guarantees} strictly exceeds
$CS(\widetilde\phi,\widetilde p)$. The stated weak inequalities
therefore hold exactly when $\widetilde\phi$ is interim
Pareto efficient.
\end{proof}

\subsection[Full Matching Maximizes Social Surplus and Trade Quantity]{Full Matching Maximizes Social Surplus\texorpdfstring{\\}{ }and Trade Quantity}
\label{app:monopoly-welfare}

\begin{proof}[Proof of Proposition~\ref{prop:monopoly-welfare}]
Consider an arbitrary matching $\phi$, and let $p_\phi$ be
its induced optimal monopoly price. We first show that
\begin{equation}
\label{eq:welfare-quantity}
D_\phi(p_\phi)\leq D(p^M).
\end{equation}
If the induced maximum profit is zero, the matching generates
no transactions or consumer surplus. We may therefore restrict
attention to positive-profit matchings. Such a matching has
$p_\phi\geq\underline v$, since its demand is constant below
the valuation support.

If $p^M=\underline v$, full matching serves every consumer
and thus maximizes both social surplus and trade quantity.
Since no positive-sales matching can induce a lower optimal
price, consumer surplus cannot strictly improve either.
Henceforth suppose that $p^M$ lies in the interior of the support.

For $p_\phi<p^M$, the monopolist cannot profitably raise its
price from $p_\phi$ to $p_\phi+\varepsilon$, so
\[
(p_\phi+\varepsilon)D_\phi(p_\phi+\varepsilon)
\leq p_\phi D_\phi(p_\phi).
\]
Rearranging and using $\phi(v)\leq1$,
\[
\begin{aligned}
\varepsilon D_\phi(p_\phi)
&\leq(p_\phi+\varepsilon)
\bigl[D_\phi(p_\phi)-D_\phi(p_\phi+\varepsilon)\bigr]\\
&=(p_\phi+\varepsilon)
\int_{p_\phi}^{p_\phi+\varepsilon}\phi(v)\,dF(v)\\
&\leq(p_\phi+\varepsilon)
[F(p_\phi+\varepsilon)-F(p_\phi)].
\end{aligned}
\]
Dividing by $\varepsilon$ and letting
$\varepsilon\downarrow0$ gives
\begin{equation}
\label{eq:welfare-demand-bound}
D_\phi(p_\phi)\leq p_\phi f(p_\phi),
\end{equation}
where $f(\underline v)$ is interpreted as its right-hand limit.

Let $h=f/D$ denote the hazard rate. Log-concavity implies that
$h$ is nondecreasing, and the full-matching first-order condition
gives $p^Mh(p^M)=1$. Thus $ph(p)<1$ for every $p<p^M$ in
the interior of the support. Since $f=hD$ and $D'=-hD$,
\[
\frac{d}{dp}[pf(p)]
=D(p)\bigl[h(p)(1-ph(p))+ph'(p)\bigr]>0
\qquad\text{a.e. on }(\underline v,p^M).
\]
Since $h$ is nondecreasing and locally absolutely continuous,
this derivative is positive almost everywhere, so $pf(p)$ is
strictly increasing.\footnote{Log-concavity implies local absolute
continuity of $f$, by a standard property of concave functions.
Since $D>0$, the hazard rate $h=f/D$ is also locally absolutely
continuous.} Consequently,
\begin{equation}
\label{eq:welfare-strict-bound}
pf(p)<p^Mf(p^M)=D(p^M)
\qquad(\underline v\leq p<p^M),
\end{equation}
with the lower endpoint included by taking right limits.
Combining \eqref{eq:welfare-demand-bound} and
\eqref{eq:welfare-strict-bound}, any positive-profit matching
inducing $p_\phi<p^M$ satisfies
\begin{equation}
\label{eq:welfare-strict-quantity}
D_\phi(p_\phi)<D(p^M).
\end{equation}
For $p_\phi\geq p^M$, we instead have
\[
D_\phi(p_\phi)\leq D(p_\phi)\leq D(p^M).
\]
Thus full matching maximizes trade quantity. Since it serves
the consumers with the highest valuations, it also maximizes
social surplus. If trade quantity is strictly lower, the
social-surplus loss is at least
$p^M[D(p^M)-D_\phi(p_\phi)]>0$.

Finally, a matching that strictly increases aggregate consumer
surplus must induce $p_\phi<p^M$, since restricting access
cannot increase consumer surplus at a given price, and
consumer surplus under full matching is nonincreasing in price.
By \eqref{eq:welfare-strict-quantity}, such a matching strictly
reduces trade quantity and therefore social surplus.
\end{proof}

\section{Duopoly}\label{app:duopoly}
\subsection{Mismatching Can Improve upon Full Matching}

\label{app:local-improvement}

\begin{proof}[Proof of Proposition~\ref{prop:local-improvement}]
Since $(p^*,p^*)\in\operatorname{int}V$, choose a sufficiently small neighborhood of $(p^*,p^*)$ and fixed $0<m\le M<\infty$ such that $m\le f(v)\le M$ throughout this neighborhood.

Let $D^F(q\mid p)=D_i(q,p\mid\phi^{\mathrm{full}})$. Extend $f$ by zero outside $V$. Recall that
\[
\ell_p^F=\int_0^p f(p,y)\,dy+\int_p^{\bar v}f(t,t)\,dt.
\]
Under the maintained assumptions, $D^F(p\mid p)-p\ell_p^F$ is locally Lipschitz at $p^*$. Since the full-matching first-order condition gives $D^F(p^*\mid p^*)-p^*\ell_{p^*}^F=0$, there is $0<L<\infty$ such that $D^F(p\mid p)-p\ell_p^F\le L(p^*-p)$ for $p<p^*$ sufficiently close to $p^*$.

Fix $d_1>0$ sufficiently small, and then choose $d_0\in(0,d_1)$ sufficiently small that
\begin{equation}\label{eq:local-mismatch-bound}
\frac{p^*m}{4}\log\frac{d_1}{d_0}>L.
\end{equation}
Choose $\delta>0$ sufficiently small that $p^*>2\delta$ and $f\ge m$ on $[p^*-2\delta,p^*]^2$. Let $a=m\delta p^*/2>0$, and fix $s_0\in(0,d_0)$ such that $M s_0(d_1-d_0)<a$. For $p=p^*-\varepsilon$, consider the regions
\[
R_p=\{(p+s+d,p+s):0<s<s_0,\ d_0<d<d_1\}
\]
and their mirrors. For sufficiently small $\varepsilon>0$, these regions are contained in the chosen neighborhood. Match each consumer in these regions only to her less-preferred product with probability $\rho_\varepsilon(s,d)=\varepsilon/(2d)$, and otherwise retain full matching. All other consumers remain fully matched. Take $\varepsilon<d_0$, so these probabilities lie in $[0,1]$. Denote the resulting symmetric matching with no exclusion by $\phi_p$. At the target price, every consumer matched only to her less-preferred product still purchases, and symmetry preserves each firm's sales.

We first show that $\phi_p$ supports $(p,p)$. Consider an upward deviation to $p+x$, where $0<x\le s_0$. Since mismatching preserves each firm's sales at $p$ and weakly reduces its sales at higher prices, the profit gain from this deviation is no greater than under full matching. By the full-matching demand-loss bound \eqref{eq:full-matching-demand-loss} in Appendix~\ref{app:downward-deviations},
\[
D^F(p\mid p)-D^F(p+x\mid p)\ge\frac{p\ell_p^F}{p+x}x,
\]
the profit gain under full matching is at most $x[D^F(p\mid p)-p\ell_p^F]$. It is therefore at most $L\varepsilon x$. After the price increase, a consumer matched only to firm $i$ stops purchasing if $s<x$. Thus the expected reduction in deviation demand is at least
\[
\int_0^x\int_{d_0}^{d_1}\frac{\varepsilon}{2d}f(p+s+d,p+s)\,dd\,ds
\ge\frac{m\varepsilon x}{2}\log\frac{d_1}{d_0}.
\]
Consequently, deviation profit is reduced by at least
\[
\frac{pm\varepsilon x}{2}\log\frac{d_1}{d_0}>L\varepsilon x,
\]
where the strict inequality follows from \eqref{eq:local-mismatch-bound} for $p>p^*/2$. Hence every such upward deviation is unprofitable.

Now consider $q>p+s_0$. Mismatching weakly reduces demand from an upward deviation while preserving demand at $p$, so its deviation gain is weakly smaller than under full matching. For $p$ sufficiently close to $p^*$, $p+s_0>p^*$ and the full-matching best response to $p$ lies below $p+s_0$. By log-concavity,
\[
\Pi_i(q,p\mid\phi^{\mathrm{full}})-\Pi_i(p,p\mid\phi^{\mathrm{full}})
\le
\Pi_i(p+s_0,p\mid\phi^{\mathrm{full}})-\Pi_i(p,p\mid\phi^{\mathrm{full}}).
\]
As $p\uparrow p^*$, the right-hand side converges to $\Pi_i(p^*+s_0,p^*\mid\phi^{\mathrm{full}})-\Pi_i(p^*,p^*\mid\phi^{\mathrm{full}})<0$, because $p^*$ is the unique full-matching best response to $p^*$. Hence all such upward deviations are unprofitable for $p$ sufficiently close to $p^*$.

Downward deviations are also unprofitable. By log-concavity, no downward deviation is profitable under full matching, since the full-matching best response to $p$ lies above $p$. Mismatching preserves demand at $p$ and makes price cutting weakly less attractive. Hence no downward deviation is profitable under $\phi_p$. Therefore $\phi_p$ supports $(p,p)$.

It remains to compare welfare. A selected consumer with valuations $(p+s+d,p+s)$ obtains surplus $s+d-\varepsilon$ under full matching at $p^*$, since $\varepsilon<d_0<d$. Under $\phi_p$ at $p$, her expected surplus is
\[
(1-\rho_\varepsilon)(s+d)+\rho_\varepsilon s
=s+d-\frac{\varepsilon}{2}.
\]
She therefore gains $\varepsilon/2$. The same calculation applies to the mirror types. Everyone else remains fully matched and weakly benefits from the price reduction. Hence $\phi_p$ interim Pareto improves full matching.

Trade quantity strictly increases: mismatched consumers still purchase at $p$, and full matching at the lower price serves more consumers than at $p^*$. To compare social surplus, take $0<\varepsilon<\delta$. Types in the rectangle
\[
v_1\in[p^*-\varepsilon,p^*],\qquad
v_2\in[p^*-2\delta,p^*-\delta]
\]
do not trade at $p^*$ but buy product 1 at $p=p^*-\varepsilon$, apart from null boundary types. Their mass is at least $m\delta\varepsilon$, and each creates social surplus at least $p^*/2$. Writing $SS$ for social surplus, it follows that
\[
SS(\phi^{\mathrm{full}},(p,p))-SS(\phi^{\mathrm{full}},(p^*,p^*))
\ge a\varepsilon.
\]
The social-surplus loss from mismatching is at most
\[
2\int_0^{s_0}\int_{d_0}^{d_1}\frac{\varepsilon}{2}f(p+s+d,p+s)\,dd\,ds
\le M\varepsilon s_0(d_1-d_0).
\]
Therefore,
\[
SS(\phi_p,(p,p))-SS(\phi^{\mathrm{full}},(p^*,p^*))
\ge\varepsilon\bigl[a-Ms_0(d_1-d_0)\bigr]>0,
\]
proving the proposition.
\end{proof}

\subsection{No Downward Deviations}\label{app:consumer-optimal}

\label{app:downward-deviations}

\paragraph{Existence.}
The feasible set is nonempty because assigning every consumer the empty consideration set satisfies all upward incentive constraints. It is compact, and consumer surplus is continuous on it. Hence an optimum exists.

\begin{proof}[Proof of Lemma~\ref{claim:upward-exactness}]
Fix $0<p<p^*$ and a normalized optimum $\phi^R$ of the relaxed problem. Let $\ell=-\left.\partial_qD^F(q\mid p)\right|_{q=p}\ge0$ denote the negative of the own-price derivative of full-matching demand at $(p,p)$.

\emph{Two full-matching bounds.} We first establish two properties of full-matching demand used below. Log-concavity makes the full-matching best response $b(r)$ unique and continuous. Since $b(0)>0$ and $p^*$ is its smallest fixed point, $b(p)>p$; otherwise continuity would give an earlier fixed point. Log-concavity then implies that profit under full matching is weakly increasing at $p$. Thus the first-order condition is nonnegative, which gives
\[
D^F(p\mid p)\ge p\ell.
\]
Applying \eqref{eq:LC6} to the log-concave full-matching demand gives, for $q>p$,
\[
D^F(q\mid p)
\le D^F(p\mid p)
\exp\!\left[-\frac{\ell(q-p)}{D^F(p\mid p)}\right].
\]
Using $1-e^{-z}\ge z/(1+z)$ for $z\ge0$, together with $D^F(p\mid p)\ge p\ell$, we obtain
\begin{equation}\label{eq:full-matching-demand-loss}
D^F(p\mid p)-D^F(q\mid p)
\ge \frac{\ell(q-p)}{1+\ell(q-p)/D^F(p\mid p)}
\ge \frac{p\ell}{q}(q-p),\qquad q>p.
\end{equation}
If $\ell=0$, the bound follows from monotonicity. If $D^F(q\mid p)=0$, it follows directly from $D^F(p\mid p)\ge p\ell$. Thus \eqref{eq:full-matching-demand-loss} bounds the demand loss from an upward deviation under full matching.

\emph{Assignments under $\phi^R$.} We next record what the normalization implies about $\phi^R$. If both values are below $p$, the consumer receives the empty set. If exactly one value is above $p$, she receives either that product alone or the empty set. If both values are above $p$, she receives full consideration, only her less-preferred product, or the empty set; preferred-only matching was removed by the normalization.

We also establish one additional property of the less-preferred-only assignment. We have $s<d$ for almost every consumer mismatched to her less-preferred product only under $\phi^R$. Otherwise, if $s\ge d$, replacing less-preferred-only matching by full consideration preserves demand at $(p,p)$, weakly reduces demand from every upward deviation, and increases consumer surplus, contradicting optimality.

\emph{Price cuts are unprofitable whenever $D_i(p,p\mid\phi^R)\ge p\ell$.} The assignment structure above makes the effect of a price cut transparent. Every consumer with a nonempty consideration set already buys at $(p,p)$, except for null boundary types. A consumer assigned the empty set cannot be attracted by a price cut because no product is displayed to her. Thus the only additional customers firm $i$ can gain by cutting its price are consumers who consider both products, prefer firm $j$ at $(p,p)$, but switch to firm $i$ because of the price cut.

Define the primitive preference-gap density
\[
h_p(d)=\int_{\mathbb R}f(x,x+d)\mathbf 1\{x\ge p,\ x+d\ge p\}\,dx,
\]
where $f$ is extended by zero outside $V$. The map $(x,d)\mapsto(x,x+d)$ is affine, so log-concavity of $f$ implies that $f(x,x+d)$ is log-concave in $(x,d)$. The restrictions $x\ge p$ and $x+d\ge p$ define a convex set, so imposing them preserves log-concavity. Integrating over $x$ then implies, by Pr\'ekopa's theorem, that $h_p$ is log-concave.\footnote{See \citet[Theorem 6]{prekopa1973logarithmic}.} Symmetry of $f$ gives $h_p(d)=h_p(-d)$. Hence $h_p$ is maximized at zero.

Moreover,
\[
\ell=h_p(0)+\int_0^p f(p,y)\,dy\ge h_p(0).
\]
Therefore, for a price cut $t\in(0,p)$,
\[
D_i(p-t,p\mid\phi^R)-D_i(p,p\mid\phi^R)
\le\int_0^t h_p(d)\,dd\le t\ell.
\]
The profit gain from the cut is consequently at most
\[
\begin{aligned}
&(p-t)\bigl[D_i(p,p\mid\phi^R)+t\ell\bigr]-pD_i(p,p\mid\phi^R)\\
&\qquad=t\bigl[p\ell-D_i(p,p\mid\phi^R)\bigr]-t^2\ell,
\end{aligned}
\]
which is nonpositive whenever $D_i(p,p\mid\phi^R)\ge p\ell$.

\emph{Optimality requires $D_i(p,p\mid\phi^R)\ge p\ell$.} Suppose, toward a contradiction, that $D_i(p,p\mid\phi^R)<p\ell$. The mass of consumers excluded by $\phi^R$ who would buy from firm $i$ under full matching is $D^F(p\mid p)-D_i(p,p\mid\phi^R)$, which is at least $p\ell-D_i(p,p\mid\phi^R)>0$. Take a subset of these consumers and restore them to full consideration. Choose the total restored mass per firm to be $m$, where $0<m<p\ell-D_i(p,p\mid\phi^R)$. We select consumers according to their reservation prices under full matching, restoring those with lower reservation prices first until the restored mass reaches $m$, and proceed symmetrically for the two firms. Atomlessness allows us to choose exactly this mass. Let $c$ denote the resulting reservation-price cutoff and $\phi'$ the resulting matching. Then $D_i(p,p\mid\phi')=D_i(p,p\mid\phi^R)+m<p\ell$.

We claim that, for every $p<q\le c$,
\begin{equation}\label{eq:restored-demand-loss}
D_i(p,p\mid\phi')-D_i(q,p\mid\phi')
\ge D^F(p\mid p)-D^F(q\mid p).
\end{equation}
The right-hand side is the mass of consumers who buy from firm $i$ under full matching at $(p,p)$ but stop buying from $i$ when its price rises to $q$. We show that each part of this mass generates an equal mass of demand loss for firm $i$ under $\phi'$.

Take such a consumer. If she is excluded under $\phi^R$, her full-matching reservation price is below $q\le c$, so she is restored to full consideration under $\phi'$. She therefore buys from $i$ at $p$ and stops buying from $i$ at $q$, contributing the same mass to the left-hand side of \eqref{eq:restored-demand-loss}.

Suppose instead that she has a nonempty assignment under $\phi^R$. If only her value for product $i$ exceeds $p$, normalization assigns her $\{i\}$. Her reservation price is the same as under full matching, so she is also lost from firm $i$'s demand at $q$. If both values exceed $p$ and she receives full consideration, the same conclusion holds.

Finally, suppose both values exceed $p$ and she does not receive full consideration. Since she buys from $i$ under full matching, product $i$ is her preferred product. Normalization rules out preferred-only matching, so her nonempty assignment under $\phi^R$ must be $\{j\}$, her less-preferred product. She therefore does not herself contribute to firm $i$'s demand under $\phi^R$. Instead, consider her mirror type. By symmetry, an equal mass of mirror consumers is assigned $\{i\}$. Writing the original type as $(p+s+d,p+s)$, her full-matching reservation price for firm $i$ is $p+d$, while the mirror consumer assigned $\{i\}$ has reservation price $p+s$. We established above that $s<d$. Hence, if the original consumer is lost under full matching when firm $i$ raises its price to $q$, the equal mass of mirror consumers assigned $\{i\}$ is also lost under $\phi'$.

Thus every unit of firm $i$'s demand lost between $p$ and $q$ under full matching corresponds to an equal unit of demand that is also lost under $\phi'$. This proves \eqref{eq:restored-demand-loss}.

We now combine \eqref{eq:restored-demand-loss} with the full-matching demand-loss bound \eqref{eq:full-matching-demand-loss}. For $q\in(p,c]$, the profit gain from deviating to $q$ is
\begin{equation}\label{eq:restored-profit-gain}
\begin{aligned}
&qD_i(q,p\mid\phi')-pD_i(p,p\mid\phi')\\
&\quad=(q-p)D_i(p,p\mid\phi')
-q\bigl[D_i(p,p\mid\phi')-D_i(q,p\mid\phi')\bigr]\\
&\quad\le(q-p)D_i(p,p\mid\phi')
-q\bigl[D^F(p\mid p)-D^F(q\mid p)\bigr]
&&\text{by \eqref{eq:restored-demand-loss}}\\
&\quad\le(q-p)D_i(p,p\mid\phi')-p\ell(q-p)
&&\text{by \eqref{eq:full-matching-demand-loss}}\\
&\quad=(q-p)\bigl[D_i(p,p\mid\phi')-p\ell\bigr]<0.
\end{aligned}
\end{equation}

For $q>c$, none of the restored consumers buys from firm $i$ at $q$, so $D_i(q,p\mid\phi')=D_i(q,p\mid\phi^R)$. Since $\phi^R$ satisfies all upward incentive constraints while restoration increases demand at $(p,p)$ by $m$, the deviation remains unprofitable under $\phi'$.

Hence $\phi'$ is feasible for the relaxed problem. The restored consumers obtain positive consumer surplus except on a null boundary, so $\phi'$ yields strictly greater consumer surplus than $\phi^R$, contradicting optimality. Therefore $D_i(p,p\mid\phi^R)\ge p\ell$. By the price-cut argument above, neither firm has a profitable downward deviation.
\end{proof}

\subsection[Two-Step Structure for Social-Surplus and Trade-Quantity Optimality]{Two-Step Structure for Social-Surplus\texorpdfstring{\\}{ }and Trade-Quantity Optimality}
\label{app:social-trade-two-step}

\begin{proof}[Two-step extension of Theorem~\ref{theorem:sequential-mismatching}]
Under the assumptions of Section~\ref{sec:duopoly-environment}, fix $0<p<p^*$. For either social surplus or trade quantity, we show that an optimum over $\mathcal M(p)$ admits the two steps of Theorem~\ref{theorem:sequential-mismatching}. Independence is not imposed.

First maximize the chosen objective subject only to upward price incentive constraints. When maximizing trade quantity, among all trade-quantity maximizers, choose one that also maximizes social surplus. This tie-break rules out assignments that can be rearranged to increase social surplus without changing trade quantity or violating the upward incentive constraints. The existence argument in Appendix~\ref{app:downward-deviations} applies to these continuous linear objectives. Its normalization also applies: omit products valued below $p$ and replace preferred-only assignments for consumers with both values above $p$ by full consideration. Both objectives are preserved and upward demand weakly decreases. A retained less-preferred-only assignment must have $s<d$: if $s\ge d$, full consideration preserves trade quantity, strictly raises social surplus, and weakly decreases upward demand because its reservation price is $p+d\le p+s$.

The restoration argument in Appendix~\ref{app:downward-deviations} continues to rule out downward deviations, but its optimality step must use the new objective. Let $\phi$ be the resulting normalized optimum, and let $\ell=-\left.\partial_qD^F(q\mid p)\right|_{q=p}$. Since $D^F(p\mid p)\ge p\ell$ and symmetry identifies $D^F(p\mid p)-D_i(p,p\mid\phi)$ as the excluded affordable mass per firm, if $D_i(p,p\mid\phi)<p\ell$ we can symmetrically restore a mass $0<m<p\ell-D_i(p,p\mid\phi)$ per firm of excluded affordable consumers, taking the lowest full-consideration reservation prices first, up to a cutoff $c$. Give restored consumers all their affordable products, and denote the resulting matching by $\phi'$. Thus $D_i(p,p\mid\phi')=D_i(p,p\mid\phi)+m<p\ell$. For $p<q\le c$, every full-matching customer lost at $q$ is either restored, already shown every affordable product, or represented by an equal mass of mirror consumers matched only to their less-preferred products whose reservation price $p+s$ is below $p+d$. Consequently the restored policy's demand loss is at least $D^F(p\mid p)-D^F(q\mid p)$. Equation~\eqref{eq:full-matching-demand-loss} bounds its deviation-profit gain by $(q-p)[D_i(p,p\mid\phi')-p\ell]<0$. For $q>c$, restored consumers add no deviation demand, while target profit increases. Restoration therefore preserves upward feasibility and strictly increases both social surplus and trade quantity, a contradiction. Thus $D_i(p,p\mid\phi)\ge p\ell$. Joint symmetry and log-concavity give the same price-cut bound as in Appendix~\ref{app:downward-deviations},
\[
D_i(q,p\mid\phi)-D_i(p,p\mid\phi)\le(p-q)\ell\qquad(0\le q<p),
\]
which makes every downward deviation unprofitable. This argument applies again after the rearrangements below.

At each gap $d$, exchange equal conditional masses, together with their mirror types, as in the proof of Theorem~\ref{theorem:sequential-mismatching}. Moving full consideration from lower $s$ to higher $s$, and the less-preferred-only assignment in the opposite direction, preserves both objectives: the same mass trades and the same mass incurs mismatch loss $d$. Upward demand weakly falls, since the full-consideration reservation price $p+d$ is independent of $s$, whereas the reservation price under matching only to the less-preferred product $p+s$ increases with $s$. Exchanging full consideration at lower $s$ with exclusion at higher $s$ preserves trade quantity, strictly raises social surplus, and leaves upward demand unchanged. Social optimality, or the social-surplus tie-break, excludes the latter inversions. The same conditional rearrangement therefore assigns full consideration to an upper tail of $s$ at each $d$, with a measurable cutoff; below it consumers are matched only to their less-preferred products or excluded.

Provisionally restore each excluded consumer below the cutoff to her less-preferred singleton, and each excluded sole-affordable consumer to that product. This gives the deterministic Step~1 matching $\widetilde\phi$; the inherited exclusions are Step~2. They establish feasibility of $\mathrm{CM}(D_i(\cdot,p\mid\widetilde\phi),K_i(\cdot\mid\widetilde\phi),p)$, since exclusions from one firm's singleton population never affect the other firm's demand. Averaging exclusions conditional on the assigned product value $x$ preserves every demand and both objectives.

Finally, the capacity-feasible exclusion rule in Corollary~\ref{cor:capacity-fixed-price} maximizes all three objectives. At fixed $p>0$, highest feasible target profit gives highest target sales. More strongly, that corollary's envelope minimizes the demand loss $M$ pointwise among all capacity-feasible exclusion rules supporting $p$, including those with lower target profit. Excluding singleton consumers loses quantity $M(p)$ per firm and social value
\[
\int_p^{\bar v}x\,[-dM(x)]
=pM(p)+\int_p^{\bar v}M(q)\,dq.
\]
It therefore minimizes both losses, as well as the consumer-surplus loss $\int_p^{\bar v}M(q)\,dq$; equivalently, $SS=CS+pQ$ at the target price. If the highest feasible profit is zero, incentive compatibility forces zero demand at every positive price, so maximal exclusion gives the same zero quantity and social surplus, as in the original corollary. Apply the exclusion rule independently and symmetrically to both singleton populations; conditional cutoffs in the other valuation implement its exclusion fractions deterministically.

The global result follows by the same argument as Corollary~\ref{cor:symmetric-consumer-optimum}, using Proposition~\ref{prop:local-improvement} to rule out optimal prices at or above $p^*$.\qedhere

\end{proof}

\subsection{The I.I.D. Case}\label{app:iid-case}
Throughout this subsection, valuations are i.i.d.\ with common marginal distribution $G$ and continuously differentiable, log-concave density $g$ on $[0,\bar v]$, positive on $(0,\bar v)$. Write $S=1-G$ and define $m=S/g$ on $(0,\bar v)$. Endpoint values of $m$ are understood as one-sided limits, with $m(\bar v)=0$ and $1/m(0)=g(0)$, where $1/(+\infty)=0$. Derivative formulas below are applied in the interior; endpoint inequalities are obtained by one-sided limits.
\subsubsection{Social-Optimal Matching}

\label{app:social-optimal}

\begingroup
\edef\socialsavedequation{\number\value{equation}}
\edef\socialsavedlemma{\number\value{lemma}}
\renewcommand{\theHequation}{social.\arabic{equation}}
\renewcommand{\thelemma}{I.1}
\renewcommand{\theHlemma}{social.price.\arabic{lemma}}
\newcommand{\Di}{D_1}
\newcommand{\fmat}{\phi^{\mathrm{full}}}
\newcommand{\pn}{p_n}
\newcommand{\phin}{\phi_n}
\newcommand{\LH}{\lambda_h^n}
\newcommand{\hphi}{\phi_{n,h}}
\newcommand{\vbar}{\bar v}
\newcommand{\ind}[1]{\mathbf 1\{#1\}}
\newcommand{\order}{\{v_2<v_1\}}
\newcommand{\ssur}[2]{SS(#1,(#2,#2))}
\newcommand{\step}[1]{\par\medskip\noindent\textbf{#1}\enspace}
\newcommand{\subproof}[1]{\par\medskip\noindent\textit{#1}\enspace}
\noindent\textit{Proof of Theorem~\ref{theorem:social-trade-optimal}(a).}
By symmetry, we write demand for firm 1 throughout this proof and Appendix~\ref{app:technical-lemmas}.

\step{Step 1: Restrict attention to relevant prices.}
Proposition~\ref{prop:local-improvement} establishes a symmetric equilibrium with strictly higher social surplus than full matching at $p^*$. Any matching $\phi$ and common price $p$ that improve on this benchmark must satisfy
\begin{equation*}
 p<p^*,\qquad \Di(p,p\mid\phi)>\Di(p^*,p^*\mid\fmat).
\end{equation*}
Indeed, full matching maximizes social surplus at any fixed price, so an outcome with $p\ge p^*$ cannot improve on the benchmark. Moreover, full matching at $p^*$ serves the consumers with the highest preferred valuations. An allocation with weakly fewer transactions therefore cannot generate higher social surplus. These comparisons depend only on the resulting allocations, so they also apply to the finite relaxations below.

We next bound the prices of improving equilibria away from zero. Appendix~\ref{app:trade-optimal} establishes that every symmetric equilibrium satisfies
\begin{equation*}
\Di(p,p\mid\phi)\le B(p),\qquad B(p)=p\left[g(p)+\int_p^{\vbar}g(v)^2\,dv\right].
\end{equation*}
Let $\underline p$ be the smallest positive solution of $B(p)=\Di(p^*,p^*\mid\fmat)$. Such a solution exists because $B(0)=0$ and $B$ is continuous. Equation~\eqref{eq:LC10} gives $B(p^*)-\Di(p^*,p^*\mid\fmat)=p^*g(p^*)S(p^*)>0$. By the definition of the first crossing, $B(p)\le\Di(p^*,p^*\mid\fmat)$ for $0\le p\le\underline p$. Thus no genuine symmetric equilibrium at such a price can improve on the benchmark. We may therefore restrict the price search to $[\underline p,\vbar]$ without excluding any potentially optimal genuine equilibrium.

\step{Step 2: Finite upward-deviation relaxation.}
Let $X_1\subset X_2\subset\cdots\subset(0,\vbar)$ be finite sets whose union is dense in $(0,\vbar)$. For each $n$, maximize social surplus jointly over a target price $p\in[\underline p,\vbar]$ and a symmetric matching $\phi$, subject only to the upward-deviation constraints
\begin{equation*}
 (p+x)\Di(p+x,p\mid\phi)\le p\Di(p,p\mid\phi),\qquad x\in X_n.
\end{equation*}
As in the proof of Theorem~\ref{theorem:sequential-mismatching}, we may omit products valued below $p$ and replace preferred-only assignments by full consideration whenever both products are affordable. These changes preserve demand and social surplus at the target price and weakly reduce demand from upward deviations. We may therefore restrict attention to matchings that assign the affordable product alone or the outside option when exactly one product is affordable, and full consideration, the less-preferred product alone, or the outside option when both products are affordable. Randomization among these assignments is permitted. Consumers who cannot afford either product choose not to purchase.

The compactness argument in Appendix~\ref{app:downward-deviations} gives existence of an optimum for each finite relaxation. Let $(p_n,\phi_n)$ be an optimal price and symmetric matching for the $n$-th relaxation. Proposition~\ref{prop:local-improvement} supplies the same strictly improving equilibrium, feasible for every finite relaxation. Consequently, each finite optimum strictly improves on the full-matching benchmark. By the allocation comparisons in Step 1,
\begin{equation*}
 \underline p\le p_n<p^*,\qquad
 \Di(p_n,p_n\mid\phi_n)>\Di(p^*,p^*\mid\fmat).
\end{equation*}
Steps 3--5 establish that each finite optimum can be chosen without excluding affordable consumers. Step 6 then obtains a limit that supports a genuine symmetric equilibrium and achieves the global social-surplus optimum.

\step{Step 3: Lagrange multipliers and the condition for no exclusion.}
Fix $n$ and an optimal pair $(p_n,\phi_n)$. Write $X_n=\{x_1,\ldots,x_J\}$ and $q_j=p_n+x_j$ for the tested deviation prices. At fixed $p_n$, social surplus and the finite incentive constraints are linear in matching probabilities. The set of attainable objective values and constraint values is compact and convex. A strictly feasible matching can be obtained by serving a positive mass of consumers whose only affordable valuation lies above $p_n$ but below the smallest tested deviation price. These consumers purchase at $p_n$ but not at any $q_j$, making every constraint strictly slack. Finite-dimensional separation therefore gives nonnegative multipliers $\lambda_j^n$.

Write $\lambda^n=(\lambda_1^n,\ldots,\lambda_J^n)$. For a trial price $p$ and matching $\phi$, define the Lagrangian
\begin{equation*}
\begin{aligned}
L(p,\phi,\lambda^n)
={}&\frac{SS(\phi,(p,p))}{2}\\
&+\sum_j\lambda_j^n\big[p\Di(p,p\mid\phi)-(p+x_j)\Di(p+x_j,p\mid\phi)\big].
\end{aligned}
\end{equation*}
The optimal matching $\phi_n$ maximizes $L(p_n,\phi,\lambda^n)$ over normalized matchings, without imposing the incentive constraints. For a consumer whose only affordable product has value $v\ge p_n$, showing that product gives Lagrangian score
\begin{equation*}
 h_n(v)=v+p_n\sum_j\lambda_j^n-\sum_{q_j\le v}q_j\lambda_j^n.
\end{equation*}
Define $\displaystyle\kappa_n=\sum_j\lambda_j^n\frac{q_j-p_n}{q_j}$. The score satisfies $h_n(v)\ge v(1-\kappa_n)$. Indeed, when $q_j\le v$, we have $p_n-q_j\ge-v(q_j-p_n)/q_j$, whereas a term with $q_j>v$ contributes $p_n\lambda_j^n\ge0$.

The same argument shows that, when both products are affordable, displaying only the less-preferred product has strictly positive Lagrangian score whenever $\kappa_n<1$. Since exclusion has score zero, $\kappa_n<1$ rules out exclusion of affordable consumers. We establish this inequality using optimality of the target price $p_n$.

\step{Step 4: Raising the target price.}
Since $(p_n,\phi_n)$ is jointly optimal, raising the target price and changing the matching cannot increase social surplus if the new pair is feasible. We consider a slightly higher price $p_n+h$ and construct a matching by copying the assignment probabilities of each original consumer type to a type with higher valuations. This construction limits the loss of trade from the price increase. The new matching need not satisfy the incentive constraints at $p_n+h$, so we instead compare the Lagrangian at the same multipliers at the two prices.

The increasing-hazard property recalled at the start of Appendix~\ref{app:technical-lemmas} gives that $m$ is nonincreasing. For $0<h<\bar v-p_n$, let $\theta_h=S(p_n+h)/S(p_n)$ and define $\Phi_h$ on the whole valuation interval by
\begin{equation*}
\begin{cases}
G(\Phi_h(v))=\dfrac{G(p_n+h)}{G(p_n)}G(v),&0\le v<p_n,\\[4pt]
S(\Phi_h(v))=\theta_hS(v),&p_n\le v\le\bar v.
\end{cases}
\end{equation*}
Both branches are strictly increasing and meet at $\Phi_h(p_n)=p_n+h$. Thus $\Phi_h$ is continuous and maps $[0,p_n)$ onto $[0,p_n+h)$ and $[p_n,\bar v]$ onto $[p_n+h,\bar v]$.

For $p_n\le v<\bar v$, differentiating the upper branch with respect to $h$ gives
\begin{equation*}
\frac{\partial\Phi_h(v)}{\partial h}=\frac{m(\Phi_h(v))}{m(p_n+h)}\in[0,1].
\end{equation*}
Consequently, $0\le\Phi_h(v)-v\le h$. The upper branch also weakly reduces differences between affordable valuations. Indeed, differentiating its defining identity with respect to $v$ gives $\partial_v\Phi_h(v)=m(\Phi_h(v))/m(v)\le1$, since $\Phi_h(v)\ge v$ and $m$ is nonincreasing. Extending to the endpoint by continuity, we obtain
\begin{equation*}
 0\le\Phi_h(v)-v\le h\quad(v\ge p_n),\qquad \Phi_h(a)-\Phi_h(b)\le a-b\quad(a>b\ge p_n).
\end{equation*}
For each original type $v=(v_1,v_2)$, construct $v'=(v_1',v_2')$ by $v_i'=\Phi_h(v_i)$.
Define the copied matching at the higher price by $\phi_{n,h}(S\mid v')=\phi_n(S\mid v)$ for every consideration set $S$. The transformations preserve which products are affordable, and the normalization assigns the empty set when neither product is affordable. Thus the same assignment probabilities are copied from each type $v$ to its transformed type $v'$. The copied matching is symmetric and normalized, but need not satisfy the finite incentive constraints at $p_n+h$.

For this fixed $n$, let $V(p)$ denote the maximum of $SS/2$ in the finite relaxation at target price $p$. Let $\lambda_h^n=(\lambda_{j,h}^n)_{j=1}^J$ be optimal finite multipliers at $p_n+h$, where the tested deviation prices are $q_j+h$. Since $p_n$ is jointly optimal,
\begin{equation*}
 L(p_n+h,\phi_{n,h},\lambda_h^n)\le V(p_n+h)\le V(p_n)\le L(p_n,\phi_n,\lambda_h^n).
 \tag{S.12}\label{eq:S12}
\end{equation*}
The first inequality follows from finite duality: $V(p_n+h)$ is the maximum of the Lagrangian at the optimal multipliers over all normalized matchings, including $\phi_{n,h}$. The second follows from price optimality. The third follows because $\phi_n$ satisfies the finite incentive constraints at $p_n$, so the nonnegative multipliers add nonnegative terms to its objective value $V(p_n)$. Hence
\begin{equation*}
 L(p_n+h,\phi_{n,h},\lambda_h^n)-L(p_n,\phi_n,\lambda_h^n)\le0.
 \tag{S.13}\label{eq:S13}
\end{equation*}

\step{Step 5: From price optimality to no exclusion.}
We now use \eqref{eq:S13} to establish $\kappa_n<1$. The difficulty is that we know little about the optimal matching $\phi_n$ or the individual multipliers $\lambda_j^n$. The following lemma converts the Lagrangian comparison into an inequality that substantially reduces our dependence on these unknown objects. Define
\begin{equation*}
\Gamma(p)=\sup_{q>p}\frac{q}{q-p}[qS(q)-pS(p)],\qquad p\in[\underline p,p^*).
\end{equation*}
\begin{lemma}[Price-perturbation inequality]\label{lem:social-price-perturbation}
\normalfont
There exists a choice of optimal multipliers $\lambda^n$ at $p_n$ such that
\begin{equation*}
\begin{aligned}
&(\kappa_n-1)\big[m(p_n)\Di(p_n,p_n\mid\phi_n)-\Gamma(p_n)\big]\\
&\quad\le\Gamma(p_n)-m(p_n)\Di(p^*,p^*\mid\fmat)
+p^*G(p^*)S(p^*)-\int_{p_n}^{p^*}vg(v)\,dv.
\end{aligned}
\tag{I.1}\label{eq:key}
\end{equation*}
\end{lemma}
The lemma calculates the limiting effect of raising the target price. Its proof bounds the terms involving the unknown matching $\phi_n$ and uses complementary slackness to collect the individual multipliers $\lambda_j^n$ into the single quantity $\kappa_n$. The resulting inequality depends on the matching only through its aggregate demand at $p_n$. We give the calculation in Appendix~\ref{app:technical-lemmas} so that the remaining argument can be read without its technical details.

Although we do not know this demand exactly, Step 2 establishes that it exceeds the full-matching benchmark. Thus, if $\kappa_n\ge1$, we can replace the unknown demand on the left-hand side of \eqref{eq:key} by benchmark demand, obtaining a necessary inequality involving only $\kappa_n$, the distribution, and $p_n$. We also know little about $p_n$, but Step 2 guarantees $p_n\in[\underline p,p^*)$. The following distributional inequality is sufficient to rule out $\kappa_n\ge1$ throughout this interval.

\renewcommand{\thelemma}{S.1}
\renewcommand{\theHlemma}{social.distribution.\arabic{lemma}}
\begin{lemma}[Distributional inequality]\label{lem:social-distributional}
\normalfont
Under i.i.d. log-concavity, for every $p\in[\underline p,p^*)$,
\begin{equation*}
\begin{aligned}
\Gamma(p)<{}&m(p)\Di(p^*,p^*\mid\fmat)\\
&+\min\left\{0,\int_p^{p^*}vg(v)\,dv-p^*G(p^*)S(p^*)\right\}.
\end{aligned}
\tag{I.2}\label{eq:dist}
\end{equation*}
\end{lemma}
The proof is given in Appendix~\ref{app:technical-lemmas}, under \hyperref[app:proof-social-distributional]{Proof of Lemma~\ref*{lem:social-distributional}}.

Applying Lemma~\ref{lem:social-distributional} at $p=p_n$ makes the right-hand side of \eqref{eq:key} strictly negative. The lemma also implies $\Gamma(p_n)<m(p_n)\Di(p^*,p^*\mid\fmat)$. Together with the demand comparison in Step 2, this gives
\begin{equation*}
 m(p_n)\Di(p_n,p_n\mid\phi_n)-\Gamma(p_n)>0.
\end{equation*}
Thus the coefficient of $\kappa_n-1$ on the left of \eqref{eq:key} is strictly positive, while its right-hand side is strictly negative. Consequently $\kappa_n<1$. By Step 3, every consumer with an affordable product has a purchasing assignment with strictly positive Lagrangian score, whereas exclusion has score zero. Hence $\phi_n$ excludes no affordable consumer.

\step{Step 6: Passing from the finite relaxation to a genuine equilibrium.}
Consider the sequence $(p_n,\phi_n)$ of optimal solutions defined in Step 2, with upward deviations restricted to the nested finite sets $X_n$ whose union is dense in $(0,\vbar)$. By Step 5, each $\phi_n$ can be chosen without excluding affordable consumers. We show that a limit of these solutions supports a genuine symmetric equilibrium and achieves the global social-surplus optimum.

\subproof{Taking a limit.}
Since $p_n\in[\underline p,\vbar]$, we may pass to a subsequence with $p_n\to\hat p$. Weak-* compactness of matching probabilities allows passage to a further subsequence along which $\phi_n$ converges to a symmetric matching $\phi$. The relevant demand and social-surplus integrals converge because the valuation distribution has a continuous density and the purchasing boundaries have probability zero. In particular,
\begin{equation*}
\begin{aligned}
SS(\phi_n,(p_n,p_n))&\longrightarrow SS(\phi,(\hat p,\hat p)),\\
\Di(p_n,p_n\mid\phi_n)&\longrightarrow\Di(\hat p,\hat p\mid\phi).
\end{aligned}
\tag{S.37}\label{eq:S37}
\end{equation*}
The limiting outcome strictly improves on the full-matching benchmark. Indeed, Proposition~\ref{prop:local-improvement} provides a fixed genuine equilibrium whose social surplus exceeds the benchmark by some $\delta>0$. This equilibrium is feasible for every finite relaxation, so every finite optimum achieves at least the same surplus. By \eqref{eq:S37}, this positive improvement survives in the limit. The allocation comparisons in Step 1 therefore imply $\hat p<p^*$.

Since $\phi_n$ excludes no affordable consumer, it serves every consumer who values at least one product above $p_n$. By symmetry, $\Di(p_n,p_n\mid\phi_n)=[1-G(p_n)^2]/2$. Taking limits gives
\begin{equation*}
 \Di(\hat p,\hat p\mid\phi)=\frac{1-G(\hat p)^2}{2}.
\tag{S.38}\label{eq:S38}
\end{equation*}
Under the normalization of Step 2, this is the maximum symmetric demand attainable at price $\hat p$. Hence the limiting matching also excludes no affordable consumer.

\subproof{Upward deviations.}
Fix $x\in\bigcup_nX_n$. Since the sets $X_n$ are nested, for all sufficiently large $n$,
\begin{equation*}
 (p_n+x)\Di(p_n+x,p_n\mid\phi_n)\le p_n\Di(p_n,p_n\mid\phi_n).
\end{equation*}
Passing to the limit,
\begin{equation*}
 (\hat p+x)\Di(\hat p+x,\hat p\mid\phi)\le \hat p\Di(\hat p,\hat p\mid\phi).
\tag{S.39}\label{eq:S39}
\end{equation*}
Continuity of deviation profit in the deviating price, together with density of $\bigcup_nX_n$, extends \eqref{eq:S39} to every upward deviation. Prices above the valuation support yield no demand.

\subproof{Downward deviations.}
Since the limiting matching is normalized and excludes no affordable consumer, the same argument as in the proof of Theorem~\ref{theorem:social-trade-optimal}(b) shows that no downward deviation is profitable. Together with the upward-deviation constraints, this establishes that $(\hat p,\hat p)$ is a symmetric pure-strategy equilibrium under $\phi$.

\subproof{Global optimality.}
Let $V_n=SS(\phi_n,(p_n,p_n))$ denote the maximum social surplus in the $n$-th finite relaxation. Since the constraints are nested,
\begin{equation*}
 V_1\ge V_2\ge\cdots.
\tag{S.40}\label{eq:S40}
\end{equation*}
Every genuine symmetric equilibrium that improves on the full-matching benchmark has price in $[\underline p,\vbar]$, by Step 1, and is feasible for every finite relaxation. Equilibria outside this range cannot improve on the benchmark, whereas every finite optimum strictly exceeds it. Therefore, if $V^*$ denotes the supremum of social surplus across genuine symmetric equilibria,
\begin{equation*}
 V_n\ge V^*\qquad\text{for every }n.
\tag{S.41}\label{eq:S41}
\end{equation*}
By \eqref{eq:S37},
\begin{equation*}
 SS(\phi,(\hat p,\hat p))=\lim_{n\to\infty}V_n\ge V^*.
\end{equation*}
The limiting pair is a genuine symmetric equilibrium, so the reverse inequality also holds. Thus
\begin{equation*}
 SS(\phi,(\hat p,\hat p))=V^*.
\tag{S.42}\label{eq:S42}
\end{equation*}
A globally social-surplus-optimal symmetric equilibrium therefore exists without excluding affordable consumers. Applying the symmetric cutoff rearrangement in the proof of Proposition~\ref{prop:social-trade-two-step} preserves this no-exclusion property, so an optimum can also be chosen with the Step-1 cutoff structure of Theorem~\ref{theorem:sequential-mismatching}. This proves Theorem~\ref{theorem:social-trade-optimal}(a).\nobreak\hfill\quad$\square$
\setcounter{equation}{\socialsavedequation}
\setcounter{lemma}{\socialsavedlemma}
\endgroup

\subsubsection{Trade-Optimal Matching}
\label{app:trade-optimal}

\begin{proof}[Proof of Theorem~\ref{theorem:social-trade-optimal}(b)]
\leavevmode\par
\paragraph{Step 1: Universal bounds.}
Fix an arbitrary symmetric matching $\phi$ supporting the common price $p$, allowing exclusion. Affordability and symmetry imply
\begin{equation}\label{eq:trade-affordability-bound}
D_i(p,p\mid\phi)\le\frac{1-G(p)^2}{2}.
\end{equation}
When firm $i$ raises its price from $p$ to $p+\varepsilon$, the first-order demand loss is bounded by three mutually exclusive cases.
\begin{enumerate}[label=(\roman*)]
\item Only firm $i$ is affordable at $p$. The mass lost is at most $\varepsilon g(p)G(p)+o(\varepsilon)$.
\item Both firms are affordable at $p$, but the consumer is shown only firm $i$. This includes preferred-only assignments. The mass lost is at most $\varepsilon g(p)[1-G(p)]+o(\varepsilon)$.
\item Both firms are affordable and both are considered. A consumer who initially prefers firm $i$ can be lost only if $v_i-v_j<\varepsilon$. The mass of such consumers is
\[
\int_p^{\bar v}g(v_j)
\int_{v_j}^{\min\{v_j+\varepsilon,\bar v\}}g(v_i)\,dv_i\,dv_j
=\varepsilon\int_p^{\bar v}g(v)^2\,dv+o(\varepsilon).
\]
\end{enumerate}
Consequently,
\begin{equation}\label{eq:trade-local-loss-bound}
D_i(p,p\mid\phi)-D_i(p+\varepsilon,p\mid\phi)
\le\varepsilon\left[g(p)+\int_p^{\bar v}g(v)^2\,dv\right]+o(\varepsilon).
\end{equation}
The equilibrium incentive constraint is
\[
(p+\varepsilon)D_i(p+\varepsilon,p\mid\phi)
\le pD_i(p,p\mid\phi).
\]
Rearranging this constraint and applying \eqref{eq:trade-local-loss-bound} gives
\[
\begin{aligned}
\varepsilon D_i(p,p\mid\phi)
&\le(p+\varepsilon)
\bigl[D_i(p,p\mid\phi)-D_i(p+\varepsilon,p\mid\phi)\bigr]\\
&\le(p+\varepsilon)
\left\{\varepsilon\left[g(p)+\int_p^{\bar v}g(v)^2\,dv\right]
+o(\varepsilon)\right\}.
\end{aligned}
\]
Dividing by $\varepsilon$ and letting $\varepsilon\downarrow0$, we obtain
\begin{equation}\label{eq:trade-incentive-bound}
D_i(p,p\mid\phi)\le B(p),
\end{equation}
where
\begin{equation}\label{eq:trade-B-definition}
B(p):=p\left[g(p)+\int_p^{\bar v}g(v)^2\,dv\right].
\end{equation}

\Needspace{10\baselineskip}
\paragraph{Step 2: The trade-quantity threshold.}
\begin{lemma}[Trade-Quantity Threshold Lemma]\label{lem:trade-quantity-threshold}
There is a unique $p^T\in(0,p^*)$ solving
\[
\frac{1-G(p^T)^2}{2}=B(p^T).
\]
Moreover, $B$ is strictly increasing for $0\le p\le p^T$.
\end{lemma}

\begin{proof}[Proof of Lemma~\ref{lem:trade-quantity-threshold}]
Inside this proof, write
\[
A(p)=\frac{1-G(p)^2}{2},\qquad
C(p)=g(p)+\int_p^{\bar v}g(v)^2\,dv,
\qquad B(p)=pC(p).
\]
We first establish existence. At zero, $B(0)=0<A(0)=1/2$. Equation~\eqref{eq:LC10}, together with $A(p^*)=D_1(p^*,p^*\mid\phi^{\mathrm{full}})$, gives
\[
B(p^*)-A(p^*)=p^*g(p^*)S(p^*)>0.
\]
Continuity therefore gives a crossing in $(0,p^*)$.

We next establish uniqueness directly. We show below that
\begin{equation}\label{eq:trade-C-derivative}
C'(p)\ge-\frac{C(p)^2}{A(p)},\qquad 0\le p<\bar v.
\end{equation}
At every crossing $B(p)=A(p)$, we have $p=A(p)/C(p)$, and therefore
\[
\begin{aligned}
\bigl[B(p)-A(p)\bigr]'
&=C(p)+\frac{A(p)}{C(p)}C'(p)+g(p)G(p)\\
&\ge g(p)G(p)>0.
\end{aligned}
\]
Every crossing is thus from below to above, so the crossing is unique. Before it, $B(p)<A(p)$, and \eqref{eq:trade-C-derivative} implies
\[
B'(p)=C(p)+pC'(p)
\ge C(p)\left[1-\frac{B(p)}{A(p)}\right]>0.
\]
This proves strict monotonicity up to $p^T$.

It remains to prove \eqref{eq:trade-C-derivative}. Let $S(p)=1-G(p)$. Equations~\eqref{eq:LC3} and \eqref{eq:LC4} give
\[
g'(p)\ge-\frac{g(p)^2}{S(p)},\qquad
\int_p^{\bar v}g(v)^2\,dv\ge\frac{g(p)S(p)}{2}.
\]
Define $s=S(p)$. Then
\[
C(p)\ge g(p)\left(1+\frac{s}{2}\right),\qquad
A(p)=s\left(1-\frac{s}{2}\right),
\]
and hence
\[
\frac{C(p)^2}{A(p)}
\ge\frac{g(p)^2(1+s/2)^2}{s(1-s/2)}
\ge\frac{g(p)^2(1+s)}{s}.
\]
Consequently,
\[
C'(p)=g'(p)-g(p)^2
\ge-\frac{g(p)^2(1+s)}{s}
\ge-\frac{C(p)^2}{A(p)},
\]
as required.
\end{proof}

Let $Q$ denote trade quantity under any symmetric equilibrium $(p,p)$ and matching $\phi$. If $p<p^T$, \eqref{eq:trade-incentive-bound} and Lemma~\ref{lem:trade-quantity-threshold} give
\[
Q=2D_i(p,p\mid\phi)\le2B(p)
\le2B(p^T)=1-G(p^T)^2.
\]
If $p\ge p^T$, the affordability bound \eqref{eq:trade-affordability-bound} instead gives
\[
Q\le1-G(p)^2\le1-G(p^T)^2.
\]
Thus $1-G(p^T)^2$ is a universal upper bound on trade quantity.

\paragraph{Step 3: Attainment by maximal mismatching.}
At a common price $p$, maximal mismatching preserves all affordable transactions, so
\begin{equation}\label{eq:trade-maximal-target-demand}
D_i(p,p\mid\phi_p^{\max})=\frac{1-G(p)^2}{2}.
\end{equation}
For a unilateral price $q\ge p$, demand is
\begin{equation}\label{eq:trade-maximal-upward-demand}
D_i(q,p\mid\phi_p^{\max})
=G(p)[1-G(q)]
+\int_q^{\bar v}g(v)[1-G(v+q-p)]\,dv.
\end{equation}
The first term is demand from consumers for whom only firm $i$ is affordable at the target price.\footnote{The second term follows by considering the mirror types $(p+s+d,p+s)$ and $(p+s,p+s+d)$. If $s<d$, maximal mismatching matches both consumers only to their less-preferred products, and firm $i$ retains its assigned consumer if and only if $s\ge q-p$. If $s\ge d$, both products are displayed, and firm $i$ retains its customer if and only if $d\ge q-p$. Thus retention is equivalent to $\min\{s,d\}\ge q-p$. Writing the lower valuation as $b=p+s$ and the upper valuation as $a=b+d$, this is equivalent to $b\ge q$ and $a-b\ge q-p$. The resulting ordered integral is $\int_q^{\bar v}g(b)\int_{b+q-p}^{\bar v}g(a)\,da\,db$, with no factor of two because each mirror pair yields one customer for firm $i$.}
Here and below, $G(x)=1$ for $x\ge\bar v$; deviations $q\ge\bar v$ yield zero demand. Differentiating \eqref{eq:trade-maximal-upward-demand} at $q=p$ from the right gives
\[
\begin{aligned}
\left.\partial_qD_i(q,p\mid\phi_p^{\max})\right|_{q=p+}
&=-g(p)G(p)-g(p)[1-G(p)]-\int_p^{\bar v}g(v)^2\,dv\\
&=-g(p)-\int_p^{\bar v}g(v)^2\,dv.
\end{aligned}
\]
At $p=p^T$, the defining equality for the threshold therefore gives the first-order condition
\[
\begin{aligned}
&D_i(p^T,p^T\mid\phi_{p^T}^{\max})
+p^T\left.\partial_qD_i(q,p^T\mid\phi_{p^T}^{\max})\right|_{q=p^T+}\\
&\qquad=\frac{1-G(p^T)^2}{2}
-p^T\left[g(p^T)+\int_{p^T}^{\bar v}g(v)^2\,dv\right]=0.
\end{aligned}
\]

We first rule out downward price deviations. Appendix~\ref{app:downward-deviations} establishes that the unique full-matching best response to $p^T$ lies above $p^T$, since $p^T<p^*$, and full-matching profit is increasing up to that best response by log-concavity. Maximal mismatching preserves sales at the target price and weakly decreases demand after a downward unilateral cut relative to full matching. Thus, for $0\le q<p^T$,
\[
\begin{aligned}
qD_i(q,p^T\mid\phi_{p^T}^{\max})
&\le qD_i(q,p^T\mid\phi^{\mathrm{full}})\\
&\le p^TD_i(p^T,p^T\mid\phi^{\mathrm{full}})\\
&=p^TD_i(p^T,p^T\mid\phi_{p^T}^{\max}).
\end{aligned}
\]

It remains to rule out global upward deviations. The two terms in \eqref{eq:trade-maximal-upward-demand} are separately log-concave in $q$. The first, $G(p)[1-G(q)]$, is log-concave by the survival result of \citet[Theorem 3]{bagnoli2005log}. The second is log-concave by Pr\'ekopa's theorem \citep[Theorem 6]{prekopa1973logarithmic}, integrating the jointly log-concave function
\[
g(v)[1-G(v+q-p)]\mathbf 1\{v\ge q\}
\]
over $v$, with the density extended by zero outside its support.

For $0<p<\bar v$, define directly
\[
\omega=\frac{2G(p)}{1+G(p)},\qquad
t=\frac{pg(p)}{1-G(p)},\qquad
k=2t+\frac{2p\int_p^{\bar v}g(v)^2\,dv}{[1-G(p)]^2},
\qquad z=\frac qp.
\]
At $q=p$, the two demand components have weights $\omega$ and $1-\omega$ in total demand and logarithmic derivatives $-t/p$ and $-k/p$, respectively. Applying \eqref{eq:LC6} to each component separately yields
\begin{equation}\label{eq:trade-upward-exponential-bound}
\frac{qD_i(q,p\mid\phi_p^{\max})}
{pD_i(p,p\mid\phi_p^{\max})}
\le z\left[\omega e^{-t(z-1)}+(1-\omega)e^{-k(z-1)}\right].
\end{equation}
At $p=p^T$, we have $k\ge2t$. Equation~\eqref{eq:LC2} also gives
\[
t=\frac{pg(p)}{1-G(p)}
\ge-\log[1-G(p)]
=\log\frac{2-\omega}{2(1-\omega)}.
\]
Finally, the first-order condition gives
\[
\omega t+(1-\omega)k=1.
\]
The proof therefore reduces to establishing the following technical lemma.

\noindent\textbf{Lemma~\ref{lem:weighted-average-exponential} (Weighted-Average Exponential Inequality).}
\begingroup\itshape
Let $\omega\in(0,1)$, and suppose
\[
t\ge\log\frac{2-\omega}{2(1-\omega)},\qquad
k\ge2t,\qquad
\omega t+(1-\omega)k\ge1.
\]
Then, for every $z\ge1$,
\[
\omega e^{-t(z-1)}+(1-\omega)e^{-k(z-1)}\le\frac1z.
\]
\par\endgroup

The proof is in Appendix~\ref{app:technical-lemmas}.

Applying Lemma~\ref{lem:weighted-average-exponential} to \eqref{eq:trade-upward-exponential-bound} at $p=p^T$ shows that every upward deviation $q>p^T$ is unprofitable. Together with the downward-deviation argument, this proves that $\phi_{p^T}^{\max}$ supports $(p^T,p^T)$ as an equilibrium. By \eqref{eq:trade-maximal-target-demand}, its trade quantity is $1-G(p^T)^2$, attaining the universal bound from Step~2. Hence $\phi_{p^T}^{\max}$ is symmetric trade-optimal.
\end{proof}

\subsubsection{Technical Lemmas}
\label{app:technical-lemmas}

We use the standard increasing-hazard property: $S$ is log-concave and $g/S$ is nondecreasing, so $m$ is nonincreasing \citep[Theorem 3 and Corollary 2]{bagnoli2005log}. In particular, $v/m(v)$ is strictly increasing on $[0,\bar v)$.

We first establish two preliminary results used in the technical proofs below. Lemma~\ref{lem:lc-properties} collects inequalities implied by log-concavity, while Lemma~\ref{lem:full-matching-benchmark} derives benchmark bounds from the full-matching equilibrium.

\begingroup
\edef\technicalsavedequation{\number\value{equation}}
\renewcommand{\theHequation}{technical.\arabic{equation}}
\newcommand{\Di}{D_1}
\newcommand{\fmat}{\phi^{\mathrm{full}}}
\newcommand{\pn}{p_n}
\newcommand{\phin}{\phi_n}
\newcommand{\LH}{\lambda_h^n}
\newcommand{\hphi}{\phi_{n,h}}
\newcommand{\vbar}{\bar v}
\newcommand{\ind}[1]{\mathbf 1\{#1\}}
\newcommand{\order}{\{v_2<v_1\}}
\newcommand{\ssur}[2]{SS(#1,(#2,#2))}
\newcommand{\step}[1]{\par\medskip\noindent\textbf{#1}\enspace}
\newcommand{\subproof}[1]{\par\medskip\noindent\textit{#1}\enspace}
\begingroup
\edef\commonsavedequation{\number\value{equation}}
\setcounter{equation}{0}
\renewcommand{\theequation}{LC.\arabic{equation}}
\renewcommand{\theHequation}{common.\arabic{equation}}
\begin{preliminarylemma}[Bounds for Log-Concave Distributions]\label{lem:lc-properties}
\normalfont
\noindent\textit{Survival bounds.} For every $0\le a<b<\vbar$,
\begin{equation}
e^{-(b-a)/m(b)}\le\frac{S(b)}{S(a)}\le e^{-(b-a)/m(a)}.\label{eq:LC1}
\end{equation}
In particular,
\begin{equation}
S(p)\ge e^{-p/m(p)}.\label{eq:LC2}
\end{equation}

\noindent\textit{Density and squared-density bounds.} For every $v<\vbar$,
\begin{equation}
g'(v)\ge-\frac{g(v)^2}{S(v)},\label{eq:LC3}
\end{equation}
and
\begin{equation}
\int_v^{\vbar}g(t)^2\,dt\ge\frac{g(v)S(v)}2.\label{eq:LC4}
\end{equation}

\noindent\textit{Density comparisons} \citep[Sections 3.1.1 and 3.1.3]{boyd2004convex}. For $0\le a<b<\vbar$,
\begin{equation}
g(v)\ge\min\{g(a),g(b)\},\qquad a\le v\le b.\label{eq:LC5}
\end{equation}
Moreover, if $g(b)<g(a)$, then $g$ is nonincreasing on $[b,\vbar]$.

\noindent\textit{Exponential tangent and chord bounds} \citep[Sections 3.1.3 and 3.5.1]{boyd2004convex}. More generally, if $H$ is positive and log-concave on an interval, then at every differentiability point $a$,
\begin{equation}
H(x)\le H(a)\exp\left[(x-a)\frac{H'(a)}{H(a)}\right].\label{eq:LC6}
\end{equation}
For every $a<x<b$ in that interval,
\begin{equation}
H(x)\ge H(a)^{(b-x)/(b-a)}H(b)^{(x-a)/(b-a)}.\label{eq:LC7}
\end{equation}

\end{preliminarylemma}

\begin{proof}
The increasing-hazard property gives $(\log S)'(v)=-1/m(v)$. Integrating between $a$ and $b$, since $m(a)\ge m(v)\ge m(b)$,
\[
-\frac{b-a}{m(b)}\le\log\frac{S(b)}{S(a)}\le-\frac{b-a}{m(a)}.
\]
Exponentiating gives \eqref{eq:LC1}. Setting $a=0$, using $S(0)=1$, gives \eqref{eq:LC2}.

Differentiating $m=S/g$ gives
\[
m'(v)=-1-\frac{S(v)g'(v)}{g(v)^2}\le0,
\]
which implies \eqref{eq:LC3}. Moreover, since the hazard $g/S$ is nondecreasing,
\begin{align*}
\int_v^{\vbar}g(t)^2\,dt
&=\int_v^{\vbar}\frac{g(t)}{S(t)}S(t)g(t)\,dt\\
&\ge\frac{g(v)}{S(v)}\int_v^{\vbar}S(t)g(t)\,dt\\
&=\frac{g(v)S(v)}2.
\end{align*}
This proves \eqref{eq:LC4}.

\end{proof}

\begin{preliminarylemma}[Full-Matching Duopoly Benchmark]\label{lem:full-matching-benchmark}
\normalfont
The full-matching benchmark satisfies
\begin{equation}
p^*\le m(p^*).\label{eq:LC9}
\end{equation}
Moreover,
\begin{equation}
B(p^*)-D_1(p^*,p^*\mid\fmat)
=p^*g(p^*)S(p^*)>0.\label{eq:LC10}
\end{equation}

\end{preliminarylemma}

\begin{proof}
At the full-matching equilibrium, demand is $S(p^*)[1-S(p^*)/2]$, and the first-order condition is
\[
D_1(p^*,p^*\mid\fmat)=p^*\left[
g(p^*)G(p^*)+\int_{p^*}^{\vbar}g(v)^2\,dv\right].
\]
Applying \eqref{eq:LC4},
\begin{align*}
&S(p^*)\left(1-\frac{S(p^*)}{2}\right)\\
&\qquad\ge p^*g(p^*)\left[G(p^*)+\frac{S(p^*)}{2}\right]\\
&\qquad=p^*g(p^*)\left(1-\frac{S(p^*)}{2}\right).
\end{align*}
Since the common factor is positive, $S(p^*)\ge p^*g(p^*)$, proving \eqref{eq:LC9}. Finally, subtracting the full-matching first-order condition from the definition of $B(p^*)$ gives \eqref{eq:LC10}. \end{proof}
\setcounter{equation}{\commonsavedequation}
\endgroup

\subsubsection*{Proof of Lemma~\ref{lem:social-price-perturbation}}\phantomsection\label{app:proof-social-price-perturbation}
Fix $n$. Recall that Step 4 constructs the copied matching $\phi_{n,h}$ at $p_n+h$, with optimal multipliers $\lambda_h^n$. We compare the Lagrangian at the two prices, using the price-optimality inequality \eqref{eq:S13}.

Write $L=L_U+L_B$, where $L_U$ is the contribution from consumers with exactly one affordable product and $L_B$ is the contribution from those with both products affordable. Consumers with neither product affordable contribute zero under the normalization. Write $d\nu=g(v_1)g(v_2)\,dv_1dv_2$. When integrating over ordered types, we use $v_2<v_1$ and account for their mirrors by symmetry.

\subproof{Consumers with exactly one affordable product.}
At the original target price $p_n$, these consumers have $v_2<p_n\le v_1$. Only product 1 is affordable, so their Lagrangian contribution is
\begin{equation*}
\begin{aligned}
L_U(p_n,\phi_n,\lambda_h^n)
={}&\int_{v_2<p_n\le v_1}\phi_n(\{1\}\mid v)\\
&\quad\times\underbrace{\left[v_1+p_n\sum_j\lambda_{j,h}^n
-\sum_jq_j\lambda_{j,h}^n\ind{v_1\ge q_j}\right]}_{\text{(14a)}}\,d\nu.
\end{aligned}
\tag{L.1}\label{eq:L1}
\end{equation*}
At the higher target price, the corresponding population has $v_2'<p_n+h\le v_1'$, and its Lagrangian contribution is
\begin{equation*}
\begin{aligned}
L_U(p_n+h,\phi_{n,h},\lambda_h^n)
={}&\int_{v_2'<p_n+h\le v_1'}\phi_{n,h}(\{1\}\mid v')\\
&\times\Bigg[v_1'+(p_n+h)\sum_j\lambda_{j,h}^n
-\sum_j(q_j+h)\lambda_{j,h}^n\ind{v_1'\ge q_j+h}\Bigg]\\
&\quad\times g(v_1')g(v_2')\,dv_1'dv_2'.
\end{aligned}
\end{equation*}
We express this integral in terms of the original valuations. By construction, $v_1'=\Phi_h(v_1)$, $v_2'=\Phi_h(v_2)$, and
\begin{equation*}
\begin{aligned}
\phi_{n,h}(\{1\}\mid v')&=\phi_n(\{1\}\mid v),\\
g(v_1')\,dv_1'&=\theta_hg(v_1)\,dv_1,\\
g(v_2')\,dv_2'&=\frac{G(p_n+h)}{G(p_n)}g(v_2)\,dv_2.
\end{aligned}
\end{equation*}
The density identities follow by differentiating the two branches of $\Phi_h$. Thus the transformation scales the joint density of this population by $\theta_hG(p_n+h)/G(p_n)$ while preserving its matching probabilities. Substitution gives
\begin{equation*}
\begin{aligned}
L_U(p_n+h,\phi_{n,h},\lambda_h^n)
={}&\theta_h\frac{G(p_n+h)}{G(p_n)}\int_{v_2<p_n\le v_1}\phi_n(\{1\}\mid v)\\
&\quad\times\underbrace{\left[\begin{aligned}
&\Phi_h(v_1)+(p_n+h)\sum_j\lambda_{j,h}^n\\
&\quad-\sum_j(q_j+h)\lambda_{j,h}^n\ind{\Phi_h(v_1)\ge q_j+h}
\end{aligned}\right]}_{\text{(15a)}}\,d\nu.
\end{aligned}
\tag{L.2}\label{eq:L2}
\end{equation*}
We compare the bracketed terms in \eqref{eq:L1} and \eqref{eq:L2}. Here $v_1\ge p_n$, so the upper-branch bound in Step 4 applies. If $\Phi_h(v_1)\ge q_j+h$, then $v_1+h\ge\Phi_h(v_1)\ge q_j+h$, and hence $v_1\ge q_j$. Therefore
\begin{equation*}
 \ind{\Phi_h(v_1)\ge q_j+h}\le\ind{v_1\ge q_j}.
\end{equation*}
A consumer who purchases after the transformed deviation to $q_j+h$ would therefore also have purchased after the original deviation to $q_j$. Together with $\lambda_{j,h}^n\ge0$, this gives
\begin{equation*}
\begin{aligned}
\text{(15a)}-\text{(14a)}\ge{}&\Phi_h(v_1)-v_1\\
&+h\sum_j\lambda_{j,h}^n\big[1-\ind{v_1\ge q_j}\big].
\end{aligned}
\tag{L.3}\label{eq:L3}
\end{equation*}
Applying \eqref{eq:L3} to the difference of \eqref{eq:L1} and \eqref{eq:L2} yields
\begin{equation*}
\begin{aligned}
&\frac{L_U(p_n,\phi_n,\lambda_h^n)-L_U(p_n+h,\phi_{n,h},\lambda_h^n)}{h}\\
&\quad\le \frac{1-\theta_hG(p_n+h)/G(p_n)}{h}L_U(p_n,\phi_n,\lambda_h^n)\\
&\qquad-\theta_h\frac{G(p_n+h)}{G(p_n)}\int_{v_2<p_n\le v_1}\phi_n(\{1\}\mid v)\\
&\hspace{54pt}\times\left[\frac{\Phi_h(v_1)-v_1}{h}
+\sum_j\lambda_{j,h}^n\big(1-\ind{v_1\ge q_j}\big)\right]d\nu.
\end{aligned}
\tag{L.4}\label{eq:L4}
\end{equation*}
The first term arises from the change in the density multiplier; the remaining integral records the difference between the two Lagrangian scores.

As $h\downarrow0$, the identities in Step 4 give
\begin{equation*}
\begin{aligned}
\theta_h\frac{G(p_n+h)}{G(p_n)}&\longrightarrow1,\\
\frac{1-\theta_hG(p_n+h)/G(p_n)}{h}
&\longrightarrow\frac1{m(p_n)}-\frac{g(p_n)}{G(p_n)},\\
\frac{\Phi_h(v)-v}{h}&\longrightarrow\frac{m(v)}{m(p_n)}\qquad(v\ge p_n).
\end{aligned}
\end{equation*}
The optimal multipliers also depend on $h$. For fixed $n$, however, the strict-feasibility construction in Step 3 remains uniformly strict at sufficiently close target prices. It therefore bounds the optimal multipliers $\lambda_h^n$. We may pass to a sequence $h\downarrow0$ along which $\lambda_h^n\to\lambda^n$; continuity and finite duality imply that $\lambda^n$ is optimal at $p_n$. We choose the multipliers in Lemma~\ref{lem:social-price-perturbation} to be this limit. All subsequent limits are taken along this sequence.

Taking the upper limit in \eqref{eq:L4} yields
\begin{equation*}
\begin{aligned}
&\limsup_{h\downarrow0}
\frac{L_U(p_n,\phi_n,\lambda_h^n)-L_U(p_n+h,\phi_{n,h},\lambda_h^n)}{h}\\
&\quad\le\left[\frac1{m(p_n)}-\frac{g(p_n)}{G(p_n)}\right]L_U(p_n,\phi_n,\lambda^n)\\
&\qquad-\frac1{m(p_n)}\int_{v_2<p_n\le v_1}\phi_n(\{1\}\mid v)m(v_1)\,d\nu\\
&\qquad-\sum_j\lambda_j^n\int_{v_2<p_n\le v_1}\phi_n(\{1\}\mid v)
\big[1-\ind{v_1\ge q_j}\big]d\nu.
\end{aligned}
\tag{L.5}\label{eq:L5}
\end{equation*}
This completes the bound for consumers with exactly one affordable product. We next consider consumers with both products affordable, where the valuation transformation also affects preference differences.

\subproof{Consumers with both products affordable.}
At $p_n$ consider $p_n\le v_2<v_1$. Under the normalized matching, the consumer receives full consideration, the less-preferred product alone, or the outside option. Under full consideration, she buys her preferred product at the common price and remains with it after an upward deviation of size $x_j$ precisely when $v_1-v_2\ge x_j$. With the less-preferred-only assignment, she continues purchasing after her assigned firm's price rises to $q_j$ precisely when $v_2\ge q_j$. Accounting for the mirror types, the contribution is
\begin{equation*}
\begin{aligned}
L_B(p_n,\phi_n,\lambda_h^n)
={}&\int_{p_n\le v_2<v_1}\Bigg\{\phi_n(\{1,2\}\mid v)\\
&\quad\times\underbrace{\left[\begin{aligned}
&v_1+p_n\sum_j\lambda_{j,h}^n\\
&\quad-\sum_jq_j\lambda_{j,h}^n\ind{v_1-v_2\ge x_j}
\end{aligned}\right]}_{\text{(19a)}}\\
&+\phi_n(\{2\}\mid v)
\underbrace{\left[\begin{aligned}
&v_2+p_n\sum_j\lambda_{j,h}^n\\
&\quad-\sum_jq_j\lambda_{j,h}^n\ind{v_2\ge q_j}
\end{aligned}\right]}_{\text{(19b)}}\Bigg\}d\nu.
\end{aligned}
\tag{L.6}\label{eq:L6}
\end{equation*}
At the higher target price $p_n+h$, the corresponding contribution is
\begin{equation*}
\begin{aligned}
&L_B(p_n+h,\phi_{n,h},\lambda_h^n)\\
&=\int_{p_n+h\le v_2'<v_1'}\Bigg\{
\phi_{n,h}(\{1,2\}\mid v')
\left[\begin{aligned}
&v_1'+(p_n+h)\sum_j\lambda_{j,h}^n\\
&-\sum_j(q_j+h)\lambda_{j,h}^n\ind{v_1'-v_2'\ge x_j}
\end{aligned}\right]\\
&\quad+\phi_{n,h}(\{2\}\mid v')
\left[\begin{aligned}
&v_2'+(p_n+h)\sum_j\lambda_{j,h}^n\\
&-\sum_j(q_j+h)\lambda_{j,h}^n\ind{v_2'\ge q_j+h}
\end{aligned}\right]\Bigg\}\\
&\hspace{95pt}\times g(v_1')g(v_2')\,dv_1'dv_2'.
\end{aligned}
\end{equation*}
Here both values are transformed by the upper branch of $\Phi_h$, so $v_i'=\Phi_h(v_i)$ and
\begin{equation*}
\begin{aligned}
\phi_{n,h}(\{1,2\}\mid v')&=\phi_n(\{1,2\}\mid v),\\
\phi_{n,h}(\{2\}\mid v')&=\phi_n(\{2\}\mid v),\\
g(v_i')\,dv_i'&=\theta_hg(v_i)\,dv_i.
\end{aligned}
\end{equation*}
Unlike the one-affordable population, both coordinates contribute the same factor $\theta_h$. Changing variables therefore gives
\begin{equation*}
\begin{aligned}
&L_B(p_n+h,\phi_{n,h},\lambda_h^n)\\
&=\theta_h^2\int_{p_n\le v_2<v_1}\Bigg\{\phi_n(\{1,2\}\mid v)
\underbrace{\left[\begin{aligned}
&\Phi_h(v_1)+(p_n+h)\sum_j\lambda_{j,h}^n\\
&\quad-\sum_j(q_j+h)\lambda_{j,h}^n
\ind{\Phi_h(v_1)-\Phi_h(v_2)\ge x_j}
\end{aligned}\right]}_{\text{(20a)}}\\
&\quad+\phi_n(\{2\}\mid v)
\underbrace{\left[\begin{aligned}
&\Phi_h(v_2)+(p_n+h)\sum_j\lambda_{j,h}^n\\
&\quad-\sum_j(q_j+h)\lambda_{j,h}^n
\ind{\Phi_h(v_2)\ge q_j+h}
\end{aligned}\right]}_{\text{(20b)}}\Bigg\}d\nu.
\end{aligned}
\tag{L.7}\label{eq:L7}
\end{equation*}
We compare the bracketed terms in \eqref{eq:L6} and \eqref{eq:L7}. Since $p_n\le v_2<v_1$, both values lie on the upper branch. Its contraction and displacement bounds in Step 4 imply
\begin{equation*}
\begin{aligned}
\ind{\Phi_h(v_1)-\Phi_h(v_2)\ge x_j}&\le\ind{v_1-v_2\ge x_j},\\
\ind{\Phi_h(v_2)\ge q_j+h}&\le\ind{v_2\ge q_j}.
\end{aligned}
\end{equation*}
The first inequality controls purchases under full consideration: the less-preferred product is affordable at both target prices, and $\Phi_h$ weakly reduces the preference difference. The second controls purchases under less-preferred-only matching, because the assigned value increases by at most $h$. Thus the transformation cannot create a new purchase after a tested upward deviation under either assignment. Using these inequalities and $\lambda_{j,h}^n\ge0$ yields
\begin{equation*}
\begin{aligned}
\text{(20a)}-\text{(19a)}\ge{}&\Phi_h(v_1)-v_1
+h\sum_j\lambda_{j,h}^n\big[1-\ind{v_1-v_2\ge x_j}\big],\\
\text{(20b)}-\text{(19b)}\ge{}&\Phi_h(v_2)-v_2
+h\sum_j\lambda_{j,h}^n\big[1-\ind{v_2\ge q_j}\big].
\end{aligned}
\tag{L.8}\label{eq:L8}
\end{equation*}
Applying \eqref{eq:L8} to the difference of \eqref{eq:L6} and \eqref{eq:L7}, we obtain
\begin{equation*}
\begin{aligned}
&\frac{L_B(p_n,\phi_n,\lambda_h^n)-L_B(p_n+h,\phi_{n,h},\lambda_h^n)}{h}\\
&\quad\le\frac{1-\theta_h^2}{h}L_B(p_n,\phi_n,\lambda_h^n)\\
&\qquad-\theta_h^2\int_{p_n\le v_2<v_1}\Bigg\{
\phi_n(\{1,2\}\mid v)
\left[\frac{\Phi_h(v_1)-v_1}{h}
+\sum_j\lambda_{j,h}^n(1-\ind{v_1-v_2\ge x_j})\right]\\
&\hspace{95pt}+\phi_n(\{2\}\mid v)
\left[\frac{\Phi_h(v_2)-v_2}{h}
+\sum_j\lambda_{j,h}^n(1-\ind{v_2\ge q_j})\right]
\Bigg\}d\nu.
\end{aligned}
\tag{L.9}\label{eq:L9}
\end{equation*}
Since $\theta_h^2\to1$, $(1-\theta_h^2)/h\to2/m(p_n)$, and $(\Phi_h(v)-v)/h\to m(v)/m(p_n)$ for $v\ge p_n$, taking the upper limit along the same multiplier subsequence gives
\begin{equation*}
\begin{aligned}
&\limsup_{h\downarrow0}
\frac{L_B(p_n,\phi_n,\lambda_h^n)-L_B(p_n+h,\phi_{n,h},\lambda_h^n)}{h}\\
&\quad\le\frac2{m(p_n)}L_B(p_n,\phi_n,\lambda^n)\\
&\qquad-\frac1{m(p_n)}\int_{p_n\le v_2<v_1}
\big[\phi_n(\{1,2\}\mid v)m(v_1)+\phi_n(\{2\}\mid v)m(v_2)\big]d\nu\\
&\qquad-\sum_j\lambda_j^n\int_{p_n\le v_2<v_1}
\Big[\phi_n(\{1,2\}\mid v)(1-\ind{v_1-v_2\ge x_j})\\
&\hspace{140pt}+\phi_n(\{2\}\mid v)(1-\ind{v_2\ge q_j})\Big]d\nu.
\end{aligned}
\tag{L.10}\label{eq:L10}
\end{equation*}
This completes the separate bounds for the two populations.
\subproof{Combining the two contributions.}
Since $L=L_U+L_B$, the price-optimality inequality \eqref{eq:S13} gives
\begin{equation*}
\begin{aligned}
0\le{}&\frac{L_U(p_n,\phi_n,\lambda_h^n)-L_U(p_n+h,\phi_{n,h},\lambda_h^n)}{h}\\
&+\frac{L_B(p_n,\phi_n,\lambda_h^n)-L_B(p_n+h,\phi_{n,h},\lambda_h^n)}{h}.
\end{aligned}
\end{equation*}
The sum is nonnegative for every sufficiently small $h>0$. Applying the two upper-limit bounds \eqref{eq:L5} and \eqref{eq:L10} along the common multiplier subsequence gives
\begin{equation*}
\begin{aligned}
0\le{}&\left[\frac1{m(p_n)}-\frac{g(p_n)}{G(p_n)}\right]L_U(p_n,\phi_n,\lambda^n)
+\frac2{m(p_n)}L_B(p_n,\phi_n,\lambda^n)\\
&-\frac1{m(p_n)}\int_{v_2<p_n\le v_1}\phi_n(\{1\}\mid v)m(v_1)\,d\nu\\
&-\frac1{m(p_n)}\int_{p_n\le v_2<v_1}
\big[\phi_n(\{1,2\}\mid v)m(v_1)+\phi_n(\{2\}\mid v)m(v_2)\big]d\nu\\
&-\sum_j\lambda_j^n\int_{v_2<p_n\le v_1}\phi_n(\{1\}\mid v)
(1-\ind{v_1\ge q_j})\,d\nu\\
&-\sum_j\lambda_j^n\int_{p_n\le v_2<v_1}\Big[
\phi_n(\{1,2\}\mid v)(1-\ind{v_1-v_2\ge x_j})\\
&\hspace{138pt}+\phi_n(\{2\}\mid v)(1-\ind{v_2\ge q_j})\Big]d\nu.
\end{aligned}
\tag{L.11}\label{eq:L11}
\end{equation*}
The last two terms can be expressed as demand lost after a tested upward deviation. Under the normalized matching, target demand is
\begin{equation*}
\begin{aligned}
\Di(p_n,p_n\mid\phi_n)
={}&\int_{v_2<p_n\le v_1}\phi_n(\{1\}\mid v)\,d\nu\\
&+\int_{p_n\le v_2<v_1}
[\phi_n(\{1,2\}\mid v)+\phi_n(\{2\}\mid v)]d\nu,
\end{aligned}
\end{equation*}
while demand after firm 1 raises its price to $q_j=p_n+x_j$ is
\begin{equation*}
\begin{aligned}
\Di(q_j,p_n\mid\phi_n)
={}&\int_{v_2<p_n\le v_1}\phi_n(\{1\}\mid v)\ind{v_1\ge q_j}\,d\nu\\
&+\int_{p_n\le v_2<v_1}\Big[
\phi_n(\{1,2\}\mid v)\ind{v_1-v_2\ge x_j}\\
&\hspace{100pt}+\phi_n(\{2\}\mid v)\ind{v_2\ge q_j}\Big]d\nu.
\end{aligned}
\end{equation*}
Subtracting gives
\begin{equation*}
\begin{aligned}
&\Di(p_n,p_n\mid\phi_n)-\Di(q_j,p_n\mid\phi_n)\\
&\quad=\int_{v_2<p_n\le v_1}\phi_n(\{1\}\mid v)(1-\ind{v_1\ge q_j})d\nu\\
&\qquad+\int_{p_n\le v_2<v_1}\Big[
\phi_n(\{1,2\}\mid v)(1-\ind{v_1-v_2\ge x_j})\\
&\hspace{132pt}+\phi_n(\{2\}\mid v)(1-\ind{v_2\ge q_j})\Big]d\nu.
\end{aligned}
\end{equation*}
The last two terms of \eqref{eq:L11} therefore equal $-\sum_j\lambda_j^n[\Di(p_n,p_n\mid\phi_n)-\Di(q_j,p_n\mid\phi_n)]$. Substitution yields
\begin{equation*}
\begin{aligned}
0\le{}&\underbrace{\left[\frac1{m(p_n)}-\frac{g(p_n)}{G(p_n)}\right]L_U(p_n,\phi_n,\lambda^n)
+\frac2{m(p_n)}L_B(p_n,\phi_n,\lambda^n)}_{\text{(25a)}}\\
&-\underbrace{\frac1{m(p_n)}\Bigg\{\begin{aligned}
&\int_{v_2<p_n\le v_1}\phi_n(\{1\}\mid v)m(v_1)\,d\nu\\
&+\int_{p_n\le v_2<v_1}
[\phi_n(\{1,2\}\mid v)m(v_1)
+\phi_n(\{2\}\mid v)m(v_2)]d\nu
\end{aligned}\Bigg\}}_{\text{(25b)}}\\
&-\underbrace{\sum_j\lambda_j^n\big[\Di(p_n,p_n\mid\phi_n)-\Di(q_j,p_n\mid\phi_n)\big]}_{\text{(25c)}}.
\end{aligned}
\tag{L.12}\label{eq:L12}
\end{equation*}
We simplify (25a) and (25c) using optimality and complementary slackness. Since $\phi_n$ and $\lambda^n$ are primal and dual optima at $p_n$, finite duality gives
\begin{equation*}
 L_U(p_n,\phi_n,\lambda^n)+L_B(p_n,\phi_n,\lambda^n)
 =\frac{SS(\phi_n,(p_n,p_n))}{2}.
\end{equation*}
Hence, using $m(p_n)g(p_n)=1-G(p_n)$,
\begin{equation*}
\begin{aligned}
\text{(25a)}
={}&\frac2{m(p_n)}[L_U(p_n,\phi_n,\lambda^n)+L_B(p_n,\phi_n,\lambda^n)]\\
&-\left[\frac1{m(p_n)}+\frac{g(p_n)}{G(p_n)}\right]L_U(p_n,\phi_n,\lambda^n)\\
={}&\frac{SS(\phi_n,(p_n,p_n))}{m(p_n)}
-\frac{L_U(p_n,\phi_n,\lambda^n)}{m(p_n)G(p_n)}.
\end{aligned}
\tag{L.13}\label{eq:L13}
\end{equation*}
Complementary slackness also implies, for every $j$,
\begin{equation*}
\lambda_j^n\big[p_n\Di(p_n,p_n\mid\phi_n)-q_j\Di(q_j,p_n\mid\phi_n)\big]=0.
\end{equation*}
Thus, whenever $\lambda_j^n>0$, the corresponding upward-deviation constraint binds. Consequently
\begin{equation*}
\begin{aligned}
\text{(25c)}
&=\sum_j\lambda_j^n\big[\Di(p_n,p_n\mid\phi_n)-\Di(q_j,p_n\mid\phi_n)\big]\\
&=\Di(p_n,p_n\mid\phi_n)\sum_j\lambda_j^n\left(1-\frac{p_n}{q_j}\right)\\
&=\kappa_n\Di(p_n,p_n\mid\phi_n).
\end{aligned}
\tag{L.14}\label{eq:L14}
\end{equation*}
All individual deviation multipliers now enter this term only through $\kappa_n$. Substituting \eqref{eq:L13}--\eqref{eq:L14} into \eqref{eq:L12} and multiplying by $m(p_n)$ yields
\begin{equation*}
\begin{aligned}
&\underbrace{\begin{aligned}
&SS(\phi_n,(p_n,p_n))
-\int_{v_2<p_n\le v_1}\phi_n(\{1\}\mid v)m(v_1)\,d\nu\\
&-\int_{p_n\le v_2<v_1}
[\phi_n(\{1,2\}\mid v)m(v_1)+\phi_n(\{2\}\mid v)m(v_2)]d\nu\\
&-m(p_n)\Di(p_n,p_n\mid\phi_n)
\end{aligned}}_{\text{(28a)}}\\
&\qquad\ge\underbrace{\frac{L_U(p_n,\phi_n,\lambda^n)}{G(p_n)}}_{\text{(28b)}}
+(\kappa_n-1)m(p_n)\Di(p_n,p_n\mid\phi_n).
\end{aligned}
\tag{L.15}\label{eq:L15}
\end{equation*}
It remains to bound (28a) from above and (28b) from below. These bounds remove the detailed matching probabilities and leave only aggregate demand at $p_n$.

\subproof{Bounding the two sides of \eqref{eq:L15}.}
Under the normalized matching, social surplus is
\begin{equation*}
\begin{aligned}
SS(\phi_n,(p_n,p_n))
={}&2\int_{v_2<p_n\le v_1}\phi_n(\{1\}\mid v)v_1\,d\nu\\
&+2\int_{p_n\le v_2<v_1}
[\phi_n(\{1,2\}\mid v)v_1+\phi_n(\{2\}\mid v)v_2]d\nu.
\end{aligned}
\tag{L.16}\label{eq:L16}
\end{equation*}
Substituting \eqref{eq:L16} and the corresponding target-demand expression into (28a) gives
\begin{equation*}
\begin{aligned}
\text{(28a)}
={}&\int_{v_2<p_n\le v_1}\phi_n(\{1\}\mid v)
[2v_1-m(v_1)-m(p_n)]d\nu\\
&+\int_{p_n\le v_2<v_1}\Big\{
\phi_n(\{1,2\}\mid v)[2v_1-m(v_1)-m(p_n)]\\
&\hspace{95pt}+\phi_n(\{2\}\mid v)[2v_2-m(v_2)-m(p_n)]\Big\}d\nu.
\end{aligned}
\end{equation*}
Although we do not know which consumers are matched with which products, the increasing-hazard property allows us to bound this expression independently of their detailed assignments. The calculation proceeds in small steps:
\begin{equation*}
\begin{aligned}
\text{(28a)}
&\le\int_{v_2<p_n\le v_1}\phi_n(\{1\}\mid v)[2v_1-m(v_1)-m(p_n)]d\nu\\
&\quad+\int_{p_n\le v_2<v_1}[\phi_n(\{1,2\}\mid v)+\phi_n(\{2\}\mid v)]
[2v_1-m(v_1)-m(p_n)]d\nu\\[4pt]
&=[2p^*-m(p^*)-m(p_n)]\Di(p_n,p_n\mid\phi_n)\\
&\quad+\int_{v_2<p_n\le v_1}\phi_n(\{1\}\mid v)
[2(v_1-p^*)-m(v_1)+m(p^*)]d\nu\\
&\quad+\int_{p_n\le v_2<v_1}[\phi_n(\{1,2\}\mid v)+\phi_n(\{2\}\mid v)]\\
&\hspace{75pt}\times[2(v_1-p^*)-m(v_1)+m(p^*)]d\nu\\[4pt]
&\le [2p^*-m(p^*)-m(p_n)]\Di(p_n,p_n\mid\phi_n)\\
&\quad+\int_{\substack{v_2<v_1\\v_1\ge p^*}}
[2(v_1-p^*)-m(v_1)+m(p^*)]d\nu\\[4pt]
&=[2p^*-m(p^*)-m(p_n)]\Di(p_n,p_n\mid\phi_n)\\
&\quad+\int_{p^*}^{\vbar}[2(v-p^*)-m(v)+m(p^*)]g(v)G(v)\,dv\\[4pt]
&=[2p^*-m(p^*)-m(p_n)]
[\Di(p_n,p_n\mid\phi_n)-\Di(p^*,p^*\mid\fmat)]\\
&\quad+\int_{p^*}^{\vbar}[2v-m(v)-m(p_n)]g(v)G(v)\,dv\\[4pt]
&\le\int_{p^*}^{\vbar}[2v-m(v)-m(p_n)]g(v)G(v)\,dv\\[4pt]
&=\int_{p^*}^{\vbar}[2vg(v)G(v)-S(v)G(v)]\,dv
-m(p_n)\Di(p^*,p^*\mid\fmat)\\[4pt]
&=\int_{p^*}^{\vbar}vg(v)\,dv-[vG(v)S(v)]_{p^*}^{\vbar}
-m(p_n)\Di(p^*,p^*\mid\fmat)\\[4pt]
&=\int_{p^*}^{\vbar}vg(v)\,dv+p^*G(p^*)S(p^*)
-m(p_n)\Di(p^*,p^*\mid\fmat).
\end{aligned}
\tag{L.17}\label{eq:L17}
\end{equation*}
We justify the steps in this chain. Since $m$ is nonincreasing by the increasing-hazard property, $2v-m(v)$ is increasing. Thus replacing a purchased less-preferred value $v_2$ by the higher value $v_1$ gives the first inequality. Next, separate the benchmark coefficient $2p^*-m(p^*)-m(p_n)$. The residual function $2(v_1-p^*)-m(v_1)+m(p^*)$ is negative below $p^*$ and nonnegative above it. Dropping negative contributions and bounding purchasing probabilities by one gives the next inequality.

Integrating over $v_2$ gives $g(v)G(v)$. Under full matching, firm 1 sells to the higher-valued consumers above $p^*$, so its demand is $\int_{p^*}^{\vbar}g(v)G(v)\,dv$. By \eqref{eq:LC9} and the increasing-hazard property, $p^*\le m(p^*)\le m(p_n)$ because $p_n<p^*$. Thus $2p^*-m(p^*)-m(p_n)\le0$. Together with $\Di(p_n,p_n\mid\phi_n)>\Di(p^*,p^*\mid\fmat)$ from Step 2, this justifies dropping the benchmark-demand difference.

Finally, using $S(v)=1-G(v)$ and $m(v)g(v)=S(v)$, the last lines follow from
\begin{equation*}
 \frac{d}{dv}[vG(v)S(v)]=G(v)S(v)+vg(v)[1-2G(v)]
\end{equation*}
and $S(\vbar)=0$.

For the lower bound on (28b), write $M(p)=\int_p^{\vbar}vg(v)\,dv$. Recall the definition of $\Gamma$ before Lemma~\ref{lem:social-price-perturbation}. By \eqref{eq:LC9} and the increasing-hazard property, $p_n<p^*\le m(p^*)\le m(p_n)$. Thus the monopoly revenue function has a positive derivative at $p_n$. Thus $\Gamma(p_n)>0$, and its supremum is unchanged if restricted to $p_n<q\le\vbar$.

Recall from Step 3 that $h_n(v)=v+p_n\sum_j\lambda_j^n-\sum_jq_j\lambda_j^n\ind{v\ge q_j}$. The optimal matching maximizes the Lagrangian pointwise, so the contribution from a consumer with exactly one affordable product is the positive part of its purchasing score. Consequently,
\begin{equation*}
\begin{aligned}
\text{(28b)}
&=\frac{L_U(p_n,\phi_n,\lambda^n)}{G(p_n)}\\
&=\int_{p_n}^{\vbar}[h_n(v)]_+g(v)\,dv\\
&\ge\int_{p_n}^{\vbar}h_n(v)g(v)\,dv\\
&=M(p_n)+\sum_j\lambda_j^n[p_nS(p_n)-q_jS(q_j)]\\
&\ge M(p_n)-\Gamma(p_n)\sum_j\lambda_j^n\frac{q_j-p_n}{q_j}\\
&=M(p_n)-\kappa_n\Gamma(p_n).
\end{aligned}
\tag{L.18}\label{eq:L18}
\end{equation*}
The first inequality uses $[h_n(v)]_+\ge h_n(v)$. The second follows from the definition of $\Gamma(p_n)$, which implies
\begin{equation*}
 q_jS(q_j)-p_nS(p_n)\le\Gamma(p_n)\frac{q_j-p_n}{q_j}.
\end{equation*}
This removes the individual multipliers in favor of $\kappa_n$. Combining the upper bound \eqref{eq:L17} with the lower bound \eqref{eq:L18} in \eqref{eq:L15} gives
\begin{equation*}
\begin{aligned}
&\int_{p^*}^{\vbar}vg(v)\,dv+p^*G(p^*)S(p^*)
-m(p_n)\Di(p^*,p^*\mid\fmat)\\
&\quad\ge M(p_n)-\kappa_n\Gamma(p_n)
+(\kappa_n-1)m(p_n)\Di(p_n,p_n\mid\phi_n).
\end{aligned}
\end{equation*}
Since $M(p_n)-\int_{p^*}^{\vbar}vg(v)\,dv=\int_{p_n}^{p^*}vg(v)\,dv$, rearranging yields
\begin{equation*}
\begin{aligned}
&(\kappa_n-1)[m(p_n)\Di(p_n,p_n\mid\phi_n)-\Gamma(p_n)]\\
&\quad\le\Gamma(p_n)-m(p_n)\Di(p^*,p^*\mid\fmat)
+p^*G(p^*)S(p^*)-\int_{p_n}^{p^*}vg(v)\,dv.
\end{aligned}
\end{equation*}
This is exactly \eqref{eq:key}, proving Lemma~\ref{lem:social-price-perturbation}.\nobreak\hfill\quad$\square$

\subsubsection*{Proof of Lemma~\ref{lem:social-distributional}}\phantomsection\label{app:proof-social-distributional}
\begingroup
\edef\sonesavedequation{\number\value{equation}}
\setcounter{equation}{0}
\renewcommand{\theequation}{S.1.\arabic{equation}}
\renewcommand{\theHequation}{distributional.\arabic{equation}}
\begin{proof}[\unskip\nopunct]
\begingroup
\def\D{D}\def\C{C}\def\dd{\,\mathrm d}
\def\Df{D_1(p^*,p^*\mid\phi^{\mathrm{full}})}\def\pf{\underline p}
Maintain the manuscript's assumptions: $g$ is a continuously differentiable, log-concave density on $[0,\bar v]$, positive on $(0,\bar v)$, $G$ is its distribution function, $S=1-G$, and $m(v)=S(v)/g(v)$. The benchmark $p^*$ is the full-matching duopoly price. Recall
\[
 \Gamma(p)=\sup_{q>p}\frac{q}{q-p}[qS(q)-pS(p)],\qquad
 B(v)=v\left[g(v)+\int_v^{\bar v}g(w)^2\dd w\right],
\]
where $S(q)=0$ beyond the support, and the price floor satisfies $B(\pf)=\Df$.

\step{Normalize the two comparisons}
Fix $p\in[\pf,p^*)$. The lemma requires $\Gamma(p)$ to be smaller than two quantities: the benchmark term $m(p)\Df$, and that term plus the integral adjustment
$\int_p^{p^*}vg(v)\dd v-p^*G(p^*)S(p^*)$.

Write
\[
 r=\frac{p}{m(p)},\qquad
 \gamma=\frac{\Gamma(p)}{S(p)m(p)},
\]
and define the two normalized benchmarks
\[
 \D=\frac{\Df}{S(p)},\qquad
 \C=\frac{m(p)\Df+\int_p^{p^*}vg(v)\dd v-p^*G(p^*)S(p^*)}{S(p)m(p)}.
\]
Here $0<r<1$ by $p<p^*\le m(p^*)\le m(p)$, using \eqref{eq:LC9} and monotonicity of $m$, and $S(p)m(p)>0$. Since $(qS(q))'|_{q=p}=S(p)(1-r)>0$, nearby upward deviations have positive gain. Prices $q\ge\bar v$ have zero demand and negative gain, so the supremum defining $\Gamma(p)$ can equivalently be restricted to $p<q\le\bar v$.

Dividing the two required inequalities by $S(p)m(p)$, our goal becomes
\begin{equation}\label{polishedS1:goal}
 \gamma<\min\{\D,\C\}.
\end{equation}
It is enough to show
\begin{equation}\label{polishedS1:rows}
\begin{array}{ll}
 \gamma\le1/4<\min\{\D,\C\},&0<r\le1/6,\\[3pt]
 \gamma\le Q(r)<\min\{\D,\C\},&1/6\le r<1,
\end{array}
\end{equation}
where $Q(r)=1/e-21r/100$.

Throughout the proof, write $u=p^*/m(p)$ and $y=S(p^*)/S(p)$.

Part A proves $\D>1/4$ for $0<r\le1/6$ and $\D>Q(r)$ for $1/6\le r<1$. Part B proves $\gamma\le1/4$ and $\gamma\le Q(r)$, and Part C proves $\C>1/4$ and $\C>Q(r)$, in these respective ranges. These bounds imply \eqref{polishedS1:rows}.

\step{A. Bound the unadjusted benchmark}
We first show a few results that will be useful later.

\subproof{Claim A.1.} Let $\alpha(r)=e^{r-1}-e^{2r-2}/2$. We have
\begin{align}
 &0<r<u\le1,\quad 0<S(p)<1,\quad 0<y<1,\label{polishedS1:range}\\
 &S(p)\ge e^{-r},\qquad y\ge e^{r/u-1}\ge e^{r-1}.\label{polishedS1:common}
\end{align}
Moreover, $\alpha$ is nondecreasing on $[0,1]$.

\subproof{Proof of Claim A.1.}
Because $0<p<p^*\le m(p^*)\le m(p)$, we have $0<r<u\le1$; positivity of the density gives the other ranges. Equation~\eqref{eq:LC2} gives $S(p)\ge e^{-r}$, and \eqref{eq:LC1} gives
\[
 y\ge \exp\left(-\frac{p^*-p}{m(p^*)}\right)
 \ge \exp\left(-\frac{p^*-p}{p^*}\right)=e^{r/u-1}.
\]
Since $u\le1$, this is at least $e^{r-1}$.

Finally,
\[
 \alpha'(r)=e^{r-1}(1-e^{r-1})\in[0,1/4]\qquad(0\le r\le1),
\]
so $\alpha$ is nondecreasing.

\subproof{Claim A.2.} We have $\D\ge\alpha(r)$, and
\[
 \begin{array}{ll}
 \D>1/4,&0<r\le1/6,\\[3pt]
 \D>Q(r),&1/6\le r<1.
 \end{array}
\]

\subproof{Proof of Claim A.2.}
Using the full-matching demand formula in Lemma~\ref{lem:full-matching-benchmark},
\[
 \D=y(1-S(p)y/2)\ge y-y^2/2\ge\alpha(r).
\]
The first inequality uses $S(p)\le1$. The second follows because $y-y^2/2$ is increasing on $[0,1]$.

For $0<r\le1/6$, $\D\ge\alpha(r)\ge\alpha(0)>1/4$.

For $1/6\le r<1$, monotonicity of $\alpha$ gives $\D\ge\alpha(1/6)>1/3$, whereas $Q(r)\le Q(1/6)<1/3$.

\step{B. Bound the deviation gain}
\subproof{Claim B.1.} For every $0<r<1$,
\begin{equation}\label{polishedS1:gain}
 \gamma\le Q(r).
\end{equation}

\subproof{Proof of Claim B.1.}
For an upward deviation $p<q<\bar v$, write $x=(q-p)/m(p)>0$ and $T=S(q)/S(p)$. By \eqref{eq:LC1}, $T\le e^{-x}$. The normalized deviation gain before taking the supremum is
\begin{equation}\label{polishedS1:kernel}
 \gamma(r,x,T)=\frac{r+x}{x}[(r+x)T-r]
 =xT+(2T-1)r-\frac{(1-T)r^2}{x}.
\end{equation}
It is increasing in $T$, since $\partial\gamma(r,x,T)/\partial T=(r+x)^2/x>0$. Consequently $\gamma\le\sup_{x>0}\gamma(r,x,e^{-x})$. Let $c=21/100$ and $\lambda=\log(2/(1-c))$. Expanding,
\begin{equation}\label{polishedS1:quadratic}
 \gamma(r,x,e^{-x})+cr=-\frac{1-e^{-x}}x r^2+(2e^{-x}-(1-c))r+xe^{-x}.
\end{equation}
If $x\ge\lambda$, both terms involving $r\ge0$ are nonpositive, so \eqref{polishedS1:quadratic} is at most $xe^{-x}\le1/e$.

If $0<x\le\lambda$, the expression is a concave quadratic in $r$, so its maximum over all real $r$ gives
\begin{align*}
 \gamma(r,x,e^{-x})+cr
 &\le xe^{-x}+\frac{x(2e^{-x}-(1-c))^2}{4(1-e^{-x})}\\
 &=\frac{(1-c)^2x}{4}+\frac{(1+c)^2}{4}\frac{x}{e^x-1}=:f(x).
\end{align*}
Since $x/(e^x-1)$ is convex on $x>0$, with continuous extension $1$ at zero, $f$ attains its maximum on $[0,\lambda]$ at an endpoint.\footnote{For $x>0$, the second derivative of $x/(e^x-1)$ has the sign of $(x-2)e^x+x+2$. This expression and its first derivative vanish at zero, while its second derivative is $xe^x\ge0$. The extension at zero follows from $(e^x-1)/x\to1$.} At $\lambda$, the linear coefficient in \eqref{polishedS1:quadratic} vanishes, so $f(\lambda)=\lambda e^{-\lambda}\le1/e$. At zero, $f(0)=(1+c)^2/4<1/e$.
Thus $\gamma(r,x,e^{-x})+cr\le1/e$ for every $x>0$. Taking the supremum proves \eqref{polishedS1:gain}.

\subproof{Claim B.2.} If $0<r\le1/6$, then $\gamma\le1/4$.

\subproof{Proof of Claim B.2.}
For $0\le v\le\bar v$, write $I(v)=\int_v^{\bar v}g(w)^2\dd w$. Step 1 proves $I(p)\ge S(p)g(p)$; Step 2 then uses this inequality to show $\gamma\le1/4$.

\begin{samepage}
\subproof{Step 1.} We first show
\begin{equation}\label{polishedS1:energy}
 I(p)\ge S(p)g(p).
\end{equation}
\end{samepage}

By definition, $B(v)=v[g(v)+I(v)]$ and $B(\pf)=\Df$. Differentiating and then using \eqref{eq:LC3} gives
\begin{align*}
 B'(v)&=g(v)+I(v)+vg'(v)-vg(v)^2\\
 &\ge g(v)\left[1-\frac{v}{m(v)}(1+S(v))\right]+I(v).
\end{align*}
The function $v/m(v)$ is nondecreasing, so for $v\in[\pf,p]$ we have $v/m(v)\le r\le1/6$. Since $S(v)\le1$ and $I(v)>0$, the displayed derivative is positive. Consequently $B(p)\ge B(\pf)=\Df$. Dividing this inequality by $S(p)$ and using $p/S(p)=r/g(p)$ gives
\[
 \D=\frac{B(\pf)}{S(p)}\le\frac{B(p)}{S(p)}
 =r\left(1+\frac{I(p)}{g(p)}\right).
\]
The derivative bound in Claim A.1 implies that $\alpha(r)-2r$ is decreasing. Since $\alpha(1/6)>1/3$, it follows that $\alpha(r)>2r$ for $0\le r\le1/6$. Combining this with Claim A.2 yields
\[
 2r<\alpha(r)\le\D\le r\left(1+\frac{I(p)}{g(p)}\right).
\]
Since $r>0$, we conclude $I(p)/g(p)>1\ge S(p)$, which implies \eqref{polishedS1:energy}.

\enlargethispage{2pt}
\subproof{Step 2.} If \eqref{polishedS1:energy} holds, then $\gamma\le1/4$.

Let $\xi=S^{-1}$ and $J(w)=g(\xi(w))$ for $0\le w\le1$. For $0<w<1$, differentiation gives
\[
 \xi'(w)=-\frac1{J(w)},\qquad
 J'(w)=-(\log g)'(\xi(w)).
\]
Because $\xi$ and $(\log g)'$ are both nonincreasing, their composition $(\log g)'\circ\xi$ is nondecreasing. Thus $J'$ is nonincreasing on $(0,1)$, so $J$ is concave there and, by continuity, on $[0,1]$. This uses no second derivative of $g$.\footnote{Quantile concavity is standard: see S. G. Bobkov, \emph{Extremal properties of half-spaces for log-concave distributions}, Annals of Probability 24 (1996), 35--48, Proposition A.1(a),(c), pp.45--46; \href{https://www-users.cse.umn.edu/~bobko001/papers/1996_AOP_Extremal.pdf}{author-hosted published paper}. The short verification above covers the present assumptions; the ensuing energy argument is proved here.} Substitution $w=S(v)$ also gives
\[
 I(p)=\int_0^{S(p)}J(w)\dd w,\qquad J(S(p))=g(p).
\]
Jensen's inequality on $[0,S(p)]$, followed by \eqref{polishedS1:energy}, yields
\[
 J(S(p)/2)\ge\frac1{S(p)}\int_0^{S(p)}J(w)\dd w\ge J(S(p)).
\]
The chord joining the values at $S(p)/2$ and $S(p)$ is at least $J(S(p))$. The chord joining $0$ and $S(p)/2$, using $J(0)\ge0$, is at least $2wJ(S(p))/S(p)$. Concavity therefore gives
\begin{equation}\label{polishedS1:chords}
 J(w)\ge
 \begin{cases}
 2wJ(S(p))/S(p),&0\le w\le S(p)/2,\\
 J(S(p)),&S(p)/2\le w\le S(p).
 \end{cases}
\end{equation}
These profile lower bounds give upper bounds on deviation prices through their normalized increments, which can be inserted into the normalized gain $\gamma(r,x,T)$.

For $p<q<\bar v$, write $x=(q-p)/m(p)$ and $T=S(q)/S(p)\in(0,1)$. Since $m(p)J(S(p))=S(p)$, reciprocal integration of \eqref{polishedS1:chords} gives
\begin{equation}\label{polishedS1:spacing}
 x=\frac1{m(p)}\int_{S(p)T}^{S(p)}\frac{\dd w}{J(w)}\le
 \begin{cases}
 1-T,&T\ge1/2,\\[2pt]
 \displaystyle\frac12+\frac12\log\frac1{2T},&T\le1/2.
 \end{cases}
\end{equation}
For the second branch, the part from $S(p)/2$ to $S(p)$ contributes at most $1/2$, and the part from $S(p)T$ to $S(p)/2$ contributes at most
$\frac1{m(p)}\int_{S(p)T}^{S(p)/2}S(p)/[2wJ(S(p))]\dd w=\frac12\log(1/(2T))$.

For fixed $r\ge0$ and $0<T<1$, the normalized gain \eqref{polishedS1:kernel} satisfies
\[
 \frac{\partial\gamma(r,x,T)}{\partial x}=T+\frac{(1-T)r^2}{x^2}>0.
\]
If $T\ge1/2$,
\[
 \gamma(r,x,T)\le\gamma(r,1-T,T)=(r+1-T)(T-r)\le\frac14,
\]
where the two factors sum to one.

If $0<T\le1/2$, the two terms involving $r$ in \eqref{polishedS1:kernel} are nonpositive. The second bound in \eqref{polishedS1:spacing} therefore gives
\[
 \gamma(r,x,T)\le xT\le\frac T2\left(1+\log\frac1{2T}\right)\le\frac14.
\]
For the last inequality, $T[1+\log(1/(2T))]/2$ has derivative $\frac12\log(1/(2T))\ge0$ on $(0,1/2]$ and equals $1/4$ at $T=1/2$. Taking the supremum over upward deviations proves the claim.

\step{C. Bound the adjusted benchmark in both ranges of $r$}
We first show a useful result in Claim C.1, then use it in Claim C.2 to prove $\C>1/4$ for $0<r\le1/6$ and $\C>Q(r)$ for $1/6\le r<1$.

\subproof{Claim C.1.} We have
\begin{equation}\label{polishedS1:geometric}
 \C\ge r(1-\sqrt y)+u\sqrt y+(1-2u)(y-S(p)y^2/2),
\end{equation}
as well as $y>9/25$ and $S(p)>5/6-r/2$.

\subproof{Proof of Claim C.1.}
Integration by parts gives
\[
 \int_p^{p^*}vg(v)\dd v
 =pS(p)-p^*S(p^*)+\int_p^{p^*}S(v)\dd v.
\]
Since $G(p^*)=1-S(p)y$, substituting this identity into $\C$ yields
\begin{align*}
 \C&=y(1-S(p)y/2)+r-uy+\frac1{S(p)m(p)}\int_p^{p^*}S(v)\dd v-uy(1-S(p)y)\\
 &=r+\frac1{S(p)m(p)}\int_p^{p^*}S(v)\dd v+(1-2u)y(1-S(p)y/2).
\end{align*}
The log-concavity of $S$ and \eqref{eq:LC7} give $S(p+(p^*-p)\tau)/S(p)\ge y^\tau$ for $0\le\tau\le1$. Substituting $v=p+(p^*-p)\tau$, we obtain
\[
 \frac1{S(p)m(p)}\int_p^{p^*}S(v)\dd v
 =(u-r)\int_0^1\frac{S(p+(p^*-p)\tau)}{S(p)}\dd\tau
 \ge(u-r)\int_0^1y^\tau\dd\tau.
\]
Pairing $\tau$ and $1-\tau$ and applying the arithmetic--geometric mean inequality,
\[
 \int_0^1y^\tau\dd\tau
 =\frac12\int_0^1(y^\tau+y^{1-\tau})\dd\tau\ge\sqrt y.
\]
Substituting this bound proves \eqref{polishedS1:geometric}.
By \eqref{polishedS1:common}, $y\ge e^{r/u-1}>1/e>9/25$. Also, the tangent inequality for $e^{-r}$ at $2/3$ gives
\begin{equation}\label{polishedS1:stangent}
 S(p)\ge e^{-r}\ge e^{-2/3}(5/3-r)>\frac56-\frac r2,
\end{equation}
where $e^{-2/3}>1/2$ and $5/3-r>0$.

\subproof{Claim C.2.} We have
\[
 \begin{array}{ll}
 \C>1/4,&0<r\le1/6,\\[3pt]
 \C>Q(r),&1/6\le r<1.
 \end{array}
\]

\subproof{Proof of Claim C.2.}
For $1/6\le r<1$, it is enough to show $\C\ge11/30-r/5$, since
\[
 \frac{11}{30}-\frac r5-Q(r)
 =\frac{11}{30}-\frac1e+\frac r{100}
 \ge\frac{221}{600}-\frac1e>0.
\]
For $u\le1/2$, since $S(p)\le1$ and $1-2u\ge0$, \eqref{polishedS1:geometric} gives
\[
 \C\ge(1-2u)[r(1-\sqrt y)+y-y^2/2]+2u[r(1-\sqrt y)+\sqrt y/2].
\]
\subproof{Case 1: $u\le1/2$ and $0<r\le1/6$.}
Claims A.1--A.2 and $r(1-\sqrt y)\ge0$ give
\[
 r(1-\sqrt y)+y-y^2/2\ge\alpha(r)\ge\alpha(0)>\frac14,
 \qquad r(1-\sqrt y)+\sqrt y/2>\frac14.
\]
The second bound uses $y>9/25$ from Claim C.1. Their convex combination is strictly above $1/4$.

\subproof{Case 2: $u\le1/2$ and $1/6\le r<1$.}
We have $r/u\ge1/3$. Hence $y\ge e^{r/u-1}\ge e^{-2/3}>1/2$. Monotonicity of $y-y^2/2$ on $[0,1]$ gives $r(1-\sqrt y)+y-y^2/2>3/8$. Also,
\[
 r(1-\sqrt y)+\sqrt y/2\ge\frac16(1-\sqrt y)+\frac{\sqrt y}{2}
 =\frac16+\frac{\sqrt y}{3}>\frac{11}{30}.
\]
Both expressions therefore exceed $11/30-r/5$, and their convex combination proves the bound in this case.

For $u\ge1/2$, define
\[
 F(r,y)=r+(1-r)\sqrt y-y+\left(\frac5{12}-\frac r4\right)y^2.
\]

For every $0<r<1$, Claim C.1 gives $S(p)>5/6-r/2$. Since $1-2u\le0$, \eqref{polishedS1:geometric} implies
\begin{equation}\label{polishedS1:uinterpolation}
 \C\ge(2-2u)[r(1-\sqrt y)+\sqrt y/2]+(2u-1)F(r,y).
\end{equation}
The weights are nonnegative and sum to one.

\subproof{Case 3: $u\ge1/2$ and $1/6\le r<1$.}
As in Case 2, $r(1-\sqrt y)+\sqrt y/2\ge1/6+\sqrt y/3>11/30$.

To bound $F$, we verify its three $r$-endpoints, including $r=0$ for use in Case 4. For $1/4\le y\le1$,
\begin{equation}\label{polishedS1:certificate}
\begin{aligned}
 24[F(1/6,y)-1/3]&=9y^2-24y+20\sqrt y-4\\
 &=\bigl[3(1-\sqrt y)^2+(2\sqrt y-1)^2\bigr]^2\\
 &\quad+20(2\sqrt y-1)(1-\sqrt y)^3>0.
\end{aligned}
\end{equation}
Both terms are nonnegative on $[1/4,1]$, and the square is strictly positive because $1-\sqrt y$ and $2\sqrt y-1$ cannot both vanish. Consequently
\begin{equation}\label{polishedS1:Ffirst}
\begin{aligned}
 F(1/6,y)&>\frac13,\\
 F(0,y)&=F(1/6,y)-\frac16+\frac{\sqrt y}{6}+\frac{y^2}{24}>\frac14.
\end{aligned}
\end{equation}
For the second comparison, use $y\ge1/4$. At the remaining endpoint,
\[
 F(1,y)=1-y+\frac16y^2\ge\frac16,
\]
because its derivative $-1+y/3$ is nonpositive on $[0,1]$, and its value at one is $1/6$.

Since $F$ is affine in $r$, for $1/6\le r\le1$,
\begin{align*}
 F(r,y)
 &=\frac{6(1-r)}5 F(1/6,y)+\frac{6r-1}5 F(1,y)\\
 &\ge\frac{6(1-r)}5\cdot\frac13+\frac{6r-1}5\cdot\frac16
 =\frac{11}{30}-\frac r5.
\end{align*}
The actual $y>9/25$ belongs to $[1/4,1]$, so these bounds apply. Combining them with \eqref{polishedS1:uinterpolation} proves the bound in this case.

\subproof{Case 4: $u\ge1/2$ and $0<r\le1/6$.}
For $0\le r\le1/6$, affine interpolation and \eqref{polishedS1:Ffirst} give
\[
 F(r,y)=(1-6r)F(0,y)+6rF(1/6,y)>\frac14.
\]
Both endpoint values strictly exceed $1/4$. Also $r(1-\sqrt y)+\sqrt y/2>1/4$ by $y>9/25$. Substitution into \eqref{polishedS1:uinterpolation} proves the bound in this case.

These bounds establish \eqref{polishedS1:rows}, proving \eqref{polishedS1:goal}. Multiplication by $S(p)m(p)>0$ gives the strict inequality in Lemma S.1.
\endgroup
\end{proof}
\setcounter{equation}{\sonesavedequation}
\endgroup
\setcounter{equation}{\technicalsavedequation}
\endgroup

\subsubsection*{Weighted-Average Exponential Inequality}

\begin{lemma}[Weighted-Average Exponential Inequality]\label{lem:weighted-average-exponential}
Let $\omega\in(0,1)$, and suppose
\[
t\ge\log\frac{2-\omega}{2(1-\omega)},\qquad
k\ge2t,\qquad
\omega t+(1-\omega)k\ge1.
\]
Then, for every $z\ge1$,
\[
\omega e^{-t(z-1)}+(1-\omega)e^{-k(z-1)}\le\frac1z.
\]
\end{lemma}

\begin{proof}[Proof of Lemma~\ref{lem:weighted-average-exponential}]
Equality holds at $z=1$, so fix $z>1$. For fixed $t$, the expression on the left decreases in $k$. We may therefore take the smallest admissible value,
\[
k=\max\left\{2t,\frac{1-\omega t}{1-\omega}\right\}.
\]
Along the branch $k=2t$, the expression decreases with $t$. Along the other branch, its derivative with respect to $t$ is
\[
-\omega(z-1)\left[e^{-t(z-1)}-e^{-k(z-1)}\right]\le0.
\]
The two branches agree where they meet. It therefore suffices to establish the result when
\begin{equation}\label{eq:weighted-exponential-reduced-rates}
t=\log\frac{2-\omega}{2(1-\omega)},\qquad
k=\max\left\{2t,\frac{1-\omega t}{1-\omega}\right\}.
\end{equation}

First, we show how the desired exponential bound follows from a simpler inequality. Integration gives
\[
e^{-(z-1)t}
=1-(z-1)t
+(z-1)^2\int_0^\infty e^{-(z-1)u}(t-u)_+\,du,
\]
and the same identity holds with $k$ in place of $t$. Taking their weighted average gives
\begin{equation}\label{eq:weighted-exponential-integral-identity}
\begin{aligned}
&\omega e^{-t(z-1)}+(1-\omega)e^{-k(z-1)}\\
&\quad=1-(z-1)[\omega t+(1-\omega)k]\\
&\qquad\quad +(z-1)^2\int_0^\infty e^{-(z-1)u}
\left[\omega(t-u)_++(1-\omega)(k-u)_+\right]du.
\end{aligned}
\end{equation}
Suppose we can establish
\begin{equation}\label{eq:weighted-exponential-mean-one}
\omega t+(1-\omega)k=1
\end{equation}
and
\begin{equation}\label{eq:weighted-exponential-call-bound}
\omega(t-u)_++(1-\omega)(k-u)_+\le e^{-u}
\qquad\text{for every }u\ge0.
\end{equation}
Substituting these into \eqref{eq:weighted-exponential-integral-identity}, we obtain
\[
\begin{aligned}
\omega e^{-t(z-1)}+(1-\omega)e^{-k(z-1)}
&\le1-(z-1)+(z-1)^2\int_0^\infty e^{-zu}\,du\\
&=1-(z-1)+\frac{(z-1)^2}{z}
=\frac1z.
\end{aligned}
\]
Thus \eqref{eq:weighted-exponential-mean-one} and \eqref{eq:weighted-exponential-call-bound} suffice for the desired result.

Next, we show that \eqref{eq:weighted-exponential-call-bound} follows from a simpler condition. If $0\le u\le t$, \eqref{eq:weighted-exponential-mean-one} gives
\[
\omega(t-u)+(1-\omega)(k-u)=1-u\le e^{-u}.
\]
If $u\ge k$, the left-hand side of \eqref{eq:weighted-exponential-call-bound} is zero. Finally, for $t<u<k$, consider the difference
\[
(1-\omega)(k-u)-e^{-u}.
\]
Its derivative in $u$ is $-(1-\omega)+e^{-u}$ and its second derivative is $-e^{-u}<0$. Its global maximum occurs at $u=-\log(1-\omega)$, where it equals
\[
(1-\omega)[k+\log(1-\omega)-1].
\]
Consequently, \eqref{eq:weighted-exponential-call-bound} holds whenever
\begin{equation}\label{eq:weighted-exponential-rate-condition}
k+\log(1-\omega)\le1.
\end{equation}
We have therefore reduced the proof to verifying the two properties
\[
\omega t+(1-\omega)k=1,\qquad k+\log(1-\omega)\le1.
\]
We verify them in the two cases from \eqref{eq:weighted-exponential-reduced-rates}.

\emph{Case 1: $k=2t$.} Here
\[
\omega t+(1-\omega)k=(2-\omega)t\ge1.
\]
Reduce both rates proportionally until their weighted average equals one. This only increases the exponential expression. The resulting rates are
\[
t=\frac1{2-\omega},\qquad k=\frac2{2-\omega}.
\]
The mean-one condition holds by construction, and
\[
-\log(1-\omega)\ge\omega\ge\frac{\omega}{2-\omega}
\]
implies
\[
k+\log(1-\omega)=\frac2{2-\omega}+\log(1-\omega)\le1.
\]
Hence both sufficient conditions hold for the reduced rates, proving the result in this case.

\emph{Case 2: $k=(1-\omega t)/(1-\omega)>2t$.} The weighted mean is exactly one. Substituting the expressions for $k$ and $t$ from \eqref{eq:weighted-exponential-reduced-rates}, we have
\[
\begin{aligned}
&k+\log(1-\omega)\le1\\
&\quad\Longleftrightarrow\quad
\frac{1-\omega t}{1-\omega}+\log(1-\omega)\le1\\
&\quad\Longleftrightarrow\quad
\omega t-(1-\omega)\log(1-\omega)-\omega\ge0\\
&\quad\Longleftrightarrow\quad
\omega\log\frac{2-\omega}{2}-\log(1-\omega)-\omega\ge0.
\end{aligned}
\]
Denote the last left-hand side by $\psi(\omega)$. Direct differentiation gives
\[
\psi(0)=\psi'(0)=0,\qquad
\psi''(\omega)=
\frac{\omega(\omega^2-5\omega+5)}{(1-\omega)^2(2-\omega)^2}\ge0,
\qquad 0<\omega<1.
\]
Thus $\psi'(\omega)\ge0$ and $\psi(\omega)\ge0$. This proves \eqref{eq:weighted-exponential-rate-condition}. Both sufficient conditions therefore hold in this case as well, completing the proof.
\end{proof}

\clearpage
{\small
	\addcontentsline{toc}{section}{References}
	\setlength{\bibsep}{0.25\baselineskip}
	\bibliographystyle{jpe}
	\bibliography{references_Sep17}
}

\end{document}